\documentclass[aps,prd,longbibliography,nofootinbib]{revtex4-2}

\usepackage{graphicx}
\usepackage{amsmath,amssymb,mathtools}
\usepackage{paralist}
\usepackage{comment} 
\usepackage{booktabs}
\usepackage[breaklinks=true,colorlinks=true,linkcolor=magenta,citecolor=blue,urlcolor=blue]{hyperref}

\usepackage{tikz}
\usepackage{pgfplots}
\pgfplotsset{compat=1.15}
\usepgfplotslibrary{groupplots}

\newcommand{\Mpl}{M_{\mathrm{Pl}}}

\newcommand{\dd}{\mathrm{d}}

\newcommand{\Msun}{M_{\odot}}
\newcommand{\PhN}{\Phi_{\rm N}}
\newcommand{\mulin}{\mu_{\rm lin,0}}

\begin{document}

\title{Solar System Experiments in the Search for Dark Energy and Dark Matter}

\author{Slava G. Turyshev}
\affiliation{Jet Propulsion Laboratory, California Institute of Technology,\\
4800 Oak Grove Drive, Pasadena, CA 91109-0899, USA}

\date{\today}

\begin{abstract}

We reassess the realistic discovery reach of Solar‑System experiments for dark energy (DE) and dark matter (DM), making explicit the bridge from cosmology‑level linear responses to local, screened residuals. In scalar–tensor frameworks with a universal conformal coupling $A(\phi)$ and chameleon/Vainshtein screening, we map cosmological responses $\{\mu(z,k),\Sigma(z,k)\}$ inferred by DESI and \emph{Euclid} to thin–shell or Vainshtein residuals in deep Solar potentials $\Phi_N$. We emphasize a two‑branch strategy. In a \emph{detection‑first} branch, a verified local anomaly—an Einstein equivalence principle (EEP) violation, a Shapiro‑delay signal with $|\gamma-1|\sim\mathrm{few}\times 10^{-6}$, an AU‑scale Yukawa tail, or a narrowband ultralight DM (ULDM) line in clocks/atom interferometers in space (AIS)—triggers a joint refit of cosmology and Solar‑System data under a common microphysical parameterization $\{V(\phi),A(\phi)\}$. In a \emph{guardrail} branch, Solar‑System tests enforce universal constraints (EEP; PPN parameters $\gamma,\beta$; and $\dot G/G$) and close unscreened or weakly screened corners indicated by cosmology. Under realistic near‑term conditions we forecast, per conjunction, $|\gamma-1|\lesssim (2\mbox{--}5)\times 10^{-6}$ (Ka-/X‑band or optical Shapiro), $\eta_{\rm EEP}\sim (1\mbox{--}10)\times 10^{-17}$ (drag‑free AIS), $|\dot G/G|\sim(3$--$5)\times10^{-15}\,\mathrm{yr^{-1}}$ (sub-mm-class lunar laser ranging (LLR)), a uniform $\simeq 2\times$ tightening of AU‑scale Yukawa/DM‑density bounds, and $(3\mbox{--}10)\times$ improved ULDM‑coupling reach from clock networks. For a conformal benchmark, $\mu_{\rm lin,0}=0.10$ implies $\chi\simeq \sqrt{\mu_{\rm lin,0}/2}$ and a Sun thin shell $\Delta R/R\lesssim (1/3\chi)\sqrt{|\gamma-1|/2}=2.4\times 10^{-3}$ at $|\gamma-1|=5\times 10^{-6}$; Vainshtein screening at 1~AU yields $|\gamma-1|\lesssim 10^{-11}$, naturally below near‑term reach. We recommend a cost‑effective guardrail+discovery portfolio with explicit triggers for escalation to dedicated missions.

\end{abstract}

\maketitle

\small
\tableofcontents
\normalsize

\section{Introduction and scope}
\label{sec:intro} 

Cosmology and Solar System experiments probe gravity and the dark sector in complementary regimes. Large-scale surveys determine the expansion history and the growth/deflection of structure and are therefore the natural place to uncover percent-level departures from general relativity (GR) in linear response functions for clustering and lensing, commonly summarized by $\mu(z,k)$ and $\Sigma(z,k)$
as studied by cosmological probes (e.g.,  DESI\footnote{For details on the Dark Energy Spectroscopic Instrument (DESI), see \url{https://www.desi.lbl.gov/}}, Euclid\footnote{For details on the  ESA's Euclid mission, see \url{https://www.esa.int/Science_Exploration/Space_Science/Euclid}}), that jointly constrain geometry and growth  \cite{DESI2024BAO,DESI2024FS,EuclidQ1}. Solar System measurements, by contrast, operate in deep gravitational potentials with exquisite metrology and perform two tasks exceptionally well: they impose model-agnostic guardrails that any explanation of late-time acceleration or dark matter must satisfy—Einstein's equivalence principle (EEP), the parameterized post-Newtonian (PPN) parameters $\gamma$ (space curvature per unit mass) and $\beta$ (nonlinearity of superposition), and bounds on $\dot G/G$—and they open selective discovery windows for concrete hypotheses, such as ultralight fields that modulate clock frequencies or long-range Yukawa tails at astronomical-unit (AU) scales \cite{Will2014LRR,Hees2016,Wcislo2018,Talmadge1988PRL,Pitjeva2013,Pitjev2013}. 

The theoretical context that links large-scale structure to Solar System tests is screening. In scalar--tensor frameworks with a universal conformal matter coupling $A(\phi)$, environmental screening (chameleon/symmetron for potential/threshold screening; Vainshtein for derivative screening) suppresses fifth forces in deep potentials, reconciling cosmological signatures with tight local tests \cite{KhouryPRD,KhouryPRL,HinterbichlerKhouryPRL2010,BurrageSaksteinLRR,Koyama2016,BabichevDeffayet2013CQG}. Multimessenger observations of the gravitational-wave (GW) event GW170817 and the gamma-ray-burst (GRB) counterpart GRB170817A require the tensor propagation speed to match that of light to high precision, $|c_{\tt T}/c-1|\lesssim 10^{-15}$, thereby removing broad modified-gravity classes or pushing them close to the GR limit on large scales \cite{Baker2017,Creminelli2017,EzquiagaZumalacarregui2017}. Within the surviving classes we adopt a hypothesis-driven strategy: starting from a specified microphysical hypothesis (mediator and couplings), we map cosmology-level responses $(\mu,\Sigma)$ onto concrete Solar System residuals via the standard screening relations (thin shell for chameleon-like models; Vainshtein residual scalings for derivative screening). We then ask whether realistic, systematics-vetted Solar System measurements clear those residuals with margin.

The approach  used below is standard:
The EEP asserts universality of free fall; it is quantified by the Eötvös parameter $\eta$, with the MICROSCOPE mission reporting $|\eta|\sim3\times 10^{-15}$ \cite{Touboul2022}. In the PPN framework, $\gamma$ measures space curvature per unit mass and $\beta$ encodes nonlinearity; GR predicts $\gamma=\beta=1$ \cite{Will2014LRR}. The Shapiro time delay in solar conjunction measures $\gamma$ cleanly \cite{Bertotti2003}, while $\beta$ is constrained in combination with $\gamma$ by global ephemerides. Lunar Laser Ranging (LLR) constrains $\dot G/G$ and the strong-equivalence principle (SEP) and, together with planetary ephemerides, provides competitive limits on long-range departures from GR \cite{Williams2004,Pitjeva2021,Fienga2024,Turyshev2025}. Ephemerides also bound any smooth Solar System DM density and test Yukawa tails \(V(r)=-(GM/r)[1+\alpha_{\tt Y} e^{-r/\lambda}]\) at AU scales \cite{Fienga2024,Pitjeva2013,Pitjev2013}. Precision clocks and atom interferometers in space (AIS), especially with space-assisted links, search for ultralight dark matter (ULDM) through narrowband, coherence-limited modulations \cite{Hees2016,Wcislo2018,Cacciapuoti2009_ACES,Delva2017_FiberSR}. 

Operationally, the measurement program is asymmetric and reflects this complementarity. Multi-probe cosmology (e.g., DESI full-shape+BAO and Euclid weak lensing/clustering) determines the posterior in $\{w(z),\mu(z,k),\Sigma(z,k)\}$ at percent-level precision \cite{DESI2024BAO,DESI2024FS,EuclidQ1}. Screening then maps those posteriors into predicted Solar System residuals at levels set by the relevant gravitational potentials $\Phi_N$, Table~\ref{tab:potentials}. In chameleon-like models, for example, a cosmology-level excess $\mu_{\rm lin,0}\equiv \mu(z{=}0,k_{\rm fid})-1$ fixes a local slope $\chi\simeq\sqrt{\mu_{\rm lin,0}/2}$ on linear scales; the thin-shell relations then map $\{\mu_{\rm lin,0},\chi\}$ to concrete requirements on the Sun’s (or Earth’s) thin shell and hence to a target sensitivity in $\gamma$ (via the Shapiro test) or in $\eta$ (via EEP tests), with the corresponding ambient-density rescalings made explicit later. Conversely, a null result at the forecast sensitivity prunes the cosmologically viable subspace unless screening enforces sufficiently small residuals; a detection in any channel triggers a joint refit across regimes with the same microphysical parameters.

Our objective is to assemble today's leading constraints; to place realistic near-term sensitivities on a common, systematics-vetted footing; and to make the cosmology$\rightarrow$Solar System bridge explicit using standard screening relations. Concretely, we place Ka/X radio and deep-space optical links (Shapiro $\gamma$), sustained mm-class LLR ($\dot G/G$, SEP), refined ephemerides (Yukawa tails and smooth $\rho_{\rm DM}$), and clock/AIS for ULDM on the same quantitative scale.

The scope is limited to three questions: whether Solar System experiments (i) enforce model-agnostic guardrails
(EEP; PPN $\gamma,\beta$; $\dot G/G$), (ii) exclude unscreened or weakly screened regions implied by cosmology, and
(iii) open stand-alone discovery windows for dark-sector hypotheses (for example, ultralight dark matter or AU-scale
Yukawa tails). A verified Solar System anomaly in any of these channels is sufficient to trigger targeted follow-on
work irrespective of cosmology, while in the null case Solar System guardrails sharply prune theory space. We adopt
this two-branch strategy throughout: detection-first (stand-alone discovery) and guardrails (model-agnostic nulls).

The structure of this paper is as follows.  Section~\ref{sec:cosmo} summarizes DESI/Euclid constraints and the linear-response parameters $\mu(z,k)$ and $\Sigma(z,k)$ that we use as the cosmology$\rightarrow$local bridge. Section~\ref{sec:theory} sets the theoretical priors (screening, EFT/positivity, GW‑speed bound) and defines the notation; it develops the bridge from cosmological to local phenomenology, including thin-shell and Vainshtein numerics.
Section~\ref{sec:bridge} makes the connection between  cosmology and local experiments. Section~\ref{sec:observables} then reviews current Solar System bounds and near-term targets (EEP, PPN $\gamma,\beta$, $\dot G/G$, Yukawa tails, ULDM clocks/AIS) and introduces the local thin‑shell and Vainshtein mappings adjacent to the observables they constrain, thereby avoiding forward references. Section~\ref{sec:program} lays out a targeted near-term program (radio/optical conjunctions, mm-class LLR, optical clock links, and ephemerides) together with a risk register and a decision rule for dedicated missions. 
Section~\ref{sec:conclusions} closes with the asymmetric strategy---cosmology for discovery; Solar System tests for guardrails and selective DM discovery. Appendices~\ref{app:disformal}--\ref{app:joint} extend the guardrails to disformal couplings and provide a toy cross-regime likelihood. Appendix~\ref{sec:recipe} provides a step-by-step roadmap that collects the assumptions, equations, and data products used throughout. Appendix~\ref{sec:crosscheck} demonstrates that the Sun’s thin-shell behavior yields a PPN-$\gamma$ compatible with Cassini constraints. Appendix~\ref{sec:ULDM-broader} surveys an extended ultralight DM model space and the associated Solar System observables. Appendix~\ref{app:ULDM} consolidates the detailed signal templates, derivations, and quantitative coupling-reach forecasts.

\section{Recent cosmology results and modified gravity}
\label{sec:cosmo}

\subsection{Cosmological probes and linear response}
\label{sec:cosmo}

Cosmological surveys determine the background expansion and linear growth/deflection, commonly summarized by $\{\mu(z,k),\Sigma(z,k)\}$. We will use the shorthand
\[
  \mu_{\rm lin,0}\equiv \mu(z=0,k_{\rm fid})-1,\qquad 
  k_{\rm fid}\simeq 0.1\,h\,\mathrm{Mpc}^{-1}.
\]
Solar System experiments probe the screened, high‑$\Phi_N$ regime with exquisite metrology and therefore enforce model‑agnostic guardrails (EEP; PPN $\gamma,\beta$; $\dot G/G$) while opening selective discovery channels (e.g., ULDM via clocks/AIS). Screening provides the bridge from $\{\mu,\Sigma\}$ to local residuals in deep potentials; we make this mapping explicit and quantitative. Here we briefly review the recent cosmology results (e.g., DESI, Euclid). 

\subsubsection{DESI}

The Dark Energy Spectroscopic Instrument (DESI) first data release (DR1) delivers baryon acoustic oscillation (BAO) distance measurements with sub-percent precision across $0.1<z<4.2$, using bright galaxy, luminous red galaxy, emission-line galaxy, quasar, and Lyman-$\alpha$ tracers. BAO constraints on the transverse and radial distance combinations, $D_{\rm M}(z)/r_{\rm d}$ and $D_{\rm H}(z)/r_{\rm d}$, tighten the late-time expansion history; when combined with the cosmic microwave background (CMB) and Big-Bang nucleosynthesis (BBN) priors, extended models show a mild preference for\footnote{Cosmology notation: We use $w_0,w_a$ for the CPL dark-energy equation-of-state $w(a)=w_0+w_a(1-a)$, $\Omega_m$ for the present-day matter fraction, $\sigma_8$ for the rms fluctuation in $8\,h^{-1}\,\mathrm{Mpc}$ spheres, and $k$ for comoving wavenumber.} $w_0>-1$ and $w_a<0$, while $\Lambda$CDM remains statistically viable~\cite{DESIScience,DESI2024BAO,DESI2024FS}. 

Beyond BAO, DESI’s full-shape (FS) analysis of the power spectrum, including redshift-space distortions (RSD), jointly constrains geometry and growth\footnote{Here BAO denotes baryon acoustic oscillation; FS denotes full-shape; RSD denotes redshift-space distortions; CMB denotes cosmic microwave background; BBN denotes Big-Bang nucleosynthesis; MG denotes modified gravity; GR denotes General Relativity.}. In the flat $\Lambda$CDM model, DESI (FS+BAO) with a BBN prior measures $\Omega_{\rm m}=0.2962\pm0.0095$ and $\sigma_8=0.842\pm0.034$, while adding CMB data sharpens these to $\Omega_{\rm m}=0.3056\pm0.0049$ and $\sigma_8=0.8121\pm0.0053$; inclusion of external clustering+lensing (DES\,Y3) yields a $0.4\%$ determination of the Hubble parameter, 
$H_0=(68.40\pm0.27)\,\mathrm{km\,s^{-1}\,Mpc^{-1}}$~\cite{DESI2024FS}. In models with time-varying dark-energy equation of state, DESI (FS+BAO) combined with CMB and supernovae retains the DR1 BAO preference for $w_0>-1,\,w_a<0$ at similar significance~\cite{DESI2024FS}. 

DESI also reports constraints on phenomenological modified-gravity (MG) functions that rescale the Poisson equation and lensing, commonly summarized by $\mu_0$ and $\Sigma_0$ at $z=0$. DESI data alone measure $\mu_0=0.11^{+0.45}_{-0.54}$; DESI+\,CMB+\,DES\,Y3 give $\mu_0=0.04\pm0.22$ and $\Sigma_0=0.044\pm0.047$, consistent with GR~\cite{DESI2024FS}. The same analysis places an upper limit on the summed neutrino mass of $\sum m_\nu<0.071\,\mathrm{eV}$ (95\%\,CL)~\cite{DESI2024FS}. 

\subsubsection{Euclid}

The European Space Agency’s \emph{Euclid} mission Quick Data Release~1 (Q1; March 2025) provides end-to-end validated imaging and spectroscopy over three deep fields totaling $\simeq 63\,\mathrm{deg}^2$, with source catalogs that demonstrate survey-quality photometry, morphology, and point-spread function control across the VIS and NISP instruments\footnote{Here VIS denotes the visible imager; NISP denotes the near-infrared spectrometer and photometer.}~\cite{EuclidQ1,ESA_Euclid_PR}. The Q1 products include tens of millions of galaxy detections (order $10^7$) spanning look-back times to $\sim 10.5$\,Gyr, and early catalogs of strong-lensing systems and clusters that preview the science yield as area accumulates~\cite{EuclidQ1,ESA_Euclid_PR}. As the wide survey grows toward the $\gtrsim 15{,}000\,\mathrm{deg}^2$ goal, cosmology-grade weak-lensing shear and clustering data products will enable precision tests of gravitational slip $\eta\equiv\Phi/\Psi$, scale-dependent growth $f\sigma_8(k,z)$, and consistency with $\Lambda$CDM+GR at the sub-percent level in the two-point statistics~\cite{EuclidQ1}. 

One may introduce the linear slip $\eta_{\rm slip}\equiv \Phi/\Psi$, so that $\Sigma=\mu(1+\eta_{\rm slip})/2$.
In GR (negligible anisotropic stress) $\eta_{\rm slip}=1$ and $\Sigma=\mu=1$. In the conformal benchmark adopted
later, $\eta_{\rm slip}\simeq 1$ on linear scales, so $\Sigma$ tracks $\mu$ up to order-unity factors. We implement the
single-parameter bridge in Sec.~\ref{sec:bridge}.

As an example of cross-regime inference consistent with our philosophy here, \cite{Benisty:2023fRCombined} jointly fits $f(R)$ models to late-time cosmology (BAO/SNe) and to S2-orbit data at Sgr~A*, illustrating how cosmology-level posteriors can be combined with local strong-field dynamics.

\subsection{Theory and screening regimes}
\label{sec:theory}

Connecting predictions made at cosmological mean densities, $\bar\rho \sim 10^{-29}\,\mathrm{g\,cm^{-3}}$, to Solar System environments spanning (i) near-Sun interplanetary plasma along conjunction rays with  $\rho \sim 10^{-22}$--$10^{-19}\,\mathrm{g\,cm^{-3}}$, and (ii) bulk planetary/stellar matter with $\rho \sim 1$--$10^{2}\,\mathrm{g\,cm^{-3}}$, requires a controlled theoretical bridge. In scalar-tensor theories with a universal (conformal) matter coupling $A(\phi)$, one introduces the environment-dependent effective potential
\begin{equation}
V_{\rm eff}(\phi;\rho) \equiv V(\phi) + \rho\,A(\phi),
\qquad
\frac{\dd V_{\rm eff}}{\dd \phi}\bigg|_{\phi_\star} = V'(\phi_\star) + \rho\,A'(\phi_\star) = 0,
\label{eq:Veff}
\end{equation}
so that the ambient density selects $\phi_\star(\rho)$ and hence the local coupling ($'$ denotes $d/d\phi$);  $\rho$ is the Jordan-frame rest-mass density. Screening mechanisms---chameleon and symmetron (potential/threshold screening) and Vainshtein or other derivative screening---were developed to reconcile cosmic acceleration or modified growth with stringent small-scale tests~\cite{KhouryPRD,KhouryPRL,HinterbichlerKhouryPRL2010,BurrageSaksteinLRR,Koyama2016}. The multimessenger observations GW170817/GRB170817A imply that the tensor propagation speed, $c_{\tt T}$, must agree with that of light to high precision, $|c_{\tt T}/c - 1| \lesssim 10^{-15}$, thereby eliminating broad model classes or pushing them toward GR-like limits on large scales~\cite{Abbott2017GW,Abbott2017MM,Baker2017,Creminelli2017,EzquiagaZumalacarregui2017}. Throughout, a specified microphysical model means an explicit dark-sector hypothesis with fixed low-energy dynamics. In scalar-tensor frameworks this corresponds to a concrete pair $\{V(\phi),A(\phi)\}$; for DM it corresponds to an explicit coupling structure or operator.

In what follows, Solar System tests are treated as hypothesis-driven probes rather than model-blind searches. Specifically, we require (i) a specified microphysical model (a concrete mediator and couplings) that fixes $\{V(\phi),A(\phi)\}$ in scalar-tensor theories, or the corresponding coupling structure for dark matter; and (ii) at least one forecasted local signature---such as a target for the PPN parameters $\gamma-1,\ \beta-1$ ($\gamma$ measures spatial curvature per unit mass; $\beta$ encodes nonlinearity of superposition; $\gamma=\beta=1$ in GR, \cite{Will2014LRR}), tests for $\dot G/G$, a Yukawa amplitude $\alpha_{\tt Y}(\lambda)$, or a clock modulation tied to the EEP---that exceeds a credible threshold given Solar System gravitational potentials. When these conditions are met, the resulting measurements function as decisive tests.

To avoid unconstrained function fitting, we adopt minimal theoretical assumptions standard for low-energy long-range sectors: (1) effective field theory (EFT) locality and analyticity with approximate symmetries controlling small parameters; (2) radiative stability (technical naturalness) of light fields and couplings; (3) positivity and causality bounds compatible with a healthy ultraviolet (UV) completion (e.g., amplitude-positivity constraints \cite{Adams2006JHEP}); (4) consistency of the adopted screening mechanism across Newtonian gravitational potentials $\Phi_N \sim 10^{-11}$--$10^{-6}$ relevant to Solar System bodies (see Table~\ref{tab:potentials}); and (5) the GW speed constraint quoted above. 

For canonical parameters placing the solar Vainshtein radius $r_{V\odot}\sim 10^2\,\mathrm{pc}$, the fractional force residual inside the Vainshtein region scales as
$(\delta F/F)(r) \propto (r/r_V)^{3/2}$ for $r \ll r_V$~\cite{BabichevDeffayet2013CQG}, where $r$ is heliocentric distance along the probe trajectory and $\delta F/F$ is the scalar-to-Newtonian force ratio. At $r\simeq 1\,\mathrm{AU}$ this yields $(\delta F/F)\sim (1\,\mathrm{AU}/100\,\mathrm{pc})^{3/2}\sim 10^{-11}$, explaining why PPN signatures can be null even if cosmology shows percent-level growth anomalies. Under these assumptions, screening does not erase all effects; it leaves residual, environment-dependent signatures that are calculable and therefore falsifiable in Solar System settings~\cite{BurrageSaksteinLRR,Koyama2016}.
These priors restrict admissible low-energy couplings but do not uniquely determine $A(\phi)$.  

For forecasts  we adopt the canonical conformal form $A(\phi)=\exp(\chi \phi/M_{\rm Pl})$, motivated by radiative stability and a universal matter coupling; other analytic choices map to the same leading predictions with $\chi$ reinterpreted as the local slope $A'(\phi_\star)/A(\phi_\star)$. Note that here and below we take $\chi \equiv M_{\rm Pl}\,\partial\ln A/\partial\phi$ evaluated at the ambient field value $\phi_\star$ set by the local density, i.e.\ the \emph{dimensionless} coupling; thus $\partial\ln A/\partial\phi=\chi/M_{\rm Pl}$ for $A(\phi)=\exp(\chi\phi/M_{\rm Pl})$. When we write “$\chi=\mathrm{const}$”, interpret this as the local slope $A'(\phi_\star)/A(\phi_\star)$ appropriate for the observable under consideration.

\begin{table}[t]
\centering
\caption{Surface Newtonian potentials $\PhN \equiv GM/(Rc^2)$ (dimensionless) relevant for screening.}
\label{tab:potentials}
\begin{tabular}{lcc}
\hline
Body & Mass & $\Phi_N$ \\
\hline\hline
Sun & $\Msun$ & $2.12\times 10^{-6}$ \\
Jupiter & $9.55\times 10^{-4}\,\Msun$ & $1.9\times 10^{-8}$ \\
Earth & $3.00\times 10^{-6}\,\Msun$ & $6.96\times 10^{-10}$ \\
Moon & $3.69\times 10^{-8}\,\Msun$ & $3.14\times 10^{-11}$ \\
\hline
\end{tabular}
\end{table}

For chameleon-like screening, consider a spherical body of mass $M$, radius $R$, and surface Newtonian potential  $\Phi_N \equiv GM/(R c^{2})$, see Table~\ref{tab:notations} for notations. In an ambient environment where the scalar field takes the value $\phi_\infty$, while inside the body it relaxes to $\phi_c$, the object develops a thin shell of fractional thickness\footnote{\label{foot:conv}Conventions \& priors (used throughout): We adopt $c=\hbar=1$ unless noted otherwise; where dimensionless normalizations are standard (e.g., $\Phi_N\equiv GM/(Rc^2)$), we follow conventional astrophysical units. We use the reduced Planck mass $M_{\rm Pl}\equiv 1/\sqrt{8\pi G}$. Screening follows the thin‑shell (chameleon/symmetron) and Vainshtein prescriptions in Eqs.~(\ref{eq:thin-shell})--(\ref{eq:PhiN-numbers})  and (\ref{eq:rV})--(\ref{eq:gamma_Vain}); the GW‑speed bound satisfies $|c_{\tt T}/c-1|\lesssim10^{-15}$. Linear responses are $\mu(z,k)$ and $\Sigma(z,k)$ with $\mu_{\rm lin,0}\!\equiv\!\mu(z\!=\!0,k_{\rm fid})-1$, $k_{\rm fid}\!\simeq\!0.1\,h\,{\rm Mpc}^{-1}$. Unless stated, the interplanetary ambient density along conjunction rays is $\rho_\infty\in[10^{-22},10^{-19}]\,{\rm g\,cm^{-3}}$ and rescalings follow Table~\ref{tab:ambient-density}.
 
Throughout the paper the scalar field carries its canonical mass dimension, $[\phi]={\rm mass}$. We define the \emph{dimensionless} conformal matter coupling (possibly composition-dependent) evaluated at the ambient field value $\phi_\star$ as 
\begin{equation*}
\chi(\phi)\equiv M_{\rm Pl}\,\Big(\frac{\partial\ln A(\phi)}{\partial\phi}\Big)\Big|_{\phi_\star},
\end{equation*}
so that when $A(\phi)=\exp[{\chi\,\phi}/{M_{\rm Pl}}]$ with constant $\chi$, the local slope is $(\partial\ln A/\partial\phi)_{\phi_\star}=\chi/M_{\rm Pl}$. Throughout the text we use the symbol $\chi$ for this dimensionless coupling. With this convention Eqs.~(\ref{eq:thin-shell})--(\ref{eq:PhiN-numbers}) and (\ref{eq:chi_from_mu})--(\ref{eq:phi_contrast_bound}) are dimensionally consistent and coincide with the standard chameleon notation. Also, for a universal coupling one has $\chi_A=\chi_B=\chi$.}
{}
\begin{equation}
\frac{\Delta R}{R} \simeq \frac{\phi_\infty - \phi_c}{6\,\chi\,\Mpl\,\Phi_N},
\label{eq:thin-shell}
\end{equation}
where $\chi$ is the dimensionless matter coupling and $\Mpl$ is the reduced Planck mass.\footnote{From Eq.~(\ref{eq:thin-shell}) one has $\Delta R/R \propto \Phi_N^{-1}$ at fixed ambient $\phi_\infty$, so bodies with smaller surface potentials develop larger shell fractions. Using Table~\ref{tab:potentials}, the Earth and, especially, the Moon can enter the ``no thin shell'' regime ($3\,\Delta R/R \gtrsim 1$) for the same $\phi_\infty$ that yields $\Delta R/R \sim 10^{-3}$ for the Sun; conversely, present EEP/LLR nulls favor a screened-Earth branch at $1$\,AU (Sec.~\ref{sec:EEP}, Fig.~\ref{fig:eta-guardrail}). The classification also depends on the ambient density chosen for each body via $\phi_\infty=\phi_\star(\rho_\infty)$ in Eq.~(\ref{eq:phi_min_powerlaw}).}
 
\begin{table*}[t]
\centering
\caption{Notation summary (symbols used most often in the text).}
\label{tab:notations}
\begin{tabular}{ll}
\hline
Symbol & Meaning \\
\hline\hline
$\Mpl$ & reduced Planck mass \\
$A(\phi)=\exp(\chi\phi/\Mpl)$ & conformal matter coupling; $\chi \equiv \Mpl\,(\partial\ln A/\partial\phi)|_{\phi_\star}$ \\
$r_S=2GM/c^2$ & Schwarzschild radius \\
$r_V$ & Vainshtein radius, see (\ref{eq:rV}) \\
$\mu(z,k),\,\Sigma(z,k)$ & linear-response functions for clustering and lensing \\
$\mu_{\rm lin,0}$ & shorthand $\mu(z=0,k_{\rm fid})-1$ with $k_{\rm fid}\simeq 0.1\,h\,\mathrm{Mpc}^{-1}$ \\
$\alpha_{\tt Y},\,\lambda$ & Yukawa strength and range; $m=\hbar c/\lambda$ \\
ULDM & ultralight dark matter \\
$m_\phi$ & ULDM (or mediator) mass \\
$v$ & characteristic virial speed ($v/c\simeq 10^{-3}$ for the Galactic halo) \\
$t_c$ & ULDM coherence time $t_c \simeq 2\pi/(m_\phi v^2)$ \\
AU & astronomical unit \\
\hline
\end{tabular}
\end{table*}

Eq.~\eqref{eq:thin-shell} assumes a screened source with a \emph{thin} shell, $\Delta R/R\ll 1$. When $\Delta R/R$ approaches unity, the screened‑source premise fails and the
$\gamma$--mapping in Eq.~\eqref{eq:gamma_from_shell} no longer applies; such parameter points are automatically excluded by Solar System bounds (cf. Appendix~\ref{sec:crosscheck}).
For the environment-dependent dilaton, approximate analytic solutions in laboratory and spherical geometries are given in \cite{BraxFischerKaedingPitschmann2022}, which underlie several current LLR and lab constraints.

For the benchmark potential $V(\phi)=\Lambda^{4+n}\phi^{-n}$ with $A(\phi)=e^{\chi \phi/M_{\rm Pl}}$, the density-dependent minimum solves
$V'(\phi_\star)+\rho\,A'(\phi_\star)=0$ from (\ref{eq:Veff}), giving
\begin{equation}
\phi_\star(\rho)=\Big(\frac{n\,\Lambda^{4+n} M_{\rm Pl}}{\chi\,\rho}\Big)^{\!\frac{1}{n+1}},
\label{eq:phi_min_powerlaw}
\end{equation}
where we used $A(\phi_\star)\simeq 1$ for $\chi\,\phi_\star/M_{\rm Pl}\ll 1$ (equivalently, interpret $\chi$ as the local slope $A'(\phi_\star)/A(\phi_\star)$). Substituting $\phi_\infty=\phi_\star(\rho_\infty)$ and $\phi_c=\phi_\star(\rho_c)$ in (\ref{eq:thin-shell}) yields an explicit $(n,\Lambda,\chi)$
dependence for the thin shell:
\begin{equation}
\frac{\Delta R}{R}\;\simeq\;\frac{\phi_\star(\rho_\infty)-\phi_\star(\rho_c)}{6\,\chi\,M_{\rm Pl}\,\Phi_N}\,.
\label{eq:thin_shell_powerlaw}
\end{equation}
In the Sun-screened regime ($\rho_c\!\gg\!\rho_\infty$ so $\phi_c\!\ll\!\phi_\infty$), this simplifies to
$\Delta R/R \simeq \phi_\star(\rho_\infty)/\!\big(6\,\chi\,M_{\rm Pl}\,\Phi_N\big)$.

We model the near-Sun interplanetary medium along conjunction rays as log-normal, $\ln\rho_\infty\sim\mathcal{N}(\ln\rho_0,\sigma^2_{\ln\rho})$ with $\rho_0=10^{-20}\,\mathrm{g\,cm^{-3}}$ and $\sigma_{\ln\rho}=\ln 10$ (one-decade $1\sigma$), consistent with Table~\ref{tab:ambient-density}. For the power-law chameleon in (\ref{eq:phi_min_powerlaw}), 
$\Delta R/R \propto \rho_\infty^{-1/(n+1)},$
so uncertainties propagate as $\sigma_{\ln(\Delta R/R)}=\sigma_{\ln\rho}/(n+1)$. Here we adopt the conservative prior $\rho_\infty\in[10^{-22},10^{-19}]\,\mathrm{g\,cm^{-3}}$ “a few $R_\odot$” from the Sun and report sensitivities as explicit power laws in $\rho_\infty$ (Table~\ref{tab:ambient-density}).

Note that for forecasting purposes, we adopt a fiducial interplanetary medium near conjunction
$\rho_\infty \equiv \rho_{\rm IPM} \in [10^{-22},10^{-19}]~{\rm g~cm^{-3}}$ (consistent with proton densities $n_p\sim 10^{2}$--$10^{5}\,{\rm cm^{-3}}$ at a few $R_\odot$), and we present bounds as explicit power laws in $\rho_\infty$, that for the power-law chameleon $V(\phi)=\Lambda^{4+n}\phi^{-n}$ and universal conformal coupling $A(\phi)=\exp[\chi\phi/M_{\rm Pl}]$, take the form
{}
\begin{equation}
\Big(\frac{\Delta R}{R}\Big)_{\rho_\infty} \propto \rho_\infty^{-1/(n+1)},\qquad
\Lambda^{4+n}(\rho_\infty)\lesssim \frac{\chi \;\rho_\infty\,}{n\,M_{\rm Pl}}\Big(M_{\rm Pl}\Phi_{N\odot}\sqrt{|\gamma-1|_{\max}}\Big)^{n+1},
\label{eq:thin-shell-L}
\end{equation}
where $\gamma$ is the PPN parameter with the current value reported  in (\ref{eq:shapiro}). Thus, one can rescale Eqs.~(\ref{eq:thin-shell-bound})--(\ref{eq:Lambda_bound}) below to any preferred $\rho_\infty$ without re-deriving intermediate steps. As a result, this yields the compact rescaling form of the bound derived from the thin-shell relation and the Shapiro-delay null test (cf.\ Eqs.~\eqref{eq:thin-shell-bound} and~\eqref{eq:phi_contrast_bound} below).

\begin{table}[t]
\centering
\caption{Effect of the ambient-density prior on the thin-shell bound for a power-law chameleon ($n$ shown). Entries show the multiplicative rescaling of $(\Delta R/R)_{\max}$ relative to a reference $\rho_{\infty,0}=10^{-20}\,\mathrm{g\,cm^{-3}}$.
Because $\Delta R/R\propto \rho_\infty^{-1/(n+1)}$ for the power-law chameleon in (\ref{eq:phi_min_powerlaw}), the entries equal $(\rho_\infty/\rho_{\infty,0})^{-1/(n+1)}$ for the listed $n$ values.}
\label{tab:ambient-density}
\begin{tabular}{@{}cccc@{}}
\hline
$n$ & $\rho_\infty=10^{-22}$ & $10^{-20}$ (ref.) & $10^{-19}$ \\
\hline\hline
1 & $(10^{-22}/10^{-20})^{-1/2}=10$~~ & $1$ & $(10^{-19}/10^{-20})^{-1/2}\!=\!0.316$ \\
2 & $(10^{-22}/10^{-20})^{-1/3}=4.64$ & $1$ & $(10^{-19}/10^{-20})^{-1/3}\!=\!0.464$ \\
3 & $(10^{-22}/10^{-20})^{-1/4}=3.16$ & $1$ & $(10^{-19}/10^{-20})^{-1/4}\!=\!0.562$ \\
\hline
\end{tabular}
\end{table}

For conjunction analyses we adopt an explicit $b$-dependent prior,
\begin{equation}
\rho_\infty(b)=\rho_0\left(\frac{b}{b_0}\right)^{-s},\qquad
\rho_0=10^{-20}\,{\rm g\,cm^{-3}},\quad b_0=5\,R_\odot,\quad s\in[2,4]\,,
\end{equation}
which is consistent with radio-science TEC in the near-Sun corona (used later in (\ref{eq:cassini-gamma})).
For the power-law chameleon in (\ref{eq:phi_min_powerlaw}) one then has the explicit rescaling
\begin{equation}
\left(\frac{\Delta R}{R}\right)_{\!\max}(b)\propto
\left[\rho_\infty(b)\right]^{-1/(n+1)}=
\left(\frac{b}{b_0}\right)^{\,s/(n+1)}\!,
\end{equation}
so that, e.g., for $n=2$ and $s=3$, moving from $b=5\,R_\odot$ to $b=3\,R_\odot$ tightens
$(\Delta R/R)_{\max}$ by a factor $(3/5)^{3/3}\simeq 0.6$. We use this $b$-dependent prior in quoting
\emph{per-conjunction} thin-shell bounds.

For a compact source $A$ and test body $B$, the scalar--mediated force relative to Newtonian gravity may be written in a form that makes the screening of each body explicit:
\begin{equation}
\frac{F_\phi}{F_N}\;=\;2\,\chi_A\,\chi_B\;\times\;
\begin{cases}
1, & \text{$A$ and $B$ both unscreened},\\[4pt]
3\,\Delta R_A/R_A, & \text{$A$ screened,\; $B$ unscreened},\\[4pt]
3\,\Delta R_B/R_B, & \text{$A$ unscreened,\; $B$ screened},\\[4pt]
9\,(\Delta R_A/R_A)\,(\Delta R_B/R_B), & \text{$A$ and $B$ both screened}.
\end{cases}
\label{eq:force-ratio-piecewise}
\end{equation}
In particular, for an \emph{unscreened} test mass $B$ outside a screened spherical source $A$, the scalar-mediated force is suppressed relative to Newtonian gravity by
\begin{equation}
\frac{F_\phi}{F_N} \simeq 2\,\chi^{2}\,\min\!\left(1,\,\frac{3\,\Delta R}{R}\right),
\label{eq:force-ratio}
\end{equation}
with $\chi_A=\chi_B=\chi$ for a universal coupling. 
As a result,  screened objects with $\Delta R/R \ll 1$ source only a small residual fifth force \cite{KhouryPRD,KhouryPRL,BurrageSaksteinLRR}. Table \ref{tab:potentials} evaluates the relevant Newtonian potential as
\begin{equation}
\Phi_N \simeq 2.12\times 10^{-6}\ \text{(Sun)}, \qquad
6.96\times 10^{-10}\ \text{(Earth)}, \qquad
3.14\times 10^{-11}\ \text{(Moon)},
\label{eq:PhiN-numbers}
\end{equation}
which quantifies why inner--Solar System bodies are typically deep in the screened regime unless parameters lie near screening boundaries (see Appendix~\ref{sec:recipe} for more discussion on the relevant conditions.)

The screened-force relation above makes explicit how composition-dependent accelerations arise in a universal conformal coupling.  For a compact source \(S\) (e.g., Earth) with thin-shell fraction \(\Delta R_S/R_S\) and two test compositions \(A,B\), the differential acceleration in the field of \(S\) may be written in the usual Eötvös form.  Specializing the case-wise force law to a screened source and linearizing in the thin-shell parameter gives the EEP observable used later,
\begin{equation}
\eta_{\rm EEP}(A,B)\;\simeq\;2\,\chi_S\,\min\!\left\{1,\,3\,\frac{\Delta R_S}{R_S}\right\}\,\big(\chi_A-\chi_B\big),
\label{eq:eta-AIS}
\end{equation}
where \(\chi_S\) is the universal slope \(M_{\rm Pl} \partial \ln A/\partial\phi\) evaluated at the ambient value \(\phi_\star(\rho_\infty)\), and \(\chi_{A,B}\) encode the (small) composition dependence of Standard-Model masses under \(\phi\).  The source thin shell is set by the same density--minimum machinery already defined, \(\phi_\star(\rho)\) and \(\Delta R/R\) from Eqs.~(\ref{eq:thin-shell})--(\ref{eq:thin_shell_powerlaw}), so that in the screened‑source limit \(\Delta R_S/R_S \propto \phi_\star(\rho_\infty)\) and inherits the ambient--density rescaling summarized in \eqref{eq:thin-shell-L}.  Consequently, a null bound \(|\eta_{\rm EEP}|\!<\!\eta_{\max}\) maps directly into a constraint on the product \(\chi_S\,\min\!\{1,3\Delta R_S/R_S\}\) and hence on the same universal coupling \(\chi\) that controls the linear‑response excess \((\mu(z,k)-1)\) on cosmological scales.  

In chameleon‑like models the scalar Compton wavelength $\lambda_c(\rho)\!=\!m_\phi^{-1}(\rho)$ sets a mild $k$‑dependence in $\mu(z,k)$ through the transition between screened ($k\lambda_c\!\ll\!1$) and  unscreened ($k\lambda_c\!\gg\!1$) linear modes.  Our choice $k_{\rm fid}\!\simeq\!0.1\,h\,\mathrm{Mpc}^{-1}$ lies within the band where DESI/Euclid have the highest signal, and our mapping should be read as band‑limited around $k_{\rm fid}$; translating to nearby $k$ requires the standard 
linear response $\mu(z,k)$ of the chosen microphysical model.

Derivative (Vainshtein) screening suppresses fifth forces outside a characteristic Vainshtein radius $r_V$ set by nonlinear kinetic interactions. We take the cross-over scale $r_c\equiv c/H_0$ in DGP-like models (with $H_0$ the Hubble constant), so for canonical cases one finds scaling relations of the form
\begin{equation}
r_V \sim \left(r_S\,r_c^{2}\right)^{1/3} \quad \text{(DGP-like)}, 
\qquad
r_V \sim \left(\frac{r_S}{\Lambda^{3}}\right)^{1/3} \quad \text{(cubic Galileon-like)},
\label{eq:rV}
\end{equation}
where $r_S \equiv 2GM/c^{2}$ is the Schwarzschild radius, $r_c$ is the cross-over scale (of order $c/H_0$ in DGP), and $\Lambda$ is the strong-coupling scale of the Galileon sector~\cite{Koyama2016}. 

To compare, in a Vainshtein-screened model (e.g., a cubic Galileon) with Solar Vainshtein radius $r_{V\odot}$ given by \eqref{eq:rV}, the residual fractional modification inside the Vainshtein region scales as $(r/r_V)^{3/2}$ for the force-law correction~\cite{Koyama2016,BabichevDeffayet2013CQG}. At $r=1\,\mathrm{AU}$ and $r_{V\odot}\simeq 1.2\times 10^{2}\,\mathrm{pc}$, this yields $\delta F/F \sim (1\,\mathrm{AU}/100\,\mathrm{pc})^{3/2} \sim 10^{-11}$, far below current and near-term sensitivity, thereby illustrating why Solar System constraints are naturally weak for Vainshtein screening even if cosmology shows percent-level deviations (see Table~\ref{tab:Vainst}). This scaling implies $(\delta F/F)(1\,\mathrm{AU})\sim 10^{-11}$ for solar-mass sources, naturally placing Vainshtein-screened deviations below current sensitivity.

Equivalently, the near-Sun metric deviation that controls light propagation inherits the same suppression; in PPN language one may write
\begin{equation}
|\gamma-1|_{\text{Vainshtein}}(r) \sim \kappa \Big(\frac{r}{r_V}\Big)^{3/2}, \qquad \kappa=\mathcal{O}(1),
\label{eq:gamma_Vain}
\end{equation}
so that at $r=1\,\mathrm{AU}$ and $r_{V,\odot}\simeq 10^{2}\,\mathrm{pc}$ one expects $|\gamma-1|\lesssim 10^{-11}$, far below the few$\times 10^{-6}$ targets in Fig.~\ref{fig:solar-system-tests}(b); the normalization is model dependent at $\mathcal{O}(1)$ as in Eq.~(\ref{eq:gamma_from_shell}). 

\begin{table}[t!]
\centering
\caption{Illustrative DGP-like Vainshtein scan (Sun). Residual $|\,\gamma-1\,|(r)\simeq (r/r_V)^{3/2}$ from Eq.~(\ref{eq:gamma_Vain}).}
\label{tab:Vainst}
\begin{tabular}{lccc}
\hline\hline
$r_c/(c/H_0)$ & $r_{V,\odot}$ [pc] & $(1\,{\rm AU})/r_V$ & $|\,\gamma-1\,|(1\,{\rm AU})$ \\
\hline
2.0 & $\;\sim\!190$ & $3.9\times 10^{-8}$ & $7.7\times 10^{-12}$ \\
1.0 & $\;\sim\!120$ & $4.0\times 10^{-8}$ & $8.1\times 10^{-12}$ \\
0.3 & $\;\sim\!54$  & $9.0\times 10^{-8}$ & $2.7\times 10^{-11}$ \\
0.05& $\;\sim\!16$  & $3.0\times 10^{-7}$ & $1.6\times 10^{-10}$ \\
\hline
\end{tabular}
\end{table}

For a solar-mass source with $r_{S,\odot}=2.95~\mathrm{km}$ and $r_V$ from  \eqref{eq:rV} one finds $r_{V,\odot}\sim 10^2\,\mathrm{pc}$ (for DGP-like cross-over $r_c\sim c/H_0$ or a cubic Galileon with $\Lambda^3\sim H_0^2M_{\rm Pl}$), so  $|\gamma-1|_{\rm Vainshtein}(1\,\mathrm{AU})\sim\kappa\,(1\,\mathrm{AU}/r_{V,\odot})^{3/2}\lesssim10^{-11}$ even with $\kappa\sim1$. (For definiteness, a cubic Galileon sector with ${\cal L}_3 \sim (\partial\phi)^2 \Box\phi/\Lambda^3$ provides the derivative self-interaction.) The same scaling implies perihelion and range residuals below foreseeable sensitivity.  Hence, unless $r_V$ is anomalously small (e.g., by reducing $r_c\ll c/H_0$ or pushing the Galileon strong-coupling scale well above the canonical values), Solar System tests are naturally weak for derivative screening even if cosmology shows percent-level deviations. We thus present Vainshtein forecasts primarily as consistency checks—do the implied $r_V$ and $|\gamma-1|$ sit below the nuisance floors in Fig.~\ref{fig:solar-system-tests}(b)?—in parallel to the thin-shell forecasts which set actionable $\gamma$- and EEP-targets.

These screening mechanisms explain why many fully relativistic dark-energy models predict negligible parameterized post-Newtonian deviations locally, while still allowing cosmological signatures in the background expansion, growth, or lensing---subject to the gravitational-wave speed bound $|c_{\tt T}/c - 1| \lesssim 10^{-15}$~\cite{Baker2017,Creminelli2017}.

In Newtonian gauge ($\mathrm{d}s^2=-(1+2\Psi)\mathrm{d}t^2+a^2(1-2\Phi)\mathrm{d}\mathbf{x}^2$), we define
\begin{align}
-k^2\Psi &= 4\pi G\,a^2\,\mu(z,k)\,\rho_m\,\Delta_m,\\
-k^2(\Phi+\Psi) &= 8\pi G\,a^2\,\Sigma(z,k)\,\rho_m\,\Delta_m,
\label{eq:muSigma-def}
\end{align}
so that $\mu$ {rescales clustering (motion in $\Psi$)} and $\Sigma$ {rescales light deflection}. In the conformal  benchmark with negligible linear anisotropic stress (so $\Phi\simeq\Psi$), $\Sigma$ tracks $\mu$ up to $\mathcal{O}(1)$ factors; we therefore quote $\mu_{\rm lin,0}\equiv \mu(z{=}0,k_{\rm fid})-1$ as the primary cosmology$\to$local bridge parameter. In GR (including smooth fluidic DE), one has $\mu(z,k)=1$ and $\Sigma(z,k)=1$. In an unscreened, universally conformal scalar, $\mu\simeq 1+2\chi^2$ on linear scales while $\Sigma$ tracks $\mu$ up to ${\cal O}(1)$ factors (negligible linear anisotropic stress), so we use $\mu_{\rm lin,0}\equiv \mu(z{=}0,k_{\rm fid})-1$ as the cosmology$\to$local bridge.

\section{From cosmology to local residuals: screening map and targets}
\label{sec:bridge}

Throughout this section we evaluate $\mu_{\rm lin,0}\equiv \mu(z=0,k)-1$ at a representative linear scale
$k_{\rm fid}\simeq 0.1\,h\,\mathrm{Mpc}^{-1}$; mild $k$-dependence propagates through $\mu(z,k)$. Cosmological surveys probe the large-scale, low-potential regime where many modified-gravity or dark-energy scenarios imprint their primary signatures in the expansion history $w(z)$ and in growth/deflection observables summarized by $\mu(z,k)$ and $\Sigma(z,k)$ (cf.\ Sec.~\ref{sec:cosmo}). By contrast, Solar System experiments probe the high-potential, screened regime and impose model-independent guardrails through the EEP, the PPN coefficients $\gamma$ and $\beta$, and bounds on $\dot G/G$ (Sec.~\ref{sec:observables}).  We work under the tensor-speed prior summarized in Sec.~\ref{sec:theory} that forces the two regimes close to GR on both cosmological and local scales~\cite{Baker2017,Creminelli2017}. 

A quantitative workflow is two-stage: First, DESI BAO+FS and \emph{Euclid} weak-lensing and clustering determine the posterior in the cosmology-level parameter space $\{w(z),\mu(z,k),\Sigma(z,k)\}$; in a scalar--tensor embedding this corresponds to restrictions on the effective $\{V(\phi),A(\phi)\}$ subject to the $c_{\tt T}$ constraint (Sec.~\ref{sec:intro}). Second, given these posteriors, one maps to predicted local residuals using the screening relations of Sec.~\ref{sec:theory}: for chameleon-like models via the thin-shell expressions [\eqref{eq:thin-shell}--(\ref{eq:PhiN-numbers}),  (\ref{eq:gamma_from_shell})--\eqref{eq:phi_contrast_bound}] and for Vainshtein-like models via the $r_V$ and residual scalings \eqref{eq:rV}--\eqref{eq:gamma_Vain}. This yields concrete targets for Solar System tests such as $|\gamma-1|$ at solar conjunction, $\beta-1$ in ephemerides, and environment-dependent EEP/clock effects at levels set by the relevant potentials $\Phi_N$ \eqref{eq:PhiN-numbers}. A null detection at the forecast sensitivity prunes the cosmologically allowed model subspace; a detection triggers a joint re-fit across regimes with the same microphysical parameters. In our universal conformal benchmark with negligible anisotropic stress on linear scales, the metric potentials remain nearly equal ($\Phi\simeq\Psi$), so the lensing response $\Sigma(z,k)$ tracks the clustering response $\mu(z,k)$ up to $\mathcal{O}(1)$ factors. Operationally, we therefore take $\mu_{\rm lin,0}$ as the primary cosmology$\to$local bridge parameter and quote $\Sigma_0$ only for cross-checks.

\subsection{Illustrative $\gamma$ mapping}

Consider a conformally coupled scalar with $A(\phi)=\exp(\chi\phi/M_{\rm Pl})$ and a chameleon-like 
runaway potential, e.g. $V(\phi)=\Lambda^{4+n}\phi^{-n}$ with $n>0$, so that thin-shell screening applies 
(\ref{eq:thin-shell})--(\ref{eq:PhiN-numbers}). With these explicit choices, linear growth is enhanced by $\mu(z,k)\simeq 1+2\chi^2$ in the unscreened regime. Define $\mu_{\rm lin,0} \equiv \mu\big(z{=}0,k\!\sim\!0.1\,h\,{\rm Mpc}^{-1}\big)-1$, so
\begin{equation}
\chi \simeq \sqrt{\mu_{\rm lin,0}/2}. 
\label{eq:chi_from_mu}
\end{equation}

Suppose DESI+Euclid posteriors favored $\mu_0^{\rm lin}=0.10\pm0.05$ at $z\simeq 0$ (illustrative). Interpreting this in the conformal scalar model yields
$\chi \simeq \sqrt{{\mu_{\rm lin,0}}/{2}} \approx 0.224, $
indicating an $\mathcal{O}(10\%)$ unscreened enhancement of linear clustering. As noted in Sec.~\ref{sec:theory}, we adopt $k\simeq 0.1\,h\,{\rm Mpc}^{-1}$ for $\mulin$; Eq.~(\ref{eq:chi_from_mu}) holds on linear scales in the \emph{unscreened} regime ($k\lambda_c\gg1$),
with mild $k$-dependence inherited from $\mu(z,k)$ near the transition. For context, panels (b)--(c) of Fig.~\ref{fig:solar-system-tests} illustrate how a given $\mu_{\rm lin,0}$ maps into the required Sun thin-shell (via Eqs.~(\ref{eq:chi_from_mu})--(\ref{eq:phi_contrast_bound})) and the corresponding conjunction targets. Note that in Sec.~\ref{sec:cosmo}, the cosmology$\to$local map assumes a \emph{screened Sun} (thin shell), so that \eqref{eq:thin-shell} and \eqref{eq:gamma_from_shell} apply; unscreened interpretations of a nonzero $\mulin$ would instead be pruned directly by solar‑conjunction bounds on $\gamma$.

On linear scales we may write
\begin{equation}
\chi(k)\equiv \sqrt{{\textstyle\frac{1}{2}}\big(\mu(z=0,k)-1\big)}\,,\qquad
\frac{\delta\chi}{\chi}\simeq{\textstyle\frac{1}{2}}\frac{\delta\mu}{\mu-1}.
\label{eq:lin-mu}
\end{equation}
Using Eq.~(\ref{eq:lin-mu}), the null-test guardrail propagates as
\begin{equation}
\Big(\frac{\Delta R}{R}\Big)_{\!\max}(k)=\frac{1}{3\,\chi(k)}
\sqrt{{\textstyle\frac{1}{2}}|\,\gamma-1\,|_{\max}}\,,\qquad
\frac{\delta[(\Delta R/R)_{\max}]}{(\Delta R/R)_{\max}}=-\frac{\delta\chi}{\chi}\,.
\end{equation}
In Fig.~\ref{fig:mapping} we therefore recommend plotting a \emph{shaded band} obtained by evaluating
$(\Delta R/R)_{\max}(k)$ across $k\in[0.05,0.20]\,h\,{\rm Mpc}^{-1}$ with the survey’s
band-limited $\mu(z=0,k)$; the central line remains the $k_{\rm fid}\simeq0.1\,h\,{\rm Mpc}^{-1}$ value.
This explicitly displays the (mild) model dependence from the linear unscreened regime.

Locally, the Sun must be screened to respect solar-conjunction bounds on the PPN parameter $\gamma$. For a chameleon-like thin shell \eqref{eq:thin-shell}, the effective scalar charge of a screened body scales as $\alpha_\odot \simeq 3\chi\,(\Delta R/R)$, so for $|\alpha_\odot|\ll 1$  the leading metric deviation near the Sun can be estimated as\footnote{\emph{Thin-shell validity.} Eq.~(\ref{eq:gamma_from_shell}) assumes a screened Sun with $\Delta R/R\ll1$. Points with $\Delta R/R\gtrsim{\cal O}(0.1)$ violate the screened-source premise; for these, the $\gamma$--mapping does not apply and such parameter regions are excluded directly by Solar System bounds. We therefore treat Eqs.~(\ref{eq:gamma_from_shell})--(\ref{eq:phi_contrast_bound}) as guardrails within the $\Delta R/R\ll1$ domain.}
\begin{equation}
|\gamma-1| \simeq 2\,\alpha_\odot^{2} \simeq 18\,\chi^{2}\,\Big(\frac{\Delta R}{R}\Big)^{2}
\quad \Rightarrow\quad \left(\frac{\Delta R}{R}\right)_{\rm max}=\frac{1}{3\chi}\sqrt{\frac{|\gamma-1|_{\rm max}}{2}}.
\label{eq:gamma_from_shell}
\end{equation}
The coefficient in $|\gamma-1|\simeq 2\alpha_\odot^2$ is model-normalization dependent\footnote{This result matches the expression $\gamma-1=-2\alpha^2/(1+\alpha^2)$ in the small-coupling limit obtained in \cite{Damour-Esposito:1996} for tensor–scalar gravity, with $\alpha\to\alpha_\odot$. In our screened-Sun mapping the effective scalar charge is \(\alpha_\odot=3\chi\,(\Delta R/R)\),
so in the \(|\alpha_\odot|\ll1\) limit used here
\begin{equation}
|\,\gamma-1\,|\;\simeq\;2\,\alpha_\odot^2\;=\;18\,\chi^2\!\left(\frac{\Delta R}{R}\right)^{\!2},
\label{eq:gamma-1}
\end{equation}
which yields Eq.~(\ref{eq:gamma_from_shell}). Differences of \(O(1)\) in the definition of \(\alpha\) are absorbed into this coefficient.
}; with $\alpha_\odot\!=\!3\chi\,\Delta R/R$ our normalization yields (\ref{eq:gamma_from_shell}). Other conventions map by $\mathcal{O}(1)$ factors and do not affect the forecasts here. Demanding that the next-generation solar-conjunction analyses reach and do not detect $|\gamma-1|\lesssim 5\times 10^{-6}$ and solving (\ref{eq:gamma_from_shell}) for $\Delta R/R$ implies\footnote{Eq.~(\ref{eq:thin-shell-bound}) follows by combining the Sun-screened thin‑shell relation in Eq.~(\ref{eq:thin-shell}) with the Shapiro mapping of Eqs.~(\ref{eq:gamma_from_shell})–(\ref{eq:gamma-1}), using $\phi_\infty=\phi_\star(\rho_\infty)$ from Eq.~(\ref{eq:Veff}); Appendix~\ref{sec:crosscheck} gives the algebraic steps.
}
{}
\begin{equation}
\frac{\Delta R}{R}\ \lesssim\ \frac{1}{3\chi}\sqrt{\frac{|\gamma-1|}{2}}
\ \approx\ 2.4\times10^{-3}\qquad(\chi\simeq0.224,\ |\gamma-1|=5\times10^{-6}).
\label{eq:thin-shell-bound}
\end{equation}

Using the thin-shell relation \eqref{eq:thin-shell} with the Sun’s surface potential $\Phi_{N\odot}\simeq 2.12\times 10^{-6}$ [Eq.~\eqref{eq:PhiN-numbers}, see Table~\ref{tab:potentials}] then bounds the allowed field excursion across the shell:
\begin{equation}
|\phi_\infty-\phi_c| \lesssim 6\,\chi\,\Mpl\,\Phi_{N\odot}\,\left(\frac{\Delta R}{R}\right)_{\rm max}
\approx 6.7\times 10^{-9}\,\Mpl.
\label{eq:phi_contrast_bound}
\end{equation}

Eqs.~(\ref{eq:thin-shell-bound})--(\ref{eq:phi_contrast_bound}) hold for $\Delta R/R \ll 1$; when $\Delta R/R \gtrsim 0.1$ the screened‑source premise fails and the $\gamma$--mapping should not be used (we exclude such points directly by Solar System bounds).
These equations quantify how a modest, cosmology-level deviation ($\mu_0^{\rm lin}\sim 0.1$) maps to a concrete Solar System requirement: a Sun thin-shell factor $\Delta R/R \approx 2.4\times 10^{-3}$ and a correspondingly small environmental field contrast. Failure to satisfy these would produce a detectable $|\gamma-1|$ signal; conversely, a null solar-conjunction result at the $10^{-6}$ level would exclude the unscreened interpretation of $\mu_0^{\rm lin}\sim 0.1$ unless the chameleon parameters enforce $\Delta R/R \lesssim 10^{-3}$. 

Figure~\ref{fig:mapping}  maps $\mu_{\rm lin,0}$ to the maximum allowed solar thin-shell fraction $\Delta R/R$ for representative null-test sensitivities $|\gamma-1|_{\max}$ using the relations above. For a given cosmology-level excess $\mu_{\rm lin,0}$, a null solar-conjunction result at sensitivity $|\gamma-1|_{\max}$ implies a maximum allowed Sun thin-shell fraction $\Delta R/R$ via (\ref{eq:thin-shell-bound}) with curves labeled by \(|\gamma-1|_{\max}=5\times10^{-6}\) and \(1\times10^{-6}\).  For  $\mu_{\rm lin,0}\simeq0.10$, the $5\times10^{-6}$ goal requires $\Delta R/R\lesssim 2.4\times10^{-3}$,  see (\ref{eq:chi_from_mu})--(\ref{eq:phi_contrast_bound}).

Combining \eqref{eq:phi_min_powerlaw}--\eqref{eq:thin_shell_powerlaw} with (\ref{eq:thin-shell-bound}) and the screened limit $\phi_c \ll \phi_\infty$ gives
\begin{equation}
\phi_\star(\rho_\infty)\;\le\;M_{\rm Pl}\,\Phi_{N\odot}\,\sqrt{2\,|\gamma-1|_{\max}}\,,
\label{eq:phi_star_bound}
\end{equation}
so that (see (\ref{eq:thin-shell-L}))
\begin{equation}
\Lambda^{4+n}\;\le\;\frac{\chi\,\rho_\infty}{n\,M_{\rm Pl}}
\Big[M_{\rm Pl}\,\Phi_{N\odot}\,\sqrt{2\,|\gamma-1|_{\max}}\Big]^{n+1}.
\label{eq:Lambda_bound}
\end{equation}
Note that with $[\,\rho\,]=\mathrm{energy\,density}$ and $[\Lambda]=\mathrm{energy}$, (\ref{eq:Lambda_bound}) preserves $[\Lambda^{4+n}]$ on both sides; the factor $n^{-1}$ follows from $V'(\phi_\star)+\rho A'(\phi_\star)=0$ for $V(\phi)=\Lambda^{4+n}\phi^{-n}$. Eqs.~\eqref{eq:phi_star_bound}--\eqref{eq:Lambda_bound} recover the $\rho_\infty^{-1/(n+1)}$ thin‑shell scaling when eliminating $\Lambda$ against a cosmology‑level normalization, as shown explicitly in Appendix~\ref{sec:crosscheck}. 

Given $\mu_{\rm lin,0}$ (hence $\chi\simeq\sqrt{\mu_{\rm lin,0}/2}$ from (\ref{eq:chi_from_mu})) and an adopted ambient density $\rho_\infty$ appropriate to the near-Sun environment, \eqref{eq:Lambda_bound} translates the conjunction sensitivity directly into a bound on $\Lambda$ for each $n$.
For the illustrative case $\mu_{\rm lin,0}=0.10\Rightarrow\chi\simeq0.224$ and $|\gamma-1|_{\max}=5\times10^{-6}$, \eqref{eq:Lambda_bound}
gives $\Lambda^{4+n}\!\le\!({\chi\,\rho_\infty}/{n\,M_{\rm Pl}})\big[M_{\rm Pl}\,\Phi_{N\odot}\,\sqrt{10^{-5}}\big]^{n+1}$. Given $\mu_{\rm lin,0}$ we infer $\chi\simeq\sqrt{\mu_{\rm lin,0}/2}$ \eqref{eq:chi_from_mu}. 
Together with \eqref{eq:phi_min_powerlaw}--\eqref{eq:thin_shell_powerlaw} and the Solar System bounds 
\eqref{eq:thin-shell-bound}, \eqref{eq:phi_star_bound}, this makes the $(n,\Lambda)$ dependence of the forecasts explicit (see also \eqref{eq:Lambda_bound}).

Similarly, one can rescale (\ref{eq:thin-shell-bound})--(\ref{eq:Lambda_bound}) above to any preferred \(\rho_\infty\) without re-deriving intermediate steps. As a result, this yields the compact rescaling form of the bound derived from the thin-shell relation and the Shapiro-delay null test (cf.\ \eqref{eq:thin-shell-bound} and~\eqref{eq:phi_contrast_bound}). The identical \(\rho_\infty^{-1/(n+1)}\) scaling applies to the Earth thin shell that controls the EEP observable in \eqref{eq:eta-AIS}, so \(\eta\) forecasts and limits can be rescaled to any ambient-density prior using \eqref{eq:thin-shell-L} without re-deriving \(\eta\).

\subsection{Earth thin shell and EEP guardrails}

Starting from the AIS EEP observable (\ref{eq:eta-AIS-clock}),
\begin{equation}
\eta_{\rm EEP}(A,B)\;\simeq\;2\,\chi_\oplus\,
\min\!\left\{1,\,3\,\frac{\Delta R_\oplus}{R_\oplus}\right\}\,\Delta K_{\rm eff},
\label{eq:eta-EEP-worked}
\end{equation}
which is the specialization of \eqref{eq:eta-AIS-E} to Earth as the source and with the species/equipment combination
$\Delta K_{\rm eff}\equiv \sum_i \Delta K_i\,d_i$ (cf.\ Sec.~\ref{sec:DM-clocks} and \eqref{eq:clock-dm}), we now write the Earth thin shell explicitly using the same
screening machinery as for the Sun.  With the thin-shell relation \eqref{eq:thin-shell},
the density minimum \eqref{eq:phi_min_powerlaw}, and the Earth surface potential from Table~\ref{tab:potentials},
\begin{equation}
\frac{\Delta R_\oplus}{R_\oplus}\;\simeq\;\frac{\phi_\star(\rho_\infty)}{6\,\chi_\oplus\,M_{\rm Pl}\,\Phi_{N,\oplus}},
\qquad
\phi_\star(\rho)\;=\;\Big(\frac{n\,\Lambda^{4+n} M_{\rm Pl}}{\chi\,\rho}\Big)^{\!\frac{1}{n+1}}.
\label{eq:earth-shell}
\end{equation}
Two regimes follow directly:
\begin{align}
\text{(screened Earth)} \quad 3\,\frac{\Delta R_\oplus}{R_\oplus}<1
&\quad\Rightarrow \quad
\eta_{\rm EEP}\;\simeq\;\frac{\phi_\star(\rho_\infty)}{M_{\rm Pl}\Phi_{N,\oplus}} \Delta K_{\rm eff},
\label{eq:eta-screened}\\[2pt]
\text{(unscreened Earth)} \quad 3 \frac{\Delta R_\oplus}{R_\oplus}\ge 1
&\quad\Rightarrow \quad
\eta_{\rm EEP} \simeq 2 \chi_\oplus \Delta K_{\rm eff}.
\label{eq:eta-unscreened}
\end{align}
In the screened-Earth regime the factor $2\chi_\oplus\times (3\Delta R_\oplus/2R_\oplus)$ reduces to
$\phi_\star(\rho_\infty)/(M_{\rm Pl}\Phi_{N,\oplus})$, so $\eta$ directly constrains the ambient field
excursion without an explicit $\chi_\oplus$ dependence.

Eq.~\eqref{eq:eta-screened} shows the familiar thin-shell cancellation of $\chi_\oplus$: in the screened regime, a
null $\eta_{\rm max}$ constrains the environmental field excursion via
\begin{equation}
\phi_\star(\rho_\infty)\;\le\;\frac{\eta_{\rm max}}{|\Delta K_{\rm eff}|}\,M_{\rm Pl}\,\Phi_{N,\oplus},
\label{eq:phi-guardrail-earth}
\end{equation}
which is the EEP analogue of the Sun‐thin‐shell guardrail obtained from a null Shapiro test (cf.\ \eqref{eq:thin-shell-bound}--\eqref{eq:phi_star_bound} and Fig.~\ref{fig:mapping}).
Combining \eqref{eq:phi-guardrail-earth} with \eqref{eq:phi_min_powerlaw} yields the bound on the chameleon normalization:
\begin{equation}
\Lambda^{4+n}\;\le\;\frac{\chi\,\rho_\infty}{n\,M_{\rm Pl}}
\Big(\frac{\eta_{\rm max}\,M_{\rm Pl}\,\Phi_{N,\oplus}}{|\Delta K_{\rm eff}|}\Big)^{\!n+1},
\label{eq:Lambda-EEP}
\end{equation}
with the ambient-density rescaling inherited from the prior summarized in Table~\ref{tab:ambient-density} (the same
$\rho_\infty^{-\frac{1}{n+1}}$ law used for the Sun; see \eqref{eq:thin-shell-L}).
In the \textit{unscreened} case \eqref{eq:eta-unscreened}, $\eta_{\rm max}$ constrains the local slope directly as
\begin{equation}
|\chi_\oplus|\;\le\;\frac{\eta_{\rm max}}{2\,|\Delta K_{\rm eff}|},
\label{eq:chi-EEP-unscreened}
\end{equation}
which may be traded for $\mu_{\rm lin,0}$ using the cosmology bridge $\chi\simeq\sqrt{\mu_{\rm lin,0}/2}$ [\eqref{eq:chi_from_mu}] when appropriate. Eqs.~\eqref{eq:eta-screened}--\eqref{eq:chi-EEP-unscreened} therefore provide the EEP counterpart to the Sun‐thin‐shell map built from \eqref{eq:thin-shell-bound}, closing the ``\,$\gamma$ vs.\ EEP'' presentation symmetry at the same level of detail. (See also Table~\ref{tab:quick-lookup-eep} for a compact numerical summary of these guardrails across representative $\mulin$ and $|\Delta K_{\rm eff}|$.)

The boundary $3(\Delta R_\oplus/R_\oplus)=1$ depends on $\rho_\infty$ through $\phi_\star(\rho_\infty)$ in (\ref{eq:phi_min_powerlaw}). With the $b$-conditioned profile $\rho_\infty(b)$ introduced after (\ref{eq:thin-shell-L}),
the same EEP observable in Eq.~(\ref{eq:chi-EEP-unscreened}) admits a compact rescaling:
\begin{equation}
\Big(\frac{\Delta R_\oplus}{R_\oplus}\Big)(b)\propto
\big[\rho_\infty(b)\big]^{-1/(n+1)},
\end{equation}
so that lowering $b$ along a conjunction ray pushes the screened-Earth guardrail
to smaller $\Delta R_\oplus/R_\oplus$ at fixed $n$.

\subsection{Considering ULDM examples}

\emph{Higgs‐portal scalar (one‐parameter, predictive pattern).}  In the clock/comparison channel \eqref{eq:clock-dm},
a Higgs‐mixed scalar produces a correlated, one‐parameter response across species,
$(\delta\nu/\nu)_{A/B}\simeq \Delta K_H\,d_H\,\phi(t)$ with $\phi_0=\sqrt{2\rho_{\rm DM}}/m_\phi$ and coherence $t_c$ from \eqref{eq:tc-correct}.
The per‐mass coupling reach obtained by coherence‐limited stacking is the specialization of \eqref{eq:E-d-scaling}:
\begin{equation}
|d_H|\;\lesssim\;\frac{\sigma_y(\tau)}{|\Delta K_H|\,\phi_0}\,\sqrt{\frac{t_c}{T}},
\qquad
\phi_0=\frac{\sqrt{2\rho_{\rm DM}}}{m_\phi},\ \ t_c\simeq \frac{2\pi}{m_\phi v^2}.
\label{eq:reach-dH}
\end{equation}
This is the per-mass coupling detection threshold used in the forecasts; it reads directly on the stability/link targets in Table~\ref{tab:systematics-quant}  and the normalized sensitivity shown in Fig.~\ref{fig:solar-system-tests}(d).

\emph{Vector $B\!-\!L$ (oscillatory EEP with AIS).}  A massive $A'$ with $g_{B-L}$ generates a coherent field with
$|E'_0|=\sqrt{2\rho_{\rm DM}}$; the oscillatory Eötvös signal follows \eqref{eq:E-eta-osc},
$\eta_{\rm osc}(t)\simeq \big(g_{B-L}/g\big)\,\Delta(Q_{B-L}/M)\,|E'_0|\cos(m_{A'}t)$.
Using the AIS acceleration ASD and coherence stacking gives the coupling reach (the AIS specialization of \eqref{eq:E-gBL-scaling}):
\begin{equation}
g_{B-L}\;\lesssim\;g\,
\frac{S_a^{1/2}}{\Delta(Q_{B-L}/M)\,\sqrt{2\rho_{\rm DM}}}\,
\sqrt{\frac{t_c}{T}},
\label{eq:reach-gBL}
\end{equation}
where $S_a^{1/2}$ is the differential‐acceleration ASD over $\tau\lesssim t_c$ for the mission configuration.  Eqs.
\eqref{eq:reach-dH}--\eqref{eq:reach-gBL} turn the qualitative model discussion into quantitative, “is it within reach?” forecasts that
can be dropped into investment prioritization.

\begin{table}[t]
\centering
\setlength{\tabcolsep}{3pt}        
\renewcommand{\arraystretch}{1.0} 
\caption{Illustrative mapping from cosmology-level posteriors to Solar System residual targets.
Numbers are indicative to show usage of Eqs.~(\ref{eq:thin-shell})--\eqref{eq:gamma_Vain}, \eqref{eq:chi_from_mu}--\eqref{eq:phi_contrast_bound}.
\label{tab:mapping}}
\begin{tabular}{p{2.6cm}p{3.9cm}p{3.1cm}p{4.8cm}p{2.1cm}}
\hline
Model & Cosmology posterior & Mapping param & Local target(s) & Notes \\
\hline\hline
Conformal scalar & $\mulin=0.10\pm0.05$ & $\chi=\sqrt{\mulin/2}\simeq0.224$ &
$|\gamma-1|\lesssim \text{few}\times10^{-6}$; $\Delta R/R\lesssim(1.6$--$2.4)\times10^{-3}$ &
Eqs.~\eqref{eq:chi_from_mu}--\eqref{eq:phi_contrast_bound} \\
Cubic Galileon & $\Sigma_0,\,\mu_0$ consistent with GR & $r_{V\odot}\sim10^2$ pc &
$\delta F/F(1\,\mathrm{AU})\sim (r/r_V)^{3/2}\sim10^{-11}$ &
Eqs.~\eqref{eq:rV}--\eqref{eq:gamma_Vain} \\
Yukawa tail & $\alpha_{\tt Y}(\lambda)$ at $\lambda=10^{9}$--$10^{13}$ m & $m=\hbar c/\lambda$ &
$|\alpha_{\tt Y}|\ll10^{-9}$--$10^{-10}$ & Sec.~\ref{sec:DM-Yuk} \\
ULDM (scalar) & coupling $d_i$ vs. $m_\phi$ & $t_c\simeq 2\pi/(m_\phi v^2)$ &
$\delta\nu/\nu\sim K_i d_i \phi_0$; improve $\times(3$--$10)$ &
Sec.~\ref{sec:DM-clocks} \\
\hline
\end{tabular}
\end{table}

\section{Solar System observables: status and credible improvements}
\label{sec:observables} 

We summarize present leading bounds, the physics they probe, and realistic near-term gains under concrete measurement conditions (see Tables~\ref{tab:benchmarks} and \ref{tab:systematics} for details). Some of the relevant reviews are given in \cite{Turyshev:2007, Turyshev:2009}.
Throughout Sec.~\ref{sec:observables} we use ``target'' to mean a credible near-term sensitivity forecast, i.e., the level reached when dominant systematics are reduced to the quantitative budgets in Table~\ref{tab:systematics-quant} using the mitigation strategies in Table~\ref{tab:systematics}. When we cite a ``target'' we also provide the instrument or analysis reference (e.g., BepiColombo/MORE for $\gamma$, ACES/optical links for clocks).

\subsection{Einstein equivalence principle and composition dependence}
\label{sec:EEP}

The E\"otv\"os parameter $\eta_{\rm EEP}(A,B)\equiv 2\,(a_A-a_B)/(a_A+a_B)$ quantifies differential acceleration of test bodies $A$ and $B$ in the same external field. The MICROSCOPE mission compared titanium and platinum proof masses and reported~\cite{Touboul2022}
\begin{equation}
\eta_{\rm EEP}(\mathrm{Ti},\mathrm{Pt}) = \big(-1.5 \pm 2.3_{\rm stat} \pm 1.5_{\rm syst}\big)\times 10^{-15},
\label{eq:microscope-eta}
\end{equation}
which excludes many unscreened scalar--tensor couplings and severely restricts dilaton-type models. 

A purpose-built follow-on or a dedicated, drag-free, space atom interferometer such as Space-Time Explorer and QUantum Equivalence principle Space Test (STE-QUEST)-class atom-interferometer mission can credibly improve sensitivity by one order of magnitude, reaching $\eta_{\rm EEP} \sim 10^{-16}$--$10^{-17}$, using long interrogation times $T$, dual/multi-species comparisons, and common-mode rejection \cite{Battelier2021ExpAstron,Ahlers2022STEQUEST}.  
We therefore treat AIS primarily as a precision EEP instrument; however, in the \emph{positive-detection branch} (below) it becomes a direct DE/DM discovery channel, consistent with our hypothesis-driven strategy.

A complementary, explicitly DE-driven option is a \emph{tetrahedral} four-spacecraft constellation that measures the trace of the scalar force-gradient tensor in interplanetary space, directly targeting Galileon/Vainshtein phenomenology while rejecting Newtonian backgrounds \cite{TetPRD2024}. While not a traditional single-baseline AI, this DE-focused geometry can incorporate cold-atom accelerometers/gradiometers and fits our decision rule in Sec.~\ref{sec:program} when a specified model predicts a detectable local signature.

AIS enable long free-fall times, drag-free control, and multi-species comparisons. Dual-species AIS (e.g., Rb/K or Sr/Rb) test the EEP via the Eötvös parameter $\eta(A,B)\!\equiv\!2(a_A-a_B)/(a_A+a_B)$. STE-QUEST--class designs credibly reach $\eta\sim10^{-16}\text{--}10^{-17}$ under microgravity, long interrogation time $T$, and common-mode rejection \cite{Ahlers2022STEQUEST,Battelier2021ExpAstron,QTEST2015}. In scalar--tensor theories with universal conformal coupling $A(\phi)=e^{\chi\phi/M_{\rm Pl}}$, composition dependence arises when Standard-Model masses inherit distinct $\phi$-sensitivities. 

A screened source (Earth) from (\ref{eq:eta-AIS}) yields
\begin{equation}
\eta(A,B) \simeq 2 \chi_{\oplus} \Big[\min\!\Big(1, 3\frac{\Delta R_{\oplus}}{R_{\oplus}}\Big)\Big] (\chi_A-\chi_B),
\label{eq:eta-AIS-E}
\end{equation}
so AI limits map onto $(\chi_A-\chi_B)\chi_\oplus$ with the thin-shell factor from (\ref{eq:thin-shell})--(\ref{eq:PhiN-numbers}). Here $\chi_\oplus$ denotes the universal conformal slope $M_{\rm Pl} \partial\ln A/\partial\phi$ evaluated at the ambient value $\phi_\star(\rho_\infty)$ for the Earth as the source (cf.\ Sec.~\ref{sec:theory}).

Given the cosmology$\to$local bridge in Secs.~\ref{sec:theory}--\ref{sec:cosmo}, pushing $\eta$ by $1$--$2$ orders of magnitude sharpens guardrails on the same $\chi$ that controls $\mu(z,k)$ in linear growth (cf.\ Table~\ref{tab:mapping}). With $A(\phi)=e^{\chi\phi/\Mpl}$ and the Earth thin‑shell factor from \eqref{eq:thin-shell}, the AI guardrail probes the same $\chi$ that enters $\mulin$ via \eqref{eq:chi_from_mu} below, ensuring a common parameter across cosmology and local tests.

From \eqref{eq:eta-AIS-E}, a null $\eta_{\max}$ implies
\[
\bigl|\chi_\oplus\bigr| \;\lesssim\; 
\frac{\eta_{\max}}{2\,\bigl|\chi_A-\chi_B\bigr|}\;
\frac{1}{\min\!\bigl\{1,\,3\,\Delta R_\oplus/R_\oplus\bigr\}}\,,
\]
with $\Delta R_\oplus/R_\oplus$ taken from Eqs.~\eqref{eq:thin-shell}--\eqref{eq:phi_min_powerlaw} at the chosen $\rho_\infty$ via \eqref{eq:thin-shell-L}.

Considering realistic experiments, we write the differential sensitivity in terms of clock/interferometer coefficients, $\Delta K \equiv \sum_i \Delta K_i\,d_i$ from \eqref{eq:clock-dm}, then Eq. \eqref{eq:eta-AIS-E} gives
\begin{equation}
\eta(A,B) \simeq 2\,\chi_\oplus \big[\min\!\big(1,3 \frac{\Delta R_\oplus}{R_\oplus}\big)\big]\,\Delta K_{\rm eff},
\label{eq:eta-AIS-clock}
\end{equation}
where $\Delta K_{\rm eff}$ packages the relevant species sensitivity coefficients and dark-sector couplings for the \emph{instrument at hand} (clocks or AIS). For AIS EEP tests, it reduces to the composition-difference factor entering Eq.~(\ref{eq:eta-AIS-E}); for clock comparisons it is the usual
$\sum_i \Delta K_i d_i$ of Eq.~(\ref{eq:clock-dm}). This makes forecasts transparent: for a target $\eta$ and a chosen pair (e.g., Rb/K or Sr/Rb), one trades a measured (or design) $\Delta K_{\rm eff}$ and the Earth thin-shell factor from \eqref{eq:thin-shell}--\eqref{eq:thin_shell_powerlaw} for $\chi_\oplus$, in the same $\chi$ that controls $\mu_{\rm lin,0}$ via \eqref{eq:chi_from_mu}.
Beyond static EEP tests, AIS can probe time-dependent signals from ULDM: differential phases oscillate at $m_\phi$ through species-dependent $K_i$ and $d_i$ (cf.\ Sec.~\ref{sec:DM-clocks}), providing an orthogonal handle to clock networks and extending baselines with space links \cite{Badurina2021Prospects,MAGIS100,SAGE2019}. 

For representative $\Delta K_{\rm eff}=\mathcal{O}(10^{-1}\text{--}1)$ and mission parameters in
Tables~\ref{tab:systematics}--\ref{tab:systematics-quant}, $\eta \sim 10^{-16}\text{--}10^{-17}$ maps to
$|\chi_\oplus|\lesssim 10^{-16}\text{--}10^{-17}/\Delta K_{\rm eff}$ modulo the screening factor, providing a  guardrail complementary to Fig.~\ref{fig:mapping}.

Note, if an AI experiment reports a nonzero $\eta$ at significance, the inference proceeds by (i) mapping the measured $\eta(A,B)$ to the screened-source combination $\chi_\oplus \min\{1,3\Delta R_\oplus/R_\oplus\}$ via Eq.~(\ref{eq:eta-AIS-E}) and the thin-shell relations, (ii) confronting the implied local $\chi$ with cosmology-level $\mu_{\rm lin,0}$ through Eq.~(\ref{eq:chi_from_mu}) in a joint likelihood (App.~\ref{app:joint}), and (iii) prioritizing follow-up with composition-rotated AI/clock pairs to separate universal vs.\ composition-dependent couplings. This branch is explicitly included in the program in Sec.~\ref{sec:program}.

\subsection{Parameterized post-Newtonian gravity: $\gamma$ and $\beta$}
\label{sec:Shapiro}

\begin{figure}[t]
\centering
\begin{tikzpicture}
\begin{groupplot}[
  group style={
    group size=2 by 2,
    horizontal sep=0.09\columnwidth,  
    vertical sep=2.20cm               
  },
  width=0.38\columnwidth, height=4.6cm,
  scale only axis=true,
  grid=both, tick align=outside, legend cell align=left,
  every axis/.append style={
    title style={font=\small\rmfamily\mdseries, yshift=2.4ex},
    label style={font=\small\rmfamily\mdseries},
    tick label style={font=\small\rmfamily\mdseries},
    xlabel style={yshift=0.4ex}
  },
  every axis plot/.append style={line width=0.9pt, mark=none}
]

\nextgroupplot[
  xmode=log, xmin=6e9, xmax=3e14,
  ymode=log, ymin=1e-9, ymax=3,
  title={(a) $\Delta t_{\rm plasma}\propto f^{-2}$},
 title style={yshift=-1.8ex}, 
  xlabel={Carrier frequency $f$ [Hz]},
  ylabel={Residual group delay (norm.\ to X-band)},
  legend pos=south west, legend style={font=\scriptsize\rmfamily\mdseries, draw=none, fill=none},
  xtick={1e10,1e12,1e14}, xticklabels={$10^{10}$,$10^{12}$,$10^{14}$},
  ytick={1e0,1e-3,1e-6,1e-9}, yticklabels={$10^{0}$,$10^{-3}$,$10^{-6}$,$10^{-9}$}
]
\addplot+[thick] coordinates {
  (6.0e9,1.96) (8.4e9,1.0) (1.0e10,0.706) 
   (1.0e11,7.06e-3) (1.0e12,7.06e-5)
  (1.0e13,7.06e-7) (2.0e14,1.76e-9)
};
\addlegendentry{normalized to X (8.4\,GHz)}

\addplot+[only marks, mark=*]         coordinates {(8.4e9,1.0)};       \addlegendentry{X}
\addplot+[only marks, mark=triangle*] coordinates {(3.2e10,0.0689)};  \addlegendentry{Ka}
\addplot+[only marks, mark=square*]   coordinates {(2.0e14,1.76e-9)}; \addlegendentry{optical}

\nextgroupplot[
  xmode=log, xmin=3, xmax=50,
  ymode=log, ymin=7e-7, ymax=1.2e-4,  
  title={(b) $|\gamma-1|$ vs solar impact parameter},
  xlabel={$b/R_\odot$},
  title style={yshift=-1.8ex}, 
  clip=false,
  legend pos=north east, legend style={font=\scriptsize\rmfamily\mdseries, draw=none, fill=none},
  xtick={3,10,50}, xticklabels={$3$,$10$,$50$},
  ytick={1e-6,1e-5,1e-4}, yticklabels={$10^{-6}$,$10^{-5}$,$10^{-4}$}
]
\addplot+[thick] coordinates {(3,3.3e-5) (5,2.6e-5) (10,2.0e-5) (20,1.6e-5) (50,1.4e-5)};
\addlegendentry{X-band (plasma-limited)}
\addplot+[thick, dashed] coordinates {(3,1.2e-5) (5,8.0e-6) (10,5.5e-6) (20,1.6e-6) (50,1.0e-6)};
\addlegendentry{Ka/optical (calibrated)}
\addplot+[thin]             coordinates {(3,2.3e-5) (50,2.3e-5)};     \addlegendentry{Cassini $2.3\times10^{-5}$}
\addplot+[dashdotted, line width=1.1pt, draw=black]
  coordinates {(3,1.0e-6) (50,1.0e-6)};
\addlegendentry{target $1\times10^{-6}$}
\node[rotate=90, font=\small\rmfamily\mdseries, anchor=south]
  at (axis description cs:1.12,0.5) {Sensitivity to $|\gamma-1|$ (1$\sigma$)};

\nextgroupplot[
  xmin=0.03, xmax=0.20,
  ymode=log, ymin=6e-4, ymax=2e-2,
  title={(c) Null $|\gamma-1|$ $\Rightarrow$ Sun thin-shell bound},
  title style={yshift=-1.8ex}, 
  xlabel={Linear response $\mu_{\rm lin,0}$ (dimensionless)},
  ylabel={Max thin-shell $\Delta R/R$ (Sun)},
  legend pos=north east, legend style={font=\scriptsize\rmfamily\mdseries, draw=none, fill=none},
  xtick={0.03,0.10,0.20}, xticklabels={$0.03$,$0.10$,$0.20$},
  ytick={1e-3,3e-3,1e-2}, yticklabels={$10^{-3}$,,$10^{-2}$}
]
\addplot+[thick] coordinates {
  (0.03,9.27e-3) (0.05,7.15e-3) (0.10,5.06e-3) (0.15,4.13e-3) (0.20,3.58e-3)
};
\addlegendentry{$|\gamma-1|=2.3\times10^{-5}$ (Cassini)}
\addplot+[thick, dashdotted] coordinates {
  (0.03,1.93e-3) (0.05,1.49e-3) (0.10,1.06e-3) (0.15,8.64e-4) (0.20,7.47e-4)
};
\addlegendentry{target $1\times10^{-6}$}
\node[font=\scriptsize,anchor=west] at (axis cs:0.105,2.05e-3) {$\chi=\sqrt{\mu_{\rm lin,0}/2}$};

\nextgroupplot[
  xmode=log, xmin=1e-21, xmax=1e-15,
  ymode=log, ymin=5e-6, ymax=2e-3,
  title={(d) Clock sensitivity to $d_e$ vs $m_\phi$},
  title style={yshift=-1.8ex}, 
  xlabel={ULDM mass $m_\phi$ [eV]},
  clip=false,
  legend pos=north east, legend style={font=\scriptsize\rmfamily\mdseries, draw=none, fill=none},
  xtick={1e-21,1e-19,1e-17,1e-15}, xticklabels={$10^{-21}$,$10^{-19}$,$10^{-17}$,$10^{-15}$},
  ytick={1e-5,1e-4,1e-3}, yticklabels={$10^{-5}$,$10^{-4}$,$10^{-3}$}
]
\addplot+[thick, dotted] coordinates {
  (1e-21,8.0e-4) (3e-21,6.2e-4) (1e-20,4.6e-4) (3e-20,3.2e-4)
  (1e-19,2.2e-4) (3e-19,2.0e-4) (1e-18,1.9e-4) (3e-18,2.4e-4)
  (1e-17,3.6e-4) (3e-17,6.0e-4) (1e-16,1.1e-3) (1e-15,1.8e-3)
};
\addlegendentry{current}
\addplot+[thick, dashed] coordinates {
  (1e-21,2.0e-4) (3e-21,1.6e-4) (1e-20,1.1e-4) (3e-20,8.0e-5)
  (1e-19,5.5e-5) (3e-19,5.0e-5) (1e-18,4.5e-5) (3e-18,6.5e-5)
  (1e-17,1.1e-4) (3e-17,1.8e-4) (1e-16,3.8e-4) (1e-15,6.5e-4)
};
\addlegendentry{advanced links/stability}
  
\node[rotate=90, font=\small\rmfamily\mdseries, anchor=south, xshift=0.6em]
  at (axis description cs:1.10,0.5) {Clock sensitivity to $d_e$ (arb. units)};
  
\end{groupplot}
\end{tikzpicture}
\vskip -10pt 
\caption{Solar System sensitivities with advanced links and analysis. (a) Residual plasma time (group) delay, normalized to X-band (8.4\,GHz), versus $f$ [Hz],
following the $f^{-2}$ law in \eqref{eq:plasma-delay}.
(b) $1\sigma$ sensitivity to $|\gamma-1|$ versus solar impact parameter $b/R_\odot$, with Cassini ($2.3\times10^{-5}$) and a $10^{-6}$ target for reference. 
Curves assume a post-calibration residual time-delay budget $\le 0.1\,$ns for $b\gtrsim 5\,R_\odot$ (40\,ps plasma/turbulence, 30\,ps non-gravitational, 30\,ps timing/transfer).
(c) Maximum solar thin-shell fraction $\Delta R/R$ implied by a null $|\gamma-1|$ bound, as a function of the cosmology response $\mu_{\rm lin,0}$, using \eqref{eq:chi_from_mu}--\eqref{eq:thin-shell-bound}.
(d) Normalized clock sensitivity to a fine‑structure‑constant-$\alpha$-coupled coefficient $d_e$ vs $m_\phi$ from the coherence-time/bandwidth scalings, (\ref{eq:clock-dm})--\eqref{eq:tc-correct}.
}
\label{fig:solar-system-tests}
\end{figure}
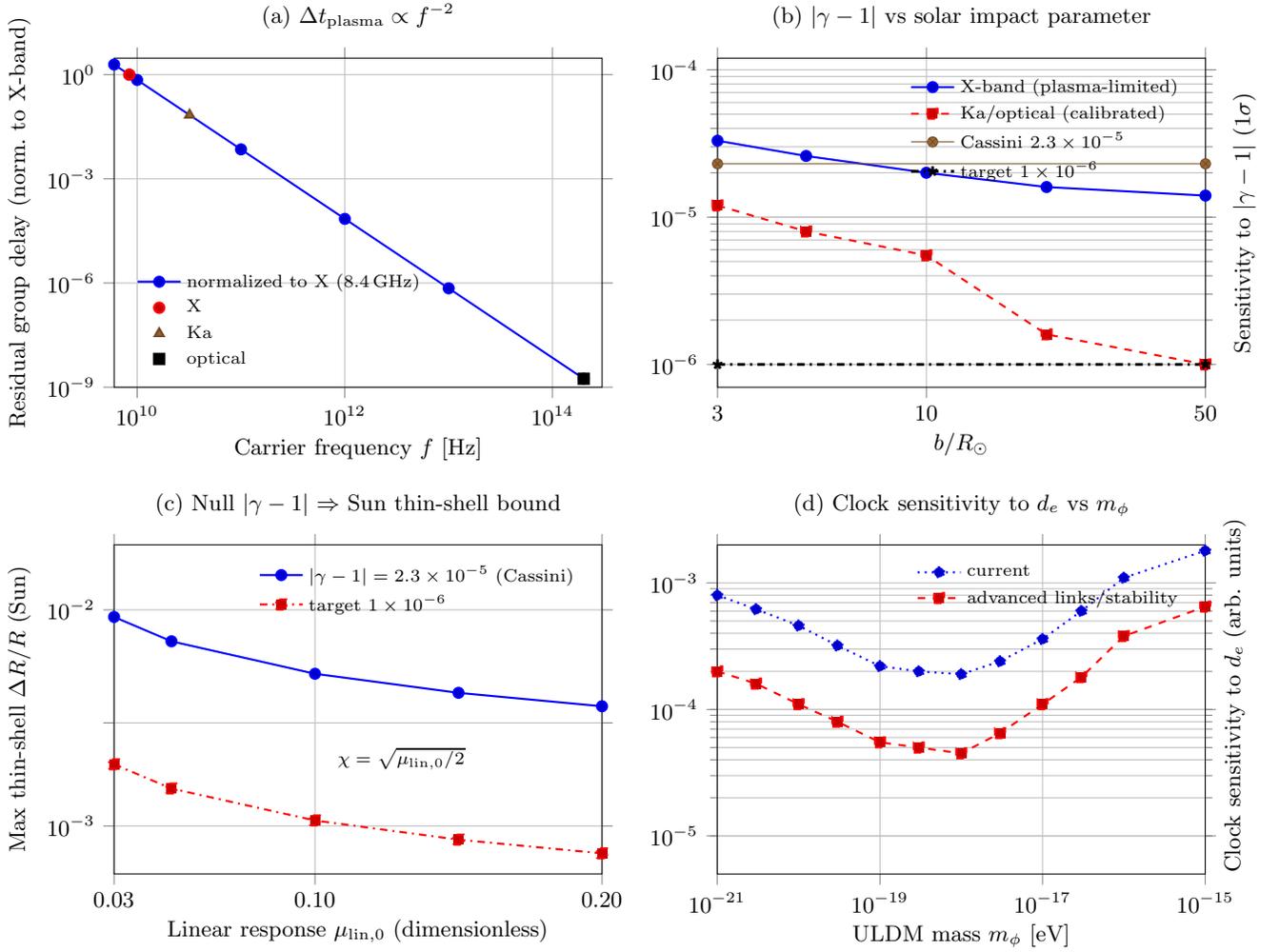

In the PPN framework, $\gamma$ measures spatial curvature per unit mass and $\beta$ encodes nonlinearity in superposition \cite{Will2014LRR}. (In GR: $\gamma=1$ and $\beta=1$.) The Shapiro time delay provides access to PPN $\gamma$ via
{}
\begin{equation}
\Delta t \simeq (1+\gamma)\,\frac{G M_\odot}{c^3}\,
\ln\!\left(\frac{4\,r_E r_R}{b^2}\right),
\label{eq:shapiro}
\end{equation}
where $r_E$ and $r_R$ are the heliocentric distances of emitter and receiver, and $b$ is the impact parameter of the radio path. The 2002 solar-conjunction experiment with Cassini spacecraft  obtained~\cite{Bertotti2003}
\begin{equation}
\gamma - 1 = (2.1 \pm 2.3)\times 10^{-5}.
\label{eq:cassini-gamma}
\end{equation}
The dependence of the 1$\sigma$ sensitivity to $|\gamma-1|$ on solar impact parameter is shown in Fig.~\ref{fig:solar-system-tests}(b), with Cassini’s $2.3\times10^{-5}$ reference and the $10^{-6}$ target.  As we show in Appendix \ref{sec:crosscheck}, for chameleon-like benchmarks, the Sun thin-shell mapping implies that a factor of $\sim$2--4 improvement over the current $|\gamma-1|$ sensitivity would directly probe the predicted residuals (see the $\rho_\infty$ scaling summarized in Table~\ref{tab:ambient-density} and the guardrail in \eqref{eq:thin-shell-bound}).

In global ephemerides, $\beta$ is constrained in combination with $\gamma$ through perihelion precession and solar-system dynamics. For a test body with $(a,e)$ one has the PPN perihelion shift per orbit
\begin{equation}
\Delta\varpi \;=\;\tfrac13({2+2\gamma-\beta})\frac{6\pi GM_\odot}{a(1-e^2)c^2},
\label{eq:PPN_perihelion}
\end{equation}
where $a$ is the semi-major axis and $e$ the orbital eccentricity of the test body. Thus, Ka/X multi-frequency conjunction arcs that tighten $\gamma$ also reduce degeneracies in \eqref{eq:PPN_perihelion}, improving $\beta$ in the subsequent global fits.

Dual-frequency Ka/X calibration mitigates coronal plasma dispersion, while accelerometry and thermal modeling bound non-gravitational forces; with optimized conjunction arcs this supports $\sigma_\gamma\!\sim\!\mathrm{few}\times 10^{-6}$. The Mercury Orbiter Radio-science Experiment (MORE) on BepiColombo mission aims at $\gamma$ at the level of a few$\times 10^{-6}$ using Ka/X multi-frequency links, improved coronal-plasma calibration, and optimized conjunction arcs, with comparable sensitivity to $\beta$ in global ephemeris fits~\cite{Iess2021,diStefano2021}. These remain among the cleanest AU-scale tests of long-range metric gravity.

Considering optical links we note that the DSOC tech demo established deep-space coherent optical links with a peak downlink of $267~\mathrm{Mb\,s^{-1}}$ at $\sim\!0.2$~AU (Dec.\ 11, 2023), and sustained $25~\mathrm{Mb\,s^{-1}}$ at $\sim\!1.5$~AU (Apr.\ 8, 2024), validating pointing, acquisition, and timing needed for precision relativistic tests
\cite{JPL_DSOC_FirstData_2023,JPL_DSOC_Record_2024,NASA_DSOC_Page}. We model coronal plasma dispersion with the standard time and group delays
{}
\begin{equation}
\Delta t_{\rm plasma}\simeq \frac{40.3\,\mathrm{m^{3}\,s^{-2}}}{c}\,
\frac{\mathrm{TEC}}{f^{2}}\;,
\qquad 
\Delta L_{\rm plasma}\equiv c\,\Delta t_{\rm plasma}
=\frac{40.3\,\mathrm{m^{3}\,s^{-2}}\,\mathrm{TEC}}{f^{2}}\,,
\label{eq:plasma-delay}
\end{equation}
where $f$ is the carrier frequency (Hz) and $\mathrm{TEC}$ the electron column density (m$^{-2}$). The normalization in Eq.~\eqref{eq:plasma-delay} follows the standard cold‑plasma dispersion law used in radio science, and underlies the $f^{-2}$ scaling.  See, e.g., \cite{Iess2021SSR,diStefano2021Cqg} for the standard radio-science normalization and calibration. The optical suppression factors quoted below follow directly from the $f^{-2}$ scaling.

At optical $f\!\sim\!2\times10^{14}$ Hz, the suppression relative to X-band ($8.4$ GHz) is
$(8.4\times10^{9}/2\times10^{14})^{2}\!\approx\!1.76\times10^{-9}$ and to Ka-band ($32$ GHz) is
$(3.2\times10^{10}/2\times10^{14})^{2}\!\approx\!2.56\times10^{-8}$, respectively. Thus optical links reduce coronal group-delay systematics by $\mathcal{O}(10^{8}\!-\!10^{9})$ vs.\ current radio bands. As summarized in Fig.~\ref{fig:solar-system-tests}(a), the residual group delay scales as $f^{-2}$; Ka and especially optical links strongly suppress coronal dispersion relative to X.

To reach $|\gamma-1|\sim\mathrm{few}\times10^{-6}$ at $b\gtrsim5\,R_\odot$ we allocate the post-calibration group-delay budget to $\le 0.1~\mathrm{ns}$  (Table~\ref{tab:systematics-quant}), split as: plasma/turbulence (40~ps; already suppressed as $f^{-2}$ per Fig.~\ref{fig:solar-system-tests}a), non-gravitational forces and thermal/attitude coupling (30~ps over a 10-day dwell), and timing/transfer chain (30~ps). This allocation tracks the frequency scaling in \eqref{eq:plasma-delay} and the pointing/thermal constraints quoted in Sec.~\ref{sec:program}, and serves as the nuisance floor drawn in Fig.~\ref{fig:solar-system-tests}(b). The $10$-day dwell assumed in Fig.~\ref{fig:solar-system-tests}(b) averages stochastic plasma and thermal/attitude fluctuations to the quoted $30$--$40$ ps contributions and thus sets the vertical offset (noise floor) against which the Shapiro slope is fit; shorter dwells would move the colored curves upward.

\subsection{Time variation of $G$ and the strong-equivalence principle}

Lunar Laser Ranging (LLR) constrains the fractional time variation of Newton's constant and the Nordtvedt parameter (a strong-equivalence-principle, SEP, violation). Early global analyses reported $|\dot G/G|\sim 10^{-13}\,\mathrm{yr^{-1}}$ and SEP at $\eta_{\rm SEP} = 4\beta - \gamma - 3 - \frac{10}{3}\xi - \alpha_1 - \frac{2}{3}\alpha_2\sim 10^{-4}$ class~\cite{Williams2004}. In minimal scalar--tensor completions the preferred-frame and Whitehead parameters are negligible, so $\eta_{\rm SEP}\!\approx\!4\beta-\gamma-3$. Ephemerides directly constrain $G\Msun$; separating $\dot G$ from solar mass loss requires solar-physics priors.

Modern ephemerides and LLR combinations give $|\dot G/G|$ 
 in the $(2$--$6)\times 10^{-14}\,\mathrm{yr}^{-1}$ range~\cite{Pitjeva2021,Fienga2024}.  Millimeter-class LLR with next-generation corner-cube retroreflectors (CCR), improved station metrology, and higher link budgets can plausibly strengthen these bounds by a factor $\sim 4$--10~\cite{Turyshev2025}.
Recent LLR global solutions already constrain $\Delta(m_g/m_i)_{\rm EM}\!=\!(-2.1\pm2.4)\times10^{-14}$ and $\dot G/G_0=(-5.0\pm9.6)\times10^{-15}\,{\rm yr^{-1}}$\cite{Biskupek2021};  next-generation infrared (IR) stations and single--CCR deployments are expected to tighten both.   See also \cite{FischerKaedingPitschmann2024} for a consolidated summary of recent screened-scalar constraints from LLR and laboratory experiments, and Cannex projections.

Note that in universally coupled, screened scalar models, Solar System bodies acquire suppressed scalar charges through the thin-shell relation (\ref{eq:thin-shell})--(\ref{eq:PhiN-numbers}). The Nordtvedt/SEP signal in the Earth--Moon system depends on the \emph{difference} of Earth/Moon charges in the Sun’s \emph{screened} field $\eta_{\rm SEP}$ and is therefore doubly thin-shell and composition suppressed. Present LLR SEP bounds are complementary but do not overtake conjunction constraints on $\gamma$ in this class; sub-mm LLR with range precision of $\sim30\,\mu$m \cite{Turyshev2025} can strengthen $\eta_{\rm SEP}$ and  $\dot G/G$ by $\sim 5$--$7\times$ to $\dot G/G_0=(3$--$7)\times10^{-15}\,{\rm yr^{-1}}$ and sharpen $\beta$ in global fits (Tables~\ref{tab:benchmarks}, \ref{tab:systematics-quant}).  Recent dilaton field-profile solutions and their application to qBOUNCE, Cannex, and LLR appear in \cite{BraxFischerKaedingPitschmann2022}.

\subsection{Solar System dark-matter density and Yukawa tails}
\label{sec:DM-Yuk}

Global ephemeris fits limit any smoothly distributed DM in the Solar System. At Saturn's orbit one finds $\rho_{\rm DM}\lesssim 1.1\times 10^{-20}\,\mathrm{g\,cm^{-3}}$, with mass enclosed within Saturn's orbit $< 8\times 10^{-11}\,M_\odot$~\cite{Talmadge1988PRL,Pitjeva2013,Pitjev2013}. The same datasets constrain Yukawa deviations of the form
\begin{equation}
V(r) = -\frac{GMm}{r}\,\Big[1+\alpha_{\tt Y}\,e^{-r/\lambda}\Big],
\label{eq:yukawa}
\end{equation}
where $\alpha_{\tt Y}$ is the strength and $\lambda$ the range. We convert range to mass via 
\[
  m \equiv \frac{\hbar c}{\lambda}
  \simeq \frac{1.97327\times 10^{-7}\ \mathrm{eV\cdot m}}{\lambda},
\]
where Table~\ref{tab:mass-range} shows representative parameters. 
From (\ref{eq:yukawa}), Yukawa acceleration takes the form
\begin{equation}
a_{\tt Y}(r) = -\frac{GM}{r^2}\,\alpha_{\tt Y}\!\left(1+\frac{r}{\lambda}\right)e^{-r/\lambda},
\end{equation}
so AU-scale sensitivities peak for $\lambda \sim 10^{9}$--$10^{13}\,\mathrm{m}$ (i.e., mediator masses $m \equiv \hbar c/\lambda \simeq 2\times10^{-16}$--$2\times10^{-20}\,\mathrm{eV}$), with small-body/spacecraft tracking now providing leading bounds in parts of this band. 

Complementary constraints on Yukawa--type deviations also follow from two-body orbital dynamics (analytic Keplerian solution in a Yukawa potential) with an explicit Solar-System bound~\cite{Benisty:2022Yukawa}.

\begin{figure}[t!]
\centering
\includegraphics[width=0.60\columnwidth]{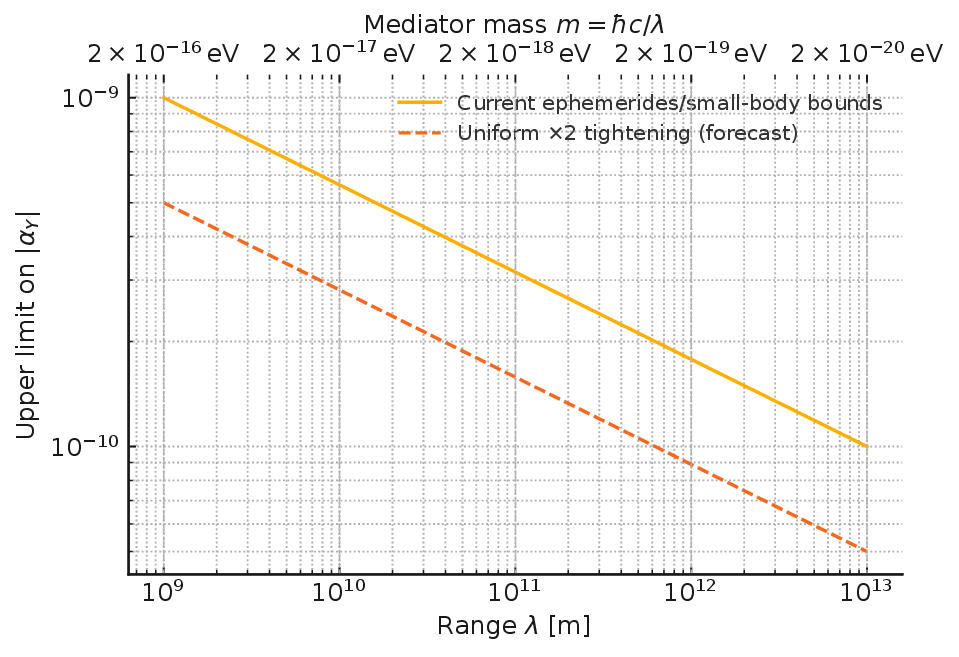}
\vskip -10pt
  \caption{Consolidated Yukawa-strength limits $|\alpha_Y|$ versus range $\lambda$.
  The solid curve is a smooth envelope anchored to representative ephemeris/small-body bounds
  $|\alpha_{\tt Y}|=10^{-9}$ at $\lambda=10^{9}\,\mathrm{m}$ and $|\alpha_{\tt Y}|=10^{-10}$ at $\lambda=10^{13}\,\mathrm{m}$;
  the dashed curve is a uniform factor-of-two tightening consistent with DSN reprocessing and improved
  asteroid catalogs (Tables~\ref{tab:mass-range} and~\ref{tab:benchmarks}). The upper axis shows $m=\hbar c/\lambda \simeq
  1.97327\times 10^{-7}\,{\rm eV\cdot m}/\lambda$.}
  \label{fig:yukawa_constraints}
\end{figure}

Across $\lambda\in[10^{9},10^{13}]$~m, current ephemerides and small-body/spacecraft tracking deliver leading bounds $|\alpha_{\tt Y}|\ll 10^{-9}$--$10^{-10}$ (depending on $\lambda$, data selection, and solar-plasma modeling~\cite{Fienga2024}), and we assume a uniform factor-of-two tightening across this band (Tables~\ref{tab:benchmarks}, \ref{tab:systematics-quant}, Fig.~\ref{fig:yukawa_constraints}).

\begin{table}[t]
\centering
\caption{Mediator mass--range map, $m=\hbar c/\lambda \simeq 1.97327\times10^{-7}\,\mathrm{eV\cdot m}/\lambda$.}
\label{tab:mass-range}
\begin{tabular}{@{}ccc@{}}
\hline
Range $\lambda$ (m) & $m$ (eV) & Comment \\
\hline\hline
$10^{9}$  & $1.97\times10^{-16}$ & inner Solar System scale \\
$10^{11}$ & $1.97\times10^{-18}$ & AU scale \\
$10^{13}$ & $1.97\times10^{-20}$ & multi--AU scale \\
\hline
\end{tabular}
\end{table}

\subsection{Ultralight dark matter (ULDM): precision clocks and interferometers}
\label{sec:DM-clocks}

\begin{table}[t]
\centering
\setlength{\tabcolsep}{6pt}
\renewcommand{\arraystretch}{1.15}
\caption{Representative Solar System observables: current bounds and  near-term targets under stated conditions.}
\label{tab:benchmarks}
\begin{tabular}{p{2.4cm}p{4.95cm}p{8.5cm}}
\hline
Observable & Current bound & Plausible near-term target (conditions) \\
\hline\hline
EEP (compositi-on-dependent $\eta$) &
$\sim 3\times10^{-15}$ (MICROSCOPE)~\cite{Touboul2022} &
$\eta \sim 10^{-16}$--$10^{-17}$ with a drag-free AIS (dual-/multi-species, long $T$, high common-mode rejection)
\cite{SAGE2019,Ahlers2022STEQUEST,Battelier2021ExpAstron,Badurina2020AION,AEDGE2020} \\
PPN $\gamma$ (light-propagation, Shapiro)
& $(2.1\pm 2.3)\times 10^{-5}$ (Cassini, solar conjunction)~\cite{Bertotti2003}
& few$\times 10^{-6}$ using BepiColombo/MORE Ka/X multi-frequency links, improved coronal-plasma calibration, optimized conjunction arcs; comparable sensitivity to $\beta$ in global fits~\cite{Iess2021,diStefano2021} \\
$\dot G/G$
& few$\times 10^{-15}\,\mathrm{yr^{-1}}$ from modern LLR+ephemerides~\cite{Pitjeva2021,Fienga2024}
& $\times(3\text{--}5)$ via sub-mm-class LLR with next-generation CCRs, improved station metrology, and higher link budgets~\cite{Turyshev2025} \\
$\rho_{\rm DM}$ at $1$--$10\,\mathrm{AU}$
& $\lesssim 10^{-20}\,\mathrm{g\,cm^{-3}}$ at Saturn; enclosed mass $< 8\times 10^{-11}\,M_\odot$~\cite{Pitjeva2013,Pitjev2013}
& $\times 2$ via long tracking arcs, NASA DSN reprocessing, and improved asteroid modeling in global ephemerides~\cite{Fienga2024} \\
ULDM (clocks)
& Leading limits for $10^{-24}$--$10^{-15}\,\mathrm{eV}$ from terrestrial and space-assisted clock comparisons~\cite{Hees2016,Wcislo2018}
& $\times(3\text{--}10)$ with space optical links and clock networks (e.g., ACES and follow-ons) to extend coherence time and suppress environmental noise~\cite{Cacciapuoti2009_ACES,Delva2017_FiberSR} \\
\hline
\end{tabular}
\end{table}

\begin{table}[t]
\centering
\setlength{\tabcolsep}{4pt}
\caption{Dominant systematics and primary mitigation for each observable.}
\label{tab:systematics}
\begin{tabular}{p{3.0cm}p{6.5cm}p{7.2cm}}
\hline
Observable & Dominant systematics & Primary mitigation \\
\hline\hline
$\gamma,\beta$ (radio science) & solar plasma; non-grav. accelerations & dual-frequency calibration; long dwell times\\
$\dot G/G,\ \eta_{\rm SEP}$ (LLR) & station timing/geometry; CCR thermal  & mm-class links; new CCRs; joint ephemerides \\
EEP (AIS) & gravity-gradient and magnetic backgrounds; wavefront/vibration noise 
& drag-free control; dual-species/common-mode rejection; gradient compensation; magnetic shielding\\
Yukawa/ephemerides & asteroid masses; SRP; thermal recoil & DSN reprocessing; asteroid catalogs; longer arcs \\
Clock DM & link noise; cavity drift; environment couplings & space--ground optical links; multi-species networks \\
\hline
\end{tabular}
\end{table}

Coherently oscillating scalar DM induces variations of fundamental constants and thus fractional frequency shifts in atomic transitions,
\begin{equation}
\frac{\delta\nu}{\nu} \simeq \sum_i K_i\,d_i\,\phi(t),
\label{eq:clock-dm}
\end{equation}
where $K_i$ are sensitivity coefficients, $d_i$ coupling parameters, and $\phi$ the DM field.  Here $d_e$ denotes the (dimensionful) coupling to the electromagnetic sector that modulates $\alpha$; we adopt $[d_i]={\rm mass}^{-1}$ so that $d_i\phi$ is dimensionless. 

For a coherently oscillating scalar $\phi(t)=\phi_0\cos(m_\phi t)$ with
$\phi_0\simeq \sqrt{2\rho_{\rm DM}}/m_\phi$, taking $\rho_{\rm DM} = 0.3\,\mathrm{GeV\,cm^{-3}}$ and $m_\phi = 10^{-18}\,\mathrm{eV}$ gives illustrative fractional modulations $\delta\nu/\nu \sim 10^{-19}$--$10^{-16}$ for \emph{representative} coupling
combinations $d_{\rm eff}\equiv \bigl|\sum_i \Delta K_i d_i\bigr|$ that satisfy existing bounds and lie within the near-term coherence-limited reach (cf. Eqs.~\eqref{eq:clock-dm}--\eqref{eq:tc-correct} and
Eq.~\eqref{eq:deff-reach} below). Values orders of magnitude smaller are fully allowed by theory; they would simply fall below the forecast sensitivity of the program (see Fig.~\ref{fig:solar-system-tests}(d)). In other words, for coherently oscillating ULDM in a virialized halo,\footnote{Normalization in (\ref{eq:clock-dm})--(\ref{eq:tc-correct}): Throughout we measure $\phi$ in energy units (mass dimension one), so the coefficients $d_i$ in \eqref{eq:clock-dm} carry inverse-energy dimension and $d_i\phi$ is dimensionless. If one prefers dimensionless couplings, define $\tilde d_i \equiv M_{\rm Pl}\,d_i$ and replace $d_i\phi \to \tilde d_i\,(\phi/M_{\rm Pl})$. None of our sensitivities depends on this convention.} the field coherence time is\footnote{Note that Eq.~(\ref{eq:tc-correct}) follows from the nonrelativistic dispersion $\omega\simeq m_\phi+p^2/(2m_\phi)$ with $p=m_\phi v$, giving a field bandwidth
$\Delta\omega\sim m_\phi v^2$ in a virialized halo; the phase coherence time is
$t_c\simeq2\pi/\Delta\omega =2\pi/(m_\phi v^2)$. Operationally, the coherent signal-to-noise ratio (SNR) saturates over $t_c$ and stacking $T/t_c$ independent bins yields ${\rm SNR}\propto\sqrt{T/t_c}$.}
{}
\begin{equation}
t_c \simeq \frac{2\pi}{m_\phi v^2}
= 4.13\times 10^{9}~\mathrm{s}\,
\left(\frac{10^{-18}\,\mathrm{eV}}{m_\phi}\right)
\left(\frac{10^{-3}}{v/c}\right)^{\!2},
\label{eq:tc-correct}
\end{equation}
where $v$ is the virial speed of the ULDM field (we adopt $v/c\simeq 10^{-3}$ as a fiducial Galactic-halo value).
Operationally, sensitivity \emph{per coherence bin} saturates at $t_c$; for a total campaign time $T$ the signal-to-noise scales as $\sqrt{N_{\rm bin}}$ with $N_{\rm bin}\!\approx\!T/t_c$ when averaging incoherently across independent bins at fixed $m_\phi$. Long-baseline optical links still enable the projected $3$--$10\times$ coupling improvements across $m_\phi\!\sim\!10^{-24}$--$10^{-15}$ eV by stabilizing the link and extending usable $T$, but the analysis should adopt the $t_c$ of \eqref{eq:tc-correct} (see \cite{VanTilburg2015PRL,StadnikFlambaum2015PRL,Arvanitaki2015PRD}.)

Setting the SNR $\simeq 1$ for a coherence-limited, narrowband search and stacking $T/t_c$ independent  bins (cf.~Eq.~(\ref{eq:tc-correct})) gives ${\rm SNR}\sim (A/\sigma_y)\sqrt{T/t_c}$. Specializing $A=|\Delta K|\,|d_{\rm eff}|\,\phi_0$ yields the per-mass coupling reach
{}
\begin{equation}
d_{\rm eff}^{\rm reach}(m_\phi) \;\equiv\;
\frac{\sigma_y(\tau)}{|\Delta K|\,\phi_0}\,
\sqrt{\frac{t_c}{T}},
\label{eq:deff-reach}
\end{equation}
where $\phi_0\simeq \sqrt{2\rho_{\rm DM}/m_\phi}$ and $t_c$ are given by
Eqs.~\eqref{eq:clock-dm}--\eqref{eq:tc-correct}, and $\sigma_y(\tau)$ is the end-to-end fractional
instability for $\tau\lesssim t_c$. Throughout the text, ``representative'' $K_i d_i$ denotes values for which
$d_{\rm eff}$ is at or within a factor of a few of $d_{\rm eff}^{\rm reach}$ under the stability targets of
Table~\ref{tab:systematics-quant} (see also Fig.~\ref{fig:solar-system-tests}(d)). This usage is detection-driven and does not impose a theoretical prior on the absolute size of the portal coefficients.

The spread of Standard-Model Yukawas is not a prior on $d_i$: these are portal couplings in a low-energy EFT. Their sizes are constrained by radiative stability and by EEP/fifth-force bounds
(Sec.~\ref{sec:EEP}) rather than by flavor structure; our ``representative'' usage is purely detection-driven. For commonly used optical transitions, $|K_\alpha|$ is typically $\mathcal{O}(10^{-1}\text{--}10^1)$ and the mass/gluon coefficients entering $\sum_i K_i d_i$ are $\mathcal{O}(1)$; the differential combination $\Delta K$ depends on the chosen pair and is documented in the clock-constraint literature (e.g., \cite{Hees2016,Wcislo2018,Cacciapuoti2009_ACES,Delva2017_FiberSR}.)

For a matched-filter search at fixed $m_\phi$, the coherent SNR saturates over $t_c$ and builds as ${\rm SNR}\propto \sqrt{T/t_c}$ when incoherently combining $N_{\rm bin}\simeq T/t_c$ independent coherence bins; hence the emphasis on extending $T$ beyond $t_c$ with stable space--ground optical links.

As an example, consider a differential comparison between transitions $A$ and $B$ sensitive to $\alpha$ with coefficients $K_\alpha^{(A)},K_\alpha^{(B)}$, for which, from (\ref{eq:clock-dm}), the fractional beat note is
\[
\Big(\frac{\delta\nu}{\nu}\Big)_{A/B} \simeq \Delta K_\alpha\,d_e\,\phi(t),\qquad \Delta K_\alpha \equiv K_\alpha^{(A)}-K_\alpha^{(B)}.
\]
Assuming a coherently oscillating field $\phi(t)=\phi_0\cos(m_\phi t)$ with $\phi_0\simeq\sqrt{2\rho_{\rm DM}}/m_\phi$ and integration time $T$, the SNR per coherence bin is
\[
\mathrm{SNR} \;\sim\; \frac{\Delta K_\alpha\,|d_e|\,\phi_0}{\sigma_y(\tau)}\,\sqrt{\frac{T}{t_c}}\,,
\]
with Allan deviation $\sigma_y(\tau)$ at interrogation time $\tau$ and $t_c\simeq 2\pi/(m_\phi v^2)$ the field coherence time. 

Terrestrial and space-assisted clock comparisons set the leading bounds over $m_\phi \sim 10^{-24}$--$10^{-15}$ eV. 

Space segments (ACES and follow-on optical links) extend the usable integration time $T$ beyond the field coherence time $t_c$ and reduce the end-to-end instability $\sigma_y(\tau\!\lesssim\!t_c)$, so the coherence-limited stacking gain ${\rm SNR}\propto\sqrt{T/t_c}$ yields the projected $3$--$10\times$ improvement in coupling reach \cite{Cacciapuoti2009_ACES,Delva2017_FiberSR} adopted below. For context, Fig.~\ref{fig:solar-system-tests}(d) summarizes the normalized sensitivity to $d_e$ across $m_\phi \in [10^{-21},10^{-15}]~\mathrm{eV}$ under current and advanced link/stability assumptions.

Figure~\ref{fig:solar-system-tests} summarizes the inputs used in the forecasts: Panel (a) shows the X-normalized residual group delay versus carrier frequency \(f\) [Hz], following the coronal dispersion law \eqref{eq:plasma-delay}. Panel (b) gives the \(1\sigma\) sensitivity to \(|\gamma-1|\) as a function of the solar impact parameter \(b/R_\odot\), set by the Shapiro time delay \eqref{eq:shapiro} against the calibrated residual noise; reference lines at Cassini \(2.3\times10^{-5}\) and the \(1\times10^{-6}\) target are indicated. Panel (c) maps a null \(\gamma\) bound into a maximum solar thin-shell fraction \(\Delta R/R\) as a function of \(\mu_{\rm lin,0}\) using \eqref{eq:chi_from_mu}--\eqref{eq:phi_contrast_bound}. Panel (d) summarizes the normalized clock sensitivity to an \(\alpha\)-coupled coefficient \(d_e\) versus \(m_\phi\) [eV] from the coherence-time/bandwidth scalings (Eqs.~(\ref{eq:clock-dm})--(\ref{eq:tc-correct})); “current” and “advanced links/stability” reflect the link-noise and integration assumptions used in the analysis. Although not plotted, the AIS EEP channel enters the same screening map via Eq.~(\ref{eq:eta-AIS}), supplying an Earth-thin-shell guardrail complementary to the Sun-thin-shell bound in panel (c).

\begin{table}[t]
\centering
\caption{Quantitative systematics targets enabling the improvements in Table~\ref{tab:benchmarks}. Allan deviation $\sigma_y$ is shown at the indicated averaging times. 
The uniform tightening factors we adopt here (e.g., $\times 2$ for AU-scale Yukawa limits or $\times(3$--$10)$ for clock-based ULDM couplings) apply to the quantitative targets listed for the dominant noise terms (cf.\ Sec.~\ref{sec:observables}).
}
\label{tab:systematics-quant}
\begin{tabular}{p{2.6cm}p{4.1cm}p{9.5cm}}
\hline
Observable & Dominant term & Quantitative target \\ \hline\hline
PPN $\gamma$ (radio) & Coronal group delay $\propto f^{-2}$ & Dual-frequency Ka/X calibration; residual group delay $\le 0.1$ ns for $b\ge5\,R_{\odot}$; long dwells (10 d); accelerometry/thermal modeling\\
PPN $\gamma$ (optical) & Coronal group delay $\propto f^{-2}$ & Optical vs.\ X: $(f_{\rm X}/f_{\rm opt})^{2}\!\simeq\!1.8\times10^{-9}$; vs.\ Ka: $(f_{\rm Ka}/f_{\rm opt})^{2}\!\simeq\!2.6\times10^{-8}$. Target: $\ge 5\times10^{8}$ (vs.\ X) and $\ge 4\times10^{7}$ (vs.\ Ka) raw reduction; add dual-freq/model margin.\\
PPN $\gamma,\beta$ (global) &
Non-grav.\ accel.\ modeling &
$\lesssim 10^{-11}\,\mathrm{m\,s^{-2}}$ bias over 10-day dwell \\
$\dot{G}/G$ (LLR) &
Normal-point precision &
$30$--$50~\mu\mathrm{m}$ with high-power LLR operations and next-gen CCRs\\
Yukawa $\alpha_{\tt Y}(\lambda)$ &
Asteroid masses/SRP/recoil &
Updated catalogs $+$ multi-year DSN arcs \\
ULDM (clocks) &
Link/cavity noise over $t_c$ &
Allan deviation $\sigma_y\!\lesssim\!10^{-18}$ at $10^5$--$10^6$ s \\
EEP (AIS) & Differential phase stability & Supports $\eta \sim 10^{-16}\!-\!10^{-17}$ over mission times 
(via long $T$, dual-species, and high common-mode rejection)\\
\hline
\end{tabular}
\end{table}

\section{What should a Solar System program look like?}
\label{sec:program}

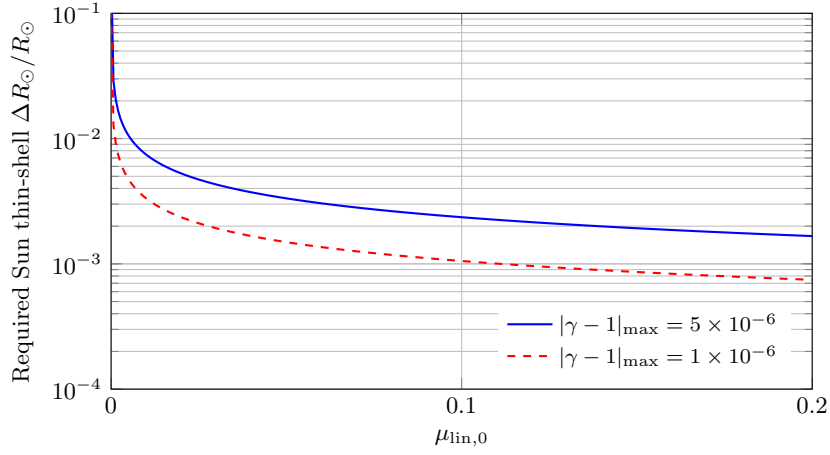
\begin{figure}[t!]
\centering
\begin{tikzpicture}
\begin{axis}[
  width=0.52\linewidth, height=5.0cm,
  scale only axis,
  xmin=0, xmax=0.20,
  ymin=1e-4, ymax=1e-1,
  ymode=log,
  xlabel={$\mulin$},
  ylabel={Required Sun thin-shell $\Delta R_\odot/R_\odot$},
  xtick={0,0.1,0.2},
  xmajorgrids=true, xminorgrids=false,
  ymajorgrids=true, yminorgrids=true,
  legend pos=south east,
  legend style={draw=none, fill=white, font=\footnotesize}
]
\addplot+[domain=1e-5:0.20, samples=300, thick, no marks] 
  {(1/(3*sqrt(x/2)))*sqrt(5e-6/2)};
\addlegendentry{$|\gamma-1|_{\max}=5\times10^{-6}$};

\addplot+[domain=1e-5:0.20, samples=300, dashed, thick, no marks] 
  {(1/(3*sqrt(x/2)))*sqrt(1e-6/2)};
\addlegendentry{$|\gamma-1|_{\max}=1\times10^{-6}$};
\end{axis}
\end{tikzpicture}
\vspace{-6pt}
\caption{Thin-shell requirement for the Sun implied by a null detection of solar-conjunction PPN $\gamma$ at sensitivity $|\gamma-1|_{\max}$ as a function of the
cosmology-level linear response $\mulin$ in the conformal scalar benchmark (adopts $k\simeq 0.1\,h\,\mathrm{Mpc}^{-1}$ for $\mulin$; see Sec.~\ref{sec:theory}.) For $\mulin\simeq 0.10$, the $|\gamma-1| \lesssim 5\times10^{-6}$ target implies $\Delta R/R \lesssim 2.4\times 10^{-3}$ [Eqs.~(\ref{eq:chi_from_mu})--(\ref{eq:phi_contrast_bound})]. (See Appendix~\ref{sec:recipe}.)
}
\label{fig:mapping}
\end{figure}

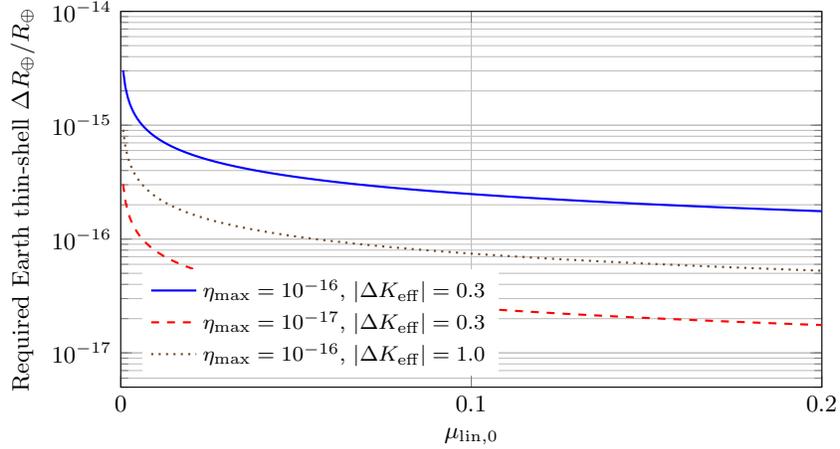
\begin{figure}[t!]
\centering
\begin{tikzpicture}
\begin{axis}[
  width=0.52\linewidth, height=5.0cm,
  scale only axis,
  xmin=0.0, xmax=0.20,
  ymin=5e-18, ymax=1e-14,
  ymode=log,
  xlabel={$\mu_{\rm lin,0}$},
  ylabel={Required Earth thin-shell $\Delta R_\oplus/R_\oplus$},
  xtick={0.0,0.10,0.20},
  xmajorgrids=true, xminorgrids=false,
  ymajorgrids=true, yminorgrids=true,
  legend pos=south west,
  legend style={draw=none, fill=white, font=\footnotesize}
]
\addplot+[domain=0.0:0.20, samples=300, no marks, thick]
  {(1/(6*sqrt(x/2)))*(1e-16/0.3)};
\addlegendentry{$\eta_{\max}=10^{-16}$,\ $|\Delta K_{\rm eff}|=0.3$};

\addplot+[domain=0.00:0.20, samples=300, dashed, no marks, thick]
  {(1/(6*sqrt(x/2)))*(1e-17/0.3)};
\addlegendentry{$\eta_{\max}=10^{-17}$,\ $|\Delta K_{\rm eff}|=0.3$};

\addplot+[domain=0.00:0.20, samples=300, dotted, no marks, thick]
  {(1/(6*sqrt(x/2)))*(1e-16/1.0)};
\addlegendentry{$\eta_{\max}=10^{-16}$,\ $|\Delta K_{\rm eff}|=1.0$};
\end{axis}
\end{tikzpicture}
\vspace{-6pt}
\caption{Null $\eta~\Rightarrow$~Earth thin-shell bound.
Uses the AIS/EEP relation \eqref{eq:eta-AIS} with the thin-shell expression \eqref{eq:thin-shell}
and the cosmology bridge $\chi=\sqrt{\mu_{\rm lin,0}/2}$ from \eqref{eq:chi_from_mu}. In the screened-source limit,
$\Delta R_\oplus/R_\oplus \le \eta_{\max}/(6\,\chi\,|\Delta K_{\rm eff}|)$ [cf.\ \eqref{eq:eta-EEP-worked}],
so stronger $\eta_{\max}$ and larger $|\Delta K_{\rm eff}|$ push the required thin shell lower.}
\label{fig:eta-guardrail}
\end{figure}

Figure~\ref{fig:solar-system-tests} assembles the measurement landscape: panel (a) shows dispersion scaling, (b) the conjunction $\gamma$ reach versus $b/R_\odot$, (c) the implied thin-shell bound, and (d) clock sensitivity to $d_e$ versus $m_\phi$ under current and advanced link assumptions; together with the thin-shell relation in Eq.~\eqref{eq:thin-shell-bound}, these curves set the decision rule we use below. Using the thin-shell relation in (\ref{eq:thin-shell-bound}), Fig.~\ref{fig:mapping} shows the resulting $\Delta R/R$ versus $\mu_{\rm lin,0}$ for two choices of $|\gamma-1|_{\max}$. Fig.~\ref{fig:eta-guardrail} shows the maximum Earth thin-shell fraction implied by a null EEP test for representative $\eta_{\max}$ and $|\Delta K_{\rm eff}|$.

For decision-making it is useful to distinguish three programmatic modes while keeping the same measurement physics. A guardrail-plus-selective-discovery mode reuses existing radio/optical links, LLR, clock networks, and ephemerides to maximize science return per cost now (Sec.~\ref{sec:observables}). A trigger-driven escalation mode—activated either by a local anomaly or by a forecast in which a specified microphysical model predicts a residual above vetted thresholds by a comfortable margin—targets the relevant observable with tailored conjunction arcs, composition/baseline rotation, or a small dedicated payload. A dedicated Solar-System--first mode pursues transformational improvements (e.g., optical deflection/near-Sun astrometry and precision gradiometry) at higher cost and risk. Our baseline assumes finite budgets and therefore emphasizes the first mode while preserving a clear trigger path to the second and, when justified by data, to the third. In an unlimited-funds scenario the first and third modes would proceed in parallel.

Based on the analysis above, we recommend a targeted, cost-effective portfolio that rides on planned space assets and existing ground infrastructure:
{}
\begin{enumerate}
\item\textit{Solar-conjunction radio science for $\gamma$ and $\beta$:} Implementation: multi-frequency Ka/X links with  BepiColombo/ MORE and future conjunction opportunities, improved coronal-plasma calibration (dual-frequency group-delay and Faraday-rotation constraints with contemporaneous solar data), optimized low-impact-parameter arcs ($b/R_\odot \simeq 3\text{--}5$) and long dwell times ($\gtrsim 10$ days) (see Fig.~\ref{fig:solar-system-tests}b), sharpening the slope fit in (\ref{eq:shapiro}). Quantitative target: $|\gamma-1| \lesssim \mathrm{few}\times 10^{-6}$ per conjunction; comparable sensitivity to $\beta-1$ from global ephemeris fits. Dominant systematics: coronal turbulence and dispersion modeling, spacecraft non-gravitational accelerations, tropospheric delay calibration~\cite{Iess2021,diStefano2021}.

\item\textit{Sustained mm-class LLR:}
Implementation: next-generation corner-cube retroreflectors with reduced thermal gradients, higher link budgets (kW-class lasers, larger apertures), improved station timing/metrology, and consistent global analysis with modern planetary ephemerides.  
Quantitative target: $|\dot G/G| \lesssim \mathrm{few}\times 10^{-15}\,\mathrm{yr^{-1}}$; factor $3$--$5\times$ tightening of SEP constraints relative to current solutions.  
Dominant systematics: station geometry and thermal control, retroreflector thermal lensing, model degeneracies with tidal parameters~\cite{Fienga2024,Turyshev2025-CCR,Turyshev2025}.

\item\textit{Global optical clock links (ground + space):}
Implementation: long-baseline optical time/frequency transfer (ACES and follow-on optical links), multi-species comparisons to decorrelate sensitivity coefficients $K_i$, and campaign lengths exceeding coherence times for the targeted ultralight--mass window. Quantitative target: improvement by a factor $3$--$10$ in scalar-coupling bounds across $m_\phi \sim 10^{-24}$--$10^{-15}\,\mathrm{eV}$; fractional stability at $10^{-18}$--$10^{-19}$ over $10^{5}$--$10^{6}$\,s typical integration.  
Dominant systematics: link noise and cycle slips, environmental couplings (temperature, magnetic fields), long-term drift of reference cavities~\cite{Cacciapuoti2009_ACES,Hees2016,Wcislo2018,Delva2017_FiberSR}.

\item\textit{Ephemerides and small-force systematics:}
Implementation: reprocess Deep Space Network tracking with updated media calibrations, extend multi-year arcs, refine asteroid catalogs and nongravitational-force models (thermal recoil, solar radiation pressure), and perform joint fits with improved solar-corona priors.  Quantitative target: factor $\sim 2$ tightening of AU-scale Yukawa strength limits $|\alpha_{\tt Y}(\lambda)|$ across $\lambda \sim 10^{9}$--$10^{13}\,\mathrm{m}$ and of smooth Solar System dark-matter density bounds (e.g., at Saturn’s orbit from $\sim 1\times 10^{-20}$ to $\sim 5\times 10^{-21}\,\mathrm{g\,cm^{-3}}$). Dominant systematics: asteroid-mass uncertainties, spacecraft thermal systematics, solar-plasma residuals in inner-planet ranges~\cite{Fienga2024,Pitjeva2013}.

\item \textit{Space atom interferometer (AIS) for EEP:}
Implementation: drag-free spacecraft, dual-/multi-species interferometers with long $T$ and common-mode rejection.
Quantitative target: $\eta \sim 10^{-16}\!-\!10^{-17}$ under realistic $T$ and vibration budgets; sensitivity to ULDM-induced, species-dependent modulations across $m_\phi\!\sim\!10^{-24}\text{--}10^{-15}\,\mathrm{eV}$ complementary to clock networks. 
Dominant systematics: gravity gradients and magnetic backgrounds; addressed by gradient compensation, magnetic shielding, and drag-free control. \emph{Notes:} Constellation options (e.g., tetrahedral trace measurements) provide an explicitly DE-driven path targeting Galileon/Vainshtein sectors \cite{Ahlers2022STEQUEST,Battelier2021ExpAstron,TetPRD2024}.
\end{enumerate}

Beyond scientific leverage, we weigh cost‑effectiveness qualitatively by (i) reusing flight‑proven infrastructure where possible (e.g., DSN Ka/X, DSOC‑class optical terminals, existing LLR stations), (ii) favoring opportunistic arcs and networked analyses that accrue \(T_{\rm dwell}\) or baseline diversity without new spacecraft, and (iii) prioritizing probes with direct parameter--signal maps (e.g., \(|\gamma-1|,\,\eta_{\rm EEP},\,\alpha_{\rm Y}(\lambda),\,d_e\)) and clean systematics budgets. Under these heuristics, the radio/optical conjunctions, sub-mm‑class LLR, clock/AIS networks, and ephemeris reprocessing deliver the highest science‑return‑per‑cost now; a dedicated mission is triggered only if the joint {cosmology\(\to\)local} map predicts (or a Solar‑System anomaly reveals) a signature clearing the vetted local sensitivity by a factor of a few (Sec.~\ref{sec:intro}  and Fig.~\ref{fig:solar-system-tests}).

Below is the systematic risk summary for a near-term program:
\begin{itemize}
\item \textit{Near-Sun optical links:} Plasma group delay scales as $f^{-2}$; for $b \lesssim 5R_\odot$ stray light and thermal drifts dominate. \emph{Require} terminal pointing jitter $\le 1~\mu\mathrm{rad}_\mathrm{RMS}$ (0.1--10 Hz), pupil temperature gradients $\le 0.5~\mathrm{K}_\mathrm{RMS}$ over $10^{2}$--$10^{3}$ s, and in-band stray-light suppression $\ge 10^{7}$ at the detector. \emph{Mitigate} with apodized baffling, narrowband filtering/FOV control, and active thermal regulation of the optical head.
\item \textit{LLR micro-metrology:} CCR thermal lensing and station timing chains can saturate gains. \emph{Require} CCR $\Delta T\le 0.1~\mathrm{K}$ over $10^{3}$ s, event-timer jitter $\le 3~\mathrm{ps}$ {RMS}, verified two-way time transfer $\le 10~\mathrm{ps}$, and station reference $\sigma_y(100\text{--}1000~\mathrm{s})\le 3\times 10^{-15}$. \emph{Mitigate} with IR wavelengths and single-CCR designs, timer cross-calibration, and redundant calibration passes. Here ``IR'' denotes infrared (e.g., 1064 nm) LLR wavelengths.
\item \textit{Ephemerides:} Asteroid mass priors and solar-plasma residuals can bias AU-scale $\alpha(\lambda)$. \emph{Require} inclusion of $\ge 300$ main-belt asteroids with $\sigma_M/M \le 20\%$, and Ka-band corona calibration with residual group delay $\le 0.1~\mathrm{ns}$ for $b\ge5R_\odot$. \emph{Mitigate} via joint fits with contemporaneous solar-wind/TEC data and periodic catalog updates/reweighting of conjunction windows.
\item \textit{Clocks/links:} Over the expected coherence time $t_c$, link and cavity noise must be subdominant. \emph{Require} end-to-end instability $\sigma_y(\tau\!=\!t_c)\le 3\times 10^{-16}$, cycle-slip probability $<10^{-6}$ per $t_c$, optical-comb integrated phase noise (1 Hz--1 kHz) $\le 0.3~\mathrm{rad}$, and multi-species comparisons with $|\Delta K_i|\ge 0.1$ to decorrelate couplings. \emph{Mitigate} via dual independent links, real-time slip detection/repair, and alternating species schedules.
\item \textit{Precision gradiometry:} Formation-keeping and scale-factor drift can masquerade as beyond-PPN signals. \emph{Require} $\le 10~\mathrm{pm}/\sqrt{\mathrm{Hz}}$ inter-satellite laser metrology and $\le 10~\mathrm{nrad}/\sqrt{\mathrm{Hz}}$ attitude jitter (1--100 mHz), baseline knowledge $\le 1~\mathrm{mm}$ (10--100 km), and bias stability $\le 1\times 10^{-12}~\mathrm{m\,s^{-2}}$ over $10^{4}$ s. \emph{Mitigate} with sign-reversal geometries, calibration slews, and thermal scale-factor tracking.
\item \textit{Space-based atom interferometry (AIS):} Vibration, wavefront, and $B$-gradient systematics dominate for long-$T$. \emph{Require} drag-free $\le 3\times 10^{-15}~\mathrm{m\,s^{-2}}/\sqrt{\mathrm{Hz}}$ (0.1--10 mHz), magnetic-field gradients $\le 1~\mathrm{nT\,m^{-1}}$, Raman/Bragg phase noise $\le 1~\mathrm{mrad}/\sqrt{\mathrm{Hz}}$, and common-mode rejection $\ge 120~\mathrm{dB}$ to reach $\eta\!\approx\!10^{-16}\text{--}10^{-17}$ with $T\!\ge\!5\text{--}10~\mathrm{s}$. \emph{Mitigate} with active vibration cancellation, magnetic shielding/trim coils, wavefront sensing, and continuous slip detection.
\end{itemize}

Further improvements are possible: 
Optical links enable PPN tests with substantially reduced coronal plasma noise (dispersion $\propto f^{-2}$),  offering $\sim 10^{8}\!-\!10^{9}$ lower group-delay systematics than X/Ka at $f \sim 2\times10^{14}$ Hz~\cite{Bertotti2003}. Interplanetary laser ranging (ILR) has already demonstrated sub-ns timing over tens of millions of km with asynchronous laser transponders, validating Shapiro-delay--grade timing on optical carriers~\cite{Smith2006_Science}. Recent deep-space optical communications (DSOC) links have shown robust high-rate operation at $\sim$0.2--1.5 AU, indicating operational readiness of narrow-beam, high-SNR optical terminals for precise time/frequency transfer~\cite{NASA_DSOC_Page}. For LLR, high-power $1064\,\mathrm{nm}$ systems and differential LLR (dLLR) \cite{Turyshev2025} with next-generation CCRs \cite{Turyshev2025-CCR} can drive normal-point precision toward the $\sim30\,\mu$m regime, tightening constraints on $\dot G/G$ and SEP (\emph{via} the Nordtvedt parameter) and improving $\beta$ through global fits~\cite{Murphy2012_CQG,Zhang2024_AA}. Concept studies of optical deflection/near-Sun astrometry (e.g., LATOR \cite{Turyshev2004_CQG,Turyshev2007_LATOR}) and optical interferometry (e.g., BEACON \cite{Turyshev2009-BEACON}) further indicate potential improvements in $\gamma$ sensitivity by orders of magnitude if stray-light and thermal-control issues are addressed~\cite{Turyshev2004_CQG}.

Decision rule for dedicated Solar System missions: Authorize a dedicated Solar System mission only if a specified microphysical model (with explicit $\{V(\phi),A(\phi)\}$ or
well-defined dark-sector couplings) \emph{predicts} at least one local signature that exceeds credible thresholds
set by Solar System potentials—e.g., violates the thin-shell guardrails implied by Eqs.~(\ref{eq:thin-shell-bound}) and \eqref{eq:phi_star_bound}--\eqref{eq:Lambda_bound} or by the bounds in Eqs.~\eqref{eq:eta-screened}--\eqref{eq:chi-EEP-unscreened}—with a margin of a few for systematics. Otherwise, prioritize opportunistic radio/optical
links, mm-class LLR, networked clocks/AIS, and ephemeris reprocessing \cite{Iess2021,diStefano2021,Fienga2024,Turyshev2025,Cacciapuoti2009_ACES,Hees2016,Wcislo2018,Delva2017_FiberSR,Pitjeva2013}. (Concept studies also consider multi-arm or tetrahedral geometries to enhance common-mode rejection and gradient control \cite{TetPRD2024}.)

\section{Conclusions}
\label{sec:conclusions}

Solar System experiments are hypothesis-driven tests under the priors of Sec.~\ref{sec:theory}: they (i) enforce universal guardrails any dark-energy or dark-matter model must satisfy, (ii) prune unscreened or weakly screened regions of theory space, and (iii) provide discovery windows for ultralight or long-range sectors.

The multimessenger bound on the gravitational-wave speed, $|c_{\tt T}/c - 1|\lesssim 10^{-15}$~\cite{Baker2017,Creminelli2017}, further couples cosmological and local regimes, making cross-checks logically tight rather than exploratory. Equally, a verified Solar-System detection is a stand-alone discovery that warrants targeted follow-ons irrespective of cosmological survey cadence.

Our recommended strategy is asymmetric but explicitly two-branched. Cosmology carries the discovery prior for late-time acceleration; DESI and \emph{Euclid} determine $\{w(z),\mu(z,k),\Sigma(z,k)\}$ at percent-level precision in two-point statistics (Sec.~\ref{sec:cosmo}).  In the detection-first branch, a verified EEP violation, a $| \gamma-1 |$ signal at the few$\times 10^{-6}$ level, a Yukawa tail, or a narrowband ULDM line triggers a joint re-fit across regimes and motivates targeted follow-on missions without waiting for cosmology; if the signal points beyond universal conformal scalars, the analysis broadens to disformal, vector, axion-like, spin-2, or gravity-only ULDM sectors (Appendices~\ref{app:disformal}, \ref{sec:ULDM-broader}). 

The Solar System program then targets the specific residuals implied by those posteriors, at forecastable levels set by Solar System potentials $\Phi_N$ and by the screening maps in Eqs.~\eqref{eq:thin-shell}--\eqref{eq:PhiN-numbers}, \eqref{eq:gamma_from_shell}--\eqref{eq:phi_contrast_bound}, and \eqref{eq:rV}--\eqref{eq:gamma_Vain}. Concretely, the near-term measurement goals in Sec.~\ref{sec:observables} and the program elements in Sec.~\ref{sec:program} deliver:
\begin{enumerate}
\item $|\gamma-1|\lesssim \mathrm{few}\times 10^{-6}$ per solar conjunction using Ka/X or optical links with improved coronal calibration (optical strongly suppresses plasma terms); comparable sensitivity to $\beta-1$ from global fits. Target the $\lesssim 1\times 10^{-8}$ regime with advanced astrometric and/or optical metrology experiments \cite{Turyshev2004_CQG,Turyshev2007_LATOR,Turyshev2009-BEACON}.  
\item $|\dot G/G|\lesssim \mathrm{few}\times 10^{-15}\,\mathrm{yr^{-1}}$ and a factor $3$--$5$ tightening of SEP (Nordtvedt) constraints via sub-mm-class LLR with next-generation CCRs and station upgrades for high-power operations. Expect a factor of 50--70 improvement with the new high-power LLR facilities and new CCRs \cite{Turyshev2025,Turyshev2025-CCR}. 
\item A factor $3$--$10$ improvement in clock-based limits on ultralight-scalar couplings across $m_\phi\sim 10^{-24}$--$10^{-15}\,\mathrm{eV}$ using long-baseline ground--space optical links. Tetrahedral spacecraft formations with optical metrology and AIS offer additional  factor of 25-50 improvements \cite{TetPRD2024}.
\item A factor $\sim 2$ strengthening of AU-scale Yukawa bounds $|\alpha_{\tt Y}(\lambda)|$ for $\lambda\sim 10^{9}$--$10^{13}\,\mathrm{m}$, and of smooth Solar System dark-matter density limits (e.g., at Saturn’s orbit from $\sim 1.1\times 10^{-20}$ to $\sim 5\times 10^{-21}\,\mathrm{g\,cm^{-3}}$) through DSN reprocessing and refined ephemerides.
\item Precision EEP: retain MICROSCOPE-level guardrails and target $\eta \!\sim\!10^{-16}\!-\!10^{-17}$ 
with a drag-free AIS (as a universal test, not a model-blind DE probe). This complements recent consolidated constraint surveys~\cite{FischerKaedingPitschmann2024}.

\end{enumerate}

The cross-regime mapping is numerically tractable. For example, an illustrative cosmology-level excess $\mu_0^{\rm lin}\simeq 0.10$ at $k\simeq 0.1\,h\,\mathrm{Mpc}^{-1}$ implies $\chi\simeq 0.224$ for a conformally coupled scalar. Solar-conjunction bounds  taking $|\gamma-1|\lesssim (2-5)\times10^{-6}$ yields $\Delta R/R \lesssim (1.6-2.4)\times10^{-3}$
and $|\phi_\infty-\phi_c| \lesssim (4.3-6.7)\times10^{-9}M_{\rm Pl}$ [Eqs.~\eqref{eq:thin-shell}--\eqref{eq:phi_contrast_bound}]. A null result at that sensitivity prunes the unscreened interpretation of the cosmological signal; a detection demands a joint re-fit of cosmology and Solar System data with the same microphysical parameters. In Vainshtein-screened models, by contrast, the residual at $1\,\mathrm{AU}$ scales as $(r/r_V)^{3/2}\sim 10^{-11}$ for $r_{V\odot}\sim 10^{2}\,\mathrm{pc}$, explaining the natural weakness of local tests even when cosmology shows percent-level deviations (Sec.~\ref{sec:theory}).

In the light of the discussion above, we recommend specific prioritized near-term portfolio:
{}
\begin{enumerate}
\item{Radio/optical solar-conjunction arcs for $\gamma$ (and $\beta$ in global fits).}
Dual-frequency Ka/X arcs and/or deep-space optical links reduce coronal dispersion by $\sim 10^{8-9}$ vs.\ X-band and enable
$|\gamma-1|\!\lesssim\!\text{few}\times 10^{-6}$ at $b\gtrsim 5\,R_\odot$ with a $\lesssim 0.1$\,ns residual group-delay budget (Fig.~\ref{fig:solar-system-tests}(a,b); Tables~\ref{tab:benchmarks}--\ref{tab:systematics-quant}). This directly sharpens the Sun thin-shell guardrail that maps to cosmology through $\chi\simeq\sqrt{\mu_{\rm lin,0}/2}$ and the conjunction null-test bounds in Eqs.~\eqref{eq:gamma_from_shell}--\eqref{eq:thin-shell-bound}.

\item{Millimeter-class LLR (new CCRs + high-power stations).}
Sustained mm-class LLR with next-generation corner-cube retroreflectors pushes $|\dot G/G|\!\lesssim\! \text{few}\times 10^{-15}\,\text{yr}^{-1}$ and tightens SEP constraints by $3$--$5\times$, complementing $\gamma$ and feeding $\beta$ in global ephemerides (Tables~\ref{tab:benchmarks}--\ref{tab:systematics-quant}).

\item{Global optical-clock links for ULDM and EEP.}
Networked optical clocks with space links deliver $3$--$10\times$ gains on scalar couplings across $m_\phi\!\sim\!10^{-24}\text{--}10^{-15}\,\text{eV}$ via coherence-limited stacking (Fig.~\ref{fig:solar-system-tests}(d); Table~\ref{tab:systematics-quant}),
and a STE-QUEST--class AIS provides a universal EEP guardrail at $\eta\!\sim\!10^{-16}\text{--}10^{-17}$, which maps to the same slope $\chi$ that appears in cosmology through Eqs.~\eqref{eq:eta-AIS-E} and \eqref{eq:eta-AIS-clock}.
\end{enumerate}

Finally, we emphasize that Solar System experiments constrain DE and DM in distinct, complementary ways:
\begin{itemize}
\item \textit{For DE}, screening in deep Solar potentials typically suppresses predicted local residuals (e.g., $\gamma-1$, $\beta-1$, $\dot G/G$) to \emph{at or just below} current sensitivity --- often within a factor of $\sim\!2$--$4$ of present bounds, with the precise target set by the ambient density $\rho_\infty$ (see \eqref{eq:thin-shell-bound} and Table~\ref{tab:ambient-density}). The role of local tests is therefore (i) to enforce universal null tests (EEP/PPN: $\gamma$, $\beta$, $\eta_{\rm SEP}$, $\dot G/G$) and (ii) to interrogate \emph{targeted, plausibly unscreened} corners that DESI/\emph{Euclid} flag via the theory bridge $V_{\rm eff}(\phi;\rho)=V(\phi)+\rho\,A(\phi)$.

\item \textit{For DM}, by contrast, Solar System probes offer \emph{selective discovery reach} with clean systematics and direct parameter--signal maps: long--baseline clock networks and AIS for ultralight fields ($m_\phi\!\sim\!10^{-24}$--$10^{-15}\,$eV; coherence--limited searches), high--precision ephemerides for AU--scale Yukawa forces ($\lambda\!\sim\!10^{9}$--$10^{13}$\,m; $\lambda\!\equiv\!\hbar/(m_\phi c)$), and, where metric couplings apply, light--propagation tests (Shapiro delay/deflection) constraining $\gamma$.
\end{itemize}

Thus, dedicated Solar System missions are warranted when a specified microphysical model with explicit $\{V(\phi),A(\phi)\}$ (or an explicit dark--sector candidate) predicts at least one \emph{local} signature that exceeds credible detection thresholds \emph{after} allocating systematic--error budgets --- preferably by a factor of a few for margin (cf.\ Eq.~\eqref{eq:thin-shell-bound}, Table~\ref{tab:ambient-density}). Otherwise, opportunistic radio/optical links, mm--class LLR, networked clocks/AIS, and ephemeris reprocessing deliver the highest science return per cost (Sec.~\ref{sec:program}). Intrinsic limitations remain: local tests primarily probe spatial modes near the AU scale ($k\!\sim\!\mathrm{AU}^{-1}$); they are systematics--limited (e.g., coronal plasma for radio, strongly mitigated by optical links; station geometry/thermal effects for LLR; asteroid masses and non--gravitational forces in ephemerides) and require an explicit theory map from cosmological posteriors to local residuals.

The cost-effectiveness rationale is explicit: under finite budgets the reuse of flight-proven microwave/optical links, LLR infrastructure, clock networks, and ephemerides provides the highest immediate science return, with escalation to dedicated missions triggered by well-specified, above-threshold residuals; with unconstrained resources, the guardrail program and dedicated Solar-System--first experiments would proceed in parallel (Sec.~\ref{sec:program}).

Within these bounds, Solar System experiments act as precision discriminants: they either reveal residual new physics consistent with cosmological hints in the targeted sectors above, or they excise model families that would otherwise remain viable from cosmology alone. 

\section*{Acknowledgments}

The author expresses gratitude to Curt J. Cutler, Jason D. Rhodes and Eric M. Huff of JPL  who provided valuable comments, encouragement, and stimulating discussions while this document was in preparation. The work described here was carried out at the Jet Propulsion Laboratory, California Institute of Technology, Pasadena, California, under a contract with the National Aeronautics and Space Administration.


\begin{thebibliography}{81}%
\makeatletter
\providecommand \@ifxundefined [1]{%
 \@ifx{#1\undefined}
}%
\providecommand \@ifnum [1]{%
 \ifnum #1\expandafter \@firstoftwo
 \else \expandafter \@secondoftwo
 \fi
}%
\providecommand \@ifx [1]{%
 \ifx #1\expandafter \@firstoftwo
 \else \expandafter \@secondoftwo
 \fi
}%
\providecommand \natexlab [1]{#1}%
\providecommand \enquote  [1]{``#1''}%
\providecommand \bibnamefont  [1]{#1}%
\providecommand \bibfnamefont [1]{#1}%
\providecommand \citenamefont [1]{#1}%
\providecommand \href@noop [0]{\@secondoftwo}%
\providecommand \href [0]{\begingroup \@sanitize@url \@href}%
\providecommand \@href[1]{\@@startlink{#1}\@@href}%
\providecommand \@@href[1]{\endgroup#1\@@endlink}%
\providecommand \@sanitize@url [0]{\catcode `\\12\catcode `\$12\catcode
  `\&12\catcode `\#12\catcode `\^12\catcode `\_12\catcode `\%12\relax}%
\providecommand \@@startlink[1]{}%
\providecommand \@@endlink[0]{}%
\providecommand \url  [0]{\begingroup\@sanitize@url \@url }%
\providecommand \@url [1]{\endgroup\@href {#1}{\urlprefix }}%
\providecommand \urlprefix  [0]{URL }%
\providecommand \Eprint [0]{\href }%
\providecommand \doibase [0]{https://doi.org/}%
\providecommand \selectlanguage [0]{\@gobble}%
\providecommand \bibinfo  [0]{\@secondoftwo}%
\providecommand \bibfield  [0]{\@secondoftwo}%
\providecommand \translation [1]{[#1]}%
\providecommand \BibitemOpen [0]{}%
\providecommand \bibitemStop [0]{}%
\providecommand \bibitemNoStop [0]{.\EOS\space}%
\providecommand \EOS [0]{\spacefactor3000\relax}%
\providecommand \BibitemShut  [1]{\csname bibitem#1\endcsname}%
\let\auto@bib@innerbib\@empty
\bibitem [{\citenamefont {{DESI
  Collaboration}}(2025{\natexlab{a}})}]{DESI2024BAO}%
  \BibitemOpen
  \bibfield  {author} {\bibinfo {author} {\bibnamefont {{DESI
  Collaboration}}},\ }\bibfield  {title} {\bibinfo {title} {{DESI 2024 VI:
  Cosmological Constraints from the Measurements of Baryon Acoustic
  Oscillations ({BAO})}},\ }\href
  {https://doi.org/10.1088/1475-7516/2025/02/021} {\bibfield  {journal}
  {\bibinfo  {journal} {JCAP}\ }\textbf {\bibinfo {volume} {2025}}\bibinfo
  {number} { (02)},\ \bibinfo {pages} {021}}\BibitemShut {NoStop}%
\bibitem [{\citenamefont {{DESI
  Collaboration}}(2025{\natexlab{b}})}]{DESI2024FS}%
  \BibitemOpen
\bibfield  {number} {  }\bibfield  {author} {\bibinfo {author} {\bibnamefont
  {{DESI Collaboration}}},\ }\bibfield  {title} {\bibinfo {title} {{DESI 2024
  VII: Cosmological Constraints from the Full-Shape Modeling of Clustering
  Measurements}},\ }\href {https://doi.org/10.1088/1475-7516/2025/07/028}
  {\bibfield  {journal} {\bibinfo  {journal} {JCAP}\ }\textbf {\bibinfo
  {volume} {2025}}\bibinfo  {number} { (07)},\ \bibinfo {pages}
  {028}}\BibitemShut {NoStop}%
\bibitem [{\citenamefont {{Euclid Collaboration}}\ \emph
  {et~al.}(2025)\citenamefont {{Euclid Collaboration}}, \citenamefont
  {{Aussel}}, \citenamefont {{Tereno}}, \citenamefont {{Schirmer}},
  \citenamefont {{Alguero}}, \citenamefont {{Altieri}}, \citenamefont
  {{Balbinot}}, \citenamefont {{de Boer}},\ and\ \citenamefont
  {et~al.}}]{EuclidQ1}%
  \BibitemOpen
\bibfield  {number} {  }\bibfield  {author} {\bibinfo {author} {\bibnamefont
  {{Euclid Collaboration}}}, \bibinfo {author} {\bibfnamefont {H.}~\bibnamefont
  {{Aussel}}}, \bibinfo {author} {\bibfnamefont {I.}~\bibnamefont {{Tereno}}},
  \bibinfo {author} {\bibfnamefont {M.}~\bibnamefont {{Schirmer}}}, \bibinfo
  {author} {\bibfnamefont {G.}~\bibnamefont {{Alguero}}}, \bibinfo {author}
  {\bibfnamefont {B.}~\bibnamefont {{Altieri}}}, \bibinfo {author}
  {\bibfnamefont {E.}~\bibnamefont {{Balbinot}}}, \bibinfo {author}
  {\bibfnamefont {T.}~\bibnamefont {{de Boer}}},\ and\ \bibinfo {author}
  {\bibnamefont {et~al.}},\ }\href@noop {} {\bibinfo {title} {{{Euclid Quick
  Data Release (Q1) -- Data release overview}}}} (\bibinfo {year} {2025}),\
  \Eprint {https://arxiv.org/abs/arXiv:2503.15302} {arXiv:2503.15302
  [astro-ph.GA]} \BibitemShut {NoStop}%
\bibitem [{\citenamefont {Will}(2014)}]{Will2014LRR}%
  \BibitemOpen
  \bibfield  {author} {\bibinfo {author} {\bibfnamefont {C.~M.}\ \bibnamefont
  {Will}},\ }\bibfield  {title} {\bibinfo {title} {{The Confrontation Between
  General Relativity and Experiment}},\ }\href
  {https://doi.org/10.12942/lrr-2014-4} {\bibfield  {journal} {\bibinfo
  {journal} {Living Reviews in Relativity}\ }\textbf {\bibinfo {volume} {17}},\
  \bibinfo {pages} {4} (\bibinfo {year} {2014})},\ \bibinfo {note} {and
  updates}\BibitemShut {NoStop}%
\bibitem [{\citenamefont {Hees}\ \emph {et~al.}(2016)\citenamefont {Hees} \emph
  {et~al.}}]{Hees2016}%
  \BibitemOpen
  \bibfield  {author} {\bibinfo {author} {\bibfnamefont {A.}~\bibnamefont
  {Hees}} \emph {et~al.},\ }\bibfield  {title} {\bibinfo {title} {{Searching
  for an oscillating massive scalar field as a dark matter candidate using
  atomic hyperfine frequency comparisons}},\ }\href
  {https://doi.org/10.1103/PhysRevLett.117.061301} {\bibfield  {journal}
  {\bibinfo  {journal} {Phys. Rev. Lett.}\ }\textbf {\bibinfo {volume} {117}},\
  \bibinfo {pages} {061301} (\bibinfo {year} {2016})}\BibitemShut {NoStop}%
\bibitem [{\citenamefont {Wcis{\l}o}\ \emph {et~al.}(2018)\citenamefont
  {Wcis{\l}o} \emph {et~al.}}]{Wcislo2018}%
  \BibitemOpen
  \bibfield  {author} {\bibinfo {author} {\bibfnamefont {P.}~\bibnamefont
  {Wcis{\l}o}} \emph {et~al.},\ }\bibfield  {title} {\bibinfo {title} {{New
  bounds on dark matter coupling from a global network of optical atomic
  clocks}},\ }\href {https://doi.org/10.1126/sciadv.aau4869} {\bibfield
  {journal} {\bibinfo  {journal} {Science Advances}\ }\textbf {\bibinfo
  {volume} {4}},\ \bibinfo {pages} {eaau4869} (\bibinfo {year}
  {2018})}\BibitemShut {NoStop}%
\bibitem [{\citenamefont {Talmadge}\ \emph {et~al.}(1988)\citenamefont
  {Talmadge}, \citenamefont {Berthias}, \citenamefont {Hellings},\ and\
  \citenamefont {Standish}}]{Talmadge1988PRL}%
  \BibitemOpen
  \bibfield  {author} {\bibinfo {author} {\bibfnamefont {C.}~\bibnamefont
  {Talmadge}}, \bibinfo {author} {\bibfnamefont {J.-P.}\ \bibnamefont
  {Berthias}}, \bibinfo {author} {\bibfnamefont {R.~W.}\ \bibnamefont
  {Hellings}},\ and\ \bibinfo {author} {\bibfnamefont {E.~M.}\ \bibnamefont
  {Standish}},\ }\bibfield  {title} {\bibinfo {title} {{Model-Independent
  Constraints on Possible Modifications of Newtonian Gravity}},\ }\href
  {https://doi.org/10.1103/PhysRevLett.61.1159} {\bibfield  {journal} {\bibinfo
   {journal} {Phys. Rev. Lett.}\ }\textbf {\bibinfo {volume} {61}},\ \bibinfo
  {pages} {1159} (\bibinfo {year} {1988})}\BibitemShut {NoStop}%
\bibitem [{\citenamefont {Pitjeva}\ and\ \citenamefont
  {Pitjev}(2013)}]{Pitjeva2013}%
  \BibitemOpen
  \bibfield  {author} {\bibinfo {author} {\bibfnamefont {E.~V.}\ \bibnamefont
  {Pitjeva}}\ and\ \bibinfo {author} {\bibfnamefont {N.~P.}\ \bibnamefont
  {Pitjev}},\ }\bibfield  {title} {\bibinfo {title} {{Relativistic effects and
  dark matter in the Solar system from observations of planets and
  spacecraft}},\ }\href {https://doi.org/10.1093/mnras/stt695} {\bibfield
  {journal} {\bibinfo  {journal} {Mon. Not. R. Astron. Soc.}\ }\textbf
  {\bibinfo {volume} {432}},\ \bibinfo {pages} {3431} (\bibinfo {year}
  {2013})}\BibitemShut {NoStop}%
\bibitem [{\citenamefont {{Pitjev}}\ and\ \citenamefont
  {{Pitjeva}}(2013)}]{Pitjev2013}%
  \BibitemOpen
  \bibfield  {author} {\bibinfo {author} {\bibfnamefont {N.~P.}\ \bibnamefont
  {{Pitjev}}}\ and\ \bibinfo {author} {\bibfnamefont {E.~V.}\ \bibnamefont
  {{Pitjeva}}},\ }\bibfield  {title} {\bibinfo {title} {{{Constraints on dark
  matter in the solar system}}},\ }\href
  {https://doi.org/10.1134/S1063773713020060} {\bibfield  {journal} {\bibinfo
  {journal} {Astronomy Letters}\ }\textbf {\bibinfo {volume} {39}},\ \bibinfo
  {pages} {141} (\bibinfo {year} {2013})}\BibitemShut {NoStop}%
\bibitem [{\citenamefont {Khoury}\ and\ \citenamefont
  {Weltman}(2004{\natexlab{a}})}]{KhouryPRD}%
  \BibitemOpen
  \bibfield  {author} {\bibinfo {author} {\bibfnamefont {J.}~\bibnamefont
  {Khoury}}\ and\ \bibinfo {author} {\bibfnamefont {A.}~\bibnamefont
  {Weltman}},\ }\bibfield  {title} {\bibinfo {title} {{Chameleon Fields}},\
  }\href {https://doi.org/10.1103/PhysRevD.69.044026} {\bibfield  {journal}
  {\bibinfo  {journal} {Phys. Rev. D}\ }\textbf {\bibinfo {volume} {69}},\
  \bibinfo {pages} {044026} (\bibinfo {year} {2004}{\natexlab{a}})}\BibitemShut
  {NoStop}%
\bibitem [{\citenamefont {Khoury}\ and\ \citenamefont
  {Weltman}(2004{\natexlab{b}})}]{KhouryPRL}%
  \BibitemOpen
  \bibfield  {author} {\bibinfo {author} {\bibfnamefont {J.}~\bibnamefont
  {Khoury}}\ and\ \bibinfo {author} {\bibfnamefont {A.}~\bibnamefont
  {Weltman}},\ }\bibfield  {title} {\bibinfo {title} {{Chameleon Fields:
  Awaiting Surprises for Tests of Gravity in Space}},\ }\href
  {https://doi.org/10.1103/PhysRevLett.93.171104} {\bibfield  {journal}
  {\bibinfo  {journal} {Phys. Rev. Lett.}\ }\textbf {\bibinfo {volume} {93}},\
  \bibinfo {pages} {171104} (\bibinfo {year} {2004}{\natexlab{b}})}\BibitemShut
  {NoStop}%
\bibitem [{\citenamefont {Hinterbichler}\ and\ \citenamefont
  {Khoury}(2010)}]{HinterbichlerKhouryPRL2010}%
  \BibitemOpen
  \bibfield  {author} {\bibinfo {author} {\bibfnamefont {K.}~\bibnamefont
  {Hinterbichler}}\ and\ \bibinfo {author} {\bibfnamefont {J.}~\bibnamefont
  {Khoury}},\ }\bibfield  {title} {\bibinfo {title} {{Symmetron Fields:
  Screening Long-Range Forces Through Local Symmetry Restoration}},\ }\href
  {https://doi.org/10.1103/PhysRevLett.104.231301} {\bibfield  {journal}
  {\bibinfo  {journal} {Phys. Rev. Lett.}\ }\textbf {\bibinfo {volume} {104}},\
  \bibinfo {pages} {231301} (\bibinfo {year} {2010})}\BibitemShut {NoStop}%
\bibitem [{\citenamefont {Burrage}\ and\ \citenamefont
  {Sakstein}(2018)}]{BurrageSaksteinLRR}%
  \BibitemOpen
  \bibfield  {author} {\bibinfo {author} {\bibfnamefont {C.}~\bibnamefont
  {Burrage}}\ and\ \bibinfo {author} {\bibfnamefont {J.}~\bibnamefont
  {Sakstein}},\ }\bibfield  {title} {\bibinfo {title} {{Tests of Chameleon
  Gravity}},\ }\href {https://doi.org/10.1007/s41114-018-0011-x} {\bibfield
  {journal} {\bibinfo  {journal} {Living Reviews in Relativity}\ }\textbf
  {\bibinfo {volume} {21}},\ \bibinfo {pages} {1} (\bibinfo {year}
  {2018})}\BibitemShut {NoStop}%
\bibitem [{\citenamefont {Koyama}(2016)}]{Koyama2016}%
  \BibitemOpen
  \bibfield  {author} {\bibinfo {author} {\bibfnamefont {K.}~\bibnamefont
  {Koyama}},\ }\bibfield  {title} {\bibinfo {title} {{Cosmological Tests of
  Modified Gravity}},\ }\href {https://doi.org/10.1088/0034-4885/79/4/046902}
  {\bibfield  {journal} {\bibinfo  {journal} {Rep. Progr. Phys.}\ }\textbf
  {\bibinfo {volume} {79}},\ \bibinfo {pages} {046902} (\bibinfo {year}
  {2016})}\BibitemShut {NoStop}%
\bibitem [{\citenamefont {Babichev}\ and\ \citenamefont
  {Deffayet}(2013)}]{BabichevDeffayet2013CQG}%
  \BibitemOpen
  \bibfield  {author} {\bibinfo {author} {\bibfnamefont {E.}~\bibnamefont
  {Babichev}}\ and\ \bibinfo {author} {\bibfnamefont {C.}~\bibnamefont
  {Deffayet}},\ }\bibfield  {title} {\bibinfo {title} {{An Introduction to the
  Vainshtein Mechanism}},\ }\href
  {https://doi.org/10.1088/0264-9381/30/18/184001} {\bibfield  {journal}
  {\bibinfo  {journal} {CQG}\ }\textbf {\bibinfo {volume} {30}},\ \bibinfo
  {pages} {184001} (\bibinfo {year} {2013})}\BibitemShut {NoStop}%
\bibitem [{\citenamefont {Baker}\ \emph {et~al.}(2017)\citenamefont {Baker},
  \citenamefont {Bellini}, \citenamefont {Ferreira}, \citenamefont {Lagos},
  \citenamefont {Noller},\ and\ \citenamefont {Sawicki}}]{Baker2017}%
  \BibitemOpen
  \bibfield  {author} {\bibinfo {author} {\bibfnamefont {T.}~\bibnamefont
  {Baker}}, \bibinfo {author} {\bibfnamefont {E.}~\bibnamefont {Bellini}},
  \bibinfo {author} {\bibfnamefont {P.~G.}\ \bibnamefont {Ferreira}}, \bibinfo
  {author} {\bibfnamefont {M.}~\bibnamefont {Lagos}}, \bibinfo {author}
  {\bibfnamefont {J.}~\bibnamefont {Noller}},\ and\ \bibinfo {author}
  {\bibfnamefont {I.}~\bibnamefont {Sawicki}},\ }\bibfield  {title} {\bibinfo
  {title} {{Strong Constraints on Cosmological Gravity from GW170817 and GRB
  170817A}},\ }\href {https://doi.org/10.1103/PhysRevLett.119.251301}
  {\bibfield  {journal} {\bibinfo  {journal} {Phys. Rev. Lett.}\ }\textbf
  {\bibinfo {volume} {119}},\ \bibinfo {pages} {251301} (\bibinfo {year}
  {2017})}\BibitemShut {NoStop}%
\bibitem [{\citenamefont {Creminelli}\ and\ \citenamefont
  {Vernizzi}(2017)}]{Creminelli2017}%
  \BibitemOpen
  \bibfield  {author} {\bibinfo {author} {\bibfnamefont {P.}~\bibnamefont
  {Creminelli}}\ and\ \bibinfo {author} {\bibfnamefont {F.}~\bibnamefont
  {Vernizzi}},\ }\bibfield  {title} {\bibinfo {title} {{Dark Energy after
  GW170817 and GRB170817A}},\ }\href
  {https://doi.org/10.1103/PhysRevLett.119.251302} {\bibfield  {journal}
  {\bibinfo  {journal} {Phys. Rev. Lett.}\ }\textbf {\bibinfo {volume} {119}},\
  \bibinfo {pages} {251302} (\bibinfo {year} {2017})}\BibitemShut {NoStop}%
\bibitem [{\citenamefont {Ezquiaga}\ and\ \citenamefont
  {Zumalac{\'a}rregui}(2017)}]{EzquiagaZumalacarregui2017}%
  \BibitemOpen
  \bibfield  {author} {\bibinfo {author} {\bibfnamefont {J.~M.}\ \bibnamefont
  {Ezquiaga}}\ and\ \bibinfo {author} {\bibfnamefont {M.}~\bibnamefont
  {Zumalac{\'a}rregui}},\ }\bibfield  {title} {\bibinfo {title} {{Dark Energy
  after GW170817: Dead Ends and the Road Ahead}},\ }\href
  {https://doi.org/10.1103/PhysRevLett.119.251304} {\bibfield  {journal}
  {\bibinfo  {journal} {Phys. Rev. Lett.}\ }\textbf {\bibinfo {volume} {119}},\
  \bibinfo {pages} {251304} (\bibinfo {year} {2017})}\BibitemShut {NoStop}%
\bibitem [{\citenamefont {Touboul}\ \emph {et~al.}(2022)\citenamefont {Touboul}
  \emph {et~al.}}]{Touboul2022}%
  \BibitemOpen
  \bibfield  {author} {\bibinfo {author} {\bibfnamefont {P.}~\bibnamefont
  {Touboul}} \emph {et~al.},\ }\bibfield  {title} {\bibinfo {title} {{Space
  test of the Equivalence Principle: MICROSCOPE final results}},\ }\href
  {https://doi.org/10.1103/PhysRevLett.129.121102} {\bibfield  {journal}
  {\bibinfo  {journal} {Phys. Rev. Lett.}\ }\textbf {\bibinfo {volume} {129}},\
  \bibinfo {pages} {121102} (\bibinfo {year} {2022})}\BibitemShut {NoStop}%
\bibitem [{\citenamefont {Bertotti}\ \emph {et~al.}(2003)\citenamefont
  {Bertotti}, \citenamefont {Iess},\ and\ \citenamefont
  {Tortora}}]{Bertotti2003}%
  \BibitemOpen
  \bibfield  {author} {\bibinfo {author} {\bibfnamefont {B.}~\bibnamefont
  {Bertotti}}, \bibinfo {author} {\bibfnamefont {L.}~\bibnamefont {Iess}},\
  and\ \bibinfo {author} {\bibfnamefont {P.}~\bibnamefont {Tortora}},\
  }\bibfield  {title} {\bibinfo {title} {{A test of general relativity using
  radio links with the Cassini spacecraft}},\ }\href
  {https://doi.org/10.1038/nature01997} {\bibfield  {journal} {\bibinfo
  {journal} {Nature}\ }\textbf {\bibinfo {volume} {425}},\ \bibinfo {pages}
  {374} (\bibinfo {year} {2003})}\BibitemShut {NoStop}%
\bibitem [{\citenamefont {Williams}\ \emph {et~al.}(2004)\citenamefont
  {Williams}, \citenamefont {Turyshev},\ and\ \citenamefont
  {Boggs}}]{Williams2004}%
  \BibitemOpen
  \bibfield  {author} {\bibinfo {author} {\bibfnamefont {J.~G.}\ \bibnamefont
  {Williams}}, \bibinfo {author} {\bibfnamefont {S.~G.}\ \bibnamefont
  {Turyshev}},\ and\ \bibinfo {author} {\bibfnamefont {D.~H.}\ \bibnamefont
  {Boggs}},\ }\bibfield  {title} {\bibinfo {title} {{Progress in Lunar Laser
  Ranging Tests of Relativistic Gravity}},\ }\href
  {https://doi.org/10.1103/PhysRevLett.93.261101} {\bibfield  {journal}
  {\bibinfo  {journal} {Phys. Rev. Lett.}\ }\textbf {\bibinfo {volume} {93}},\
  \bibinfo {pages} {261101} (\bibinfo {year} {2004})}\BibitemShut {NoStop}%
\bibitem [{\citenamefont {Pitjeva}\ \emph {et~al.}(2021)\citenamefont
  {Pitjeva}, \citenamefont {Pitjev}, \citenamefont {Pavlov},\ and\
  \citenamefont {Turygin}}]{Pitjeva2021}%
  \BibitemOpen
  \bibfield  {author} {\bibinfo {author} {\bibfnamefont {E.~V.}\ \bibnamefont
  {Pitjeva}}, \bibinfo {author} {\bibfnamefont {N.~P.}\ \bibnamefont {Pitjev}},
  \bibinfo {author} {\bibfnamefont {D.~A.}\ \bibnamefont {Pavlov}},\ and\
  \bibinfo {author} {\bibfnamefont {C.~C.}\ \bibnamefont {Turygin}},\
  }\bibfield  {title} {\bibinfo {title} {{Estimates of the change rate of solar
  mass and gravitational constant based on the dynamics of the Solar System}},\
  }\href {https://doi.org/10.1051/0004-6361/202039653} {\bibfield  {journal}
  {\bibinfo  {journal} {Astronomy \& Astrophysics}\ }\textbf {\bibinfo {volume}
  {647}},\ \bibinfo {pages} {A141} (\bibinfo {year} {2021})}\BibitemShut
  {NoStop}%
\bibitem [{\citenamefont {{Fienga}}\ and\ \citenamefont
  {{Minazzoli}}(2024)}]{Fienga2024}%
  \BibitemOpen
  \bibfield  {author} {\bibinfo {author} {\bibfnamefont {A.}~\bibnamefont
  {{Fienga}}}\ and\ \bibinfo {author} {\bibfnamefont {O.}~\bibnamefont
  {{Minazzoli}}},\ }\bibfield  {title} {\bibinfo {title} {{{Testing theories of
  gravity with planetary ephemerides}}},\ }\href
  {https://doi.org/10.1007/s41114-023-00047-0} {\bibfield  {journal} {\bibinfo
  {journal} {Living Rev. Rel.}\ }\textbf {\bibinfo {volume} {27}},\ \bibinfo
  {eid} {1} (\bibinfo {year} {2024})}\BibitemShut {NoStop}%
\bibitem [{\citenamefont {Turyshev}(2025)}]{Turyshev2025}%
  \BibitemOpen
  \bibfield  {author} {\bibinfo {author} {\bibfnamefont {S.~G.}\ \bibnamefont
  {Turyshev}},\ }\bibfield  {title} {\bibinfo {title} {{Lunar laser ranging
  with high-power continuous-wave lasers}},\ }\href
  {https://doi.org/10.1103/PhysRevApplied.23.064066} {\bibfield  {journal}
  {\bibinfo  {journal} {Physical Review Applied}\ }\textbf {\bibinfo {volume}
  {23}},\ \bibinfo {pages} {064066} (\bibinfo {year} {2025})},\ \Eprint
  {https://arxiv.org/abs/arXiv:2502.02796 [astro-ph.IM]} {arXiv:2502.02796
  [astro-ph.IM]} \BibitemShut {NoStop}%
\bibitem [{\citenamefont {Cacciapuoti}\ and\ \citenamefont
  {Salomon}(2009)}]{Cacciapuoti2009_ACES}%
  \BibitemOpen
  \bibfield  {author} {\bibinfo {author} {\bibfnamefont {L.}~\bibnamefont
  {Cacciapuoti}}\ and\ \bibinfo {author} {\bibfnamefont {C.}~\bibnamefont
  {Salomon}},\ }\bibfield  {title} {\bibinfo {title} {{Space clocks and
  fundamental tests: The ACES experiment}},\ }\href
  {https://doi.org/10.1140/epjst/e2009-01041-7} {\bibfield  {journal} {\bibinfo
   {journal} {Eur. Phys. J. Special Topics}\ }\textbf {\bibinfo {volume}
  {172}},\ \bibinfo {pages} {57} (\bibinfo {year} {2009})}\BibitemShut
  {NoStop}%
\bibitem [{\citenamefont {Delva}\ \emph {et~al.}(2017)\citenamefont {Delva},
  \citenamefont {Denker}, \citenamefont {Lion}, \citenamefont {Pottie},\ and\
  \citenamefont {et~al.}}]{Delva2017_FiberSR}%
  \BibitemOpen
  \bibfield  {author} {\bibinfo {author} {\bibfnamefont {P.}~\bibnamefont
  {Delva}}, \bibinfo {author} {\bibfnamefont {H.}~\bibnamefont {Denker}},
  \bibinfo {author} {\bibfnamefont {G.}~\bibnamefont {Lion}}, \bibinfo {author}
  {\bibfnamefont {P.-E.}\ \bibnamefont {Pottie}},\ and\ \bibinfo {author}
  {\bibnamefont {et~al.}},\ }\bibfield  {title} {\bibinfo {title} {{Test of
  Special Relativity Using a Fiber Network of Optical Clocks}},\ }\href
  {https://doi.org/10.1103/PhysRevLett.118.221102} {\bibfield  {journal}
  {\bibinfo  {journal} {Phys. Rev. Lett.}\ }\textbf {\bibinfo {volume} {118}},\
  \bibinfo {pages} {221102} (\bibinfo {year} {2017})}\BibitemShut {NoStop}%
\bibitem [{\citenamefont {{DESI Collaboration}}(2016)}]{DESIScience}%
  \BibitemOpen
  \bibfield  {author} {\bibinfo {author} {\bibnamefont {{DESI
  Collaboration}}},\ }\bibfield  {title} {\bibinfo {title} {{The DESI
  Experiment Part I: Science, Targeting, and Survey Design}},\ }\href@noop {}
  {\bibfield  {journal} {\bibinfo  {journal} {arXiv e-prints}\ } (\bibinfo
  {year} {2016})},\ \Eprint {https://arxiv.org/abs/arXiv:1611.00036
  [astro-ph.IM]} {arXiv:1611.00036 [astro-ph.IM]} \BibitemShut {NoStop}%
\bibitem [{\citenamefont {{European Space Agency}}(2025)}]{ESA_Euclid_PR}%
  \BibitemOpen
  \bibfield  {author} {\bibinfo {author} {\bibnamefont {{European Space
  Agency}}},\ }\href@noop {} {\bibinfo {title} {{Euclid opens data treasure
  trove, offers glimpse of deep fields}}},\ \bibinfo {howpublished}
  {\url{https://www.esa.int/Science_Exploration/Space_Science/Euclid/Euclid_opens_data_treasure_trove_offers_glimpse_of_deep_fields}}
  (\bibinfo {year} {2025}),\ \bibinfo {note} {press release, 19 March
  2025}\BibitemShut {NoStop}%
\bibitem [{\citenamefont {Benisty}\ \emph {et~al.}(2023)\citenamefont
  {Benisty}, \citenamefont {Mifsud}, \citenamefont {Said},\ and\ \citenamefont
  {Staicova}}]{Benisty:2023fRCombined}%
  \BibitemOpen
  \bibfield  {author} {\bibinfo {author} {\bibfnamefont {D.}~\bibnamefont
  {Benisty}}, \bibinfo {author} {\bibfnamefont {J.}~\bibnamefont {Mifsud}},
  \bibinfo {author} {\bibfnamefont {J.~L.}\ \bibnamefont {Said}},\ and\
  \bibinfo {author} {\bibfnamefont {D.}~\bibnamefont {Staicova}},\ }\bibfield
  {title} {\bibinfo {title} {Strengthening extended gravity constraints with
  combined systems: f(r) bounds from cosmology and the galactic center},\
  }\href {https://doi.org/10.1016/j.dark.2023.101344} {\bibfield  {journal}
  {\bibinfo  {journal} {Phys. Dark Univ.}\ }\textbf {\bibinfo {volume} {42}},\
  \bibinfo {pages} {101344} (\bibinfo {year} {2023})},\ \Eprint
  {https://arxiv.org/abs/2303.15040} {arXiv:2303.15040 [astro-ph.CO]}
  \BibitemShut {NoStop}%
\bibitem [{\citenamefont {{LIGO Scientific Collaboration}}\ and\ \citenamefont
  {{Virgo Collaboration}}(2017)}]{Abbott2017GW}%
  \BibitemOpen
  \bibfield  {author} {\bibinfo {author} {\bibnamefont {{LIGO Scientific
  Collaboration}}}\ and\ \bibinfo {author} {\bibnamefont {{Virgo
  Collaboration}}},\ }\bibfield  {title} {\bibinfo {title} {{GW170817:
  Observation of Gravitational Waves from a Binary Neutron Star Inspiral}},\
  }\href {https://doi.org/10.1103/PhysRevLett.119.161101} {\bibfield  {journal}
  {\bibinfo  {journal} {Phys. Rev. Lett.}\ }\textbf {\bibinfo {volume} {119}},\
  \bibinfo {pages} {161101} (\bibinfo {year} {2017})}\BibitemShut {NoStop}%
\bibitem [{\citenamefont {{LIGO Scientific Collaboration}}\ \emph
  {et~al.}(2017)\citenamefont {{LIGO Scientific Collaboration}}, \citenamefont
  {{Virgo Collaboration}} \emph {et~al.}}]{Abbott2017MM}%
  \BibitemOpen
  \bibfield  {author} {\bibinfo {author} {\bibnamefont {{LIGO Scientific
  Collaboration}}}, \bibinfo {author} {\bibnamefont {{Virgo Collaboration}}},
  \emph {et~al.},\ }\bibfield  {title} {\bibinfo {title} {{Multi-messenger
  Observations of a Binary Neutron Star Merger}},\ }\href
  {https://doi.org/10.3847/2041-8213/aa91c9} {\bibfield  {journal} {\bibinfo
  {journal} {Astrophys. J. Lett.}\ }\textbf {\bibinfo {volume} {848}},\
  \bibinfo {pages} {L12} (\bibinfo {year} {2017})}\BibitemShut {NoStop}%
\bibitem [{\citenamefont {Adams}\ \emph {et~al.}(2006)\citenamefont {Adams},
  \citenamefont {Arkani-Hamed}, \citenamefont {Dubovsky}, \citenamefont
  {Nicolis},\ and\ \citenamefont {Rattazzi}}]{Adams2006JHEP}%
  \BibitemOpen
  \bibfield  {author} {\bibinfo {author} {\bibfnamefont {A.}~\bibnamefont
  {Adams}}, \bibinfo {author} {\bibfnamefont {N.}~\bibnamefont {Arkani-Hamed}},
  \bibinfo {author} {\bibfnamefont {S.}~\bibnamefont {Dubovsky}}, \bibinfo
  {author} {\bibfnamefont {A.}~\bibnamefont {Nicolis}},\ and\ \bibinfo {author}
  {\bibfnamefont {R.}~\bibnamefont {Rattazzi}},\ }\bibfield  {title} {\bibinfo
  {title} {{Causality, Analyticity and an IR Obstruction to UV Completion}},\
  }\href {https://doi.org/10.1088/1126-6708/2006/10/014} {\bibfield  {journal}
  {\bibinfo  {journal} {JHEP}\ }\textbf {\bibinfo {volume} {2006}}\bibinfo
  {number} { (10)},\ \bibinfo {pages} {014}}\BibitemShut {NoStop}%
\bibitem [{\citenamefont {Brax}\ \emph {et~al.}(2022)\citenamefont {Brax},
  \citenamefont {Fischer}, \citenamefont {Kaeding},\ and\ \citenamefont
  {Pitschmann}}]{BraxFischerKaedingPitschmann2022}%
  \BibitemOpen
\bibfield  {number} {  }\bibfield  {author} {\bibinfo {author} {\bibfnamefont
  {P.}~\bibnamefont {Brax}}, \bibinfo {author} {\bibfnamefont {H.}~\bibnamefont
  {Fischer}}, \bibinfo {author} {\bibfnamefont {C.}~\bibnamefont {Kaeding}},\
  and\ \bibinfo {author} {\bibfnamefont {M.}~\bibnamefont {Pitschmann}},\
  }\href@noop {} {\bibinfo {title} {{Dilaton Solutions for Laboratory
  Constraints and Lunar Laser Ranging}}} (\bibinfo {year} {2022}),\ \Eprint
  {https://arxiv.org/abs/2203.12512} {arXiv:2203.12512 [gr-qc]} \BibitemShut
  {NoStop}%
\bibitem [{\citenamefont {{Damour}}\ and\ \citenamefont
  {{Esposito-Far{\`e}se}}(1996)}]{Damour-Esposito:1996}%
  \BibitemOpen
  \bibfield  {author} {\bibinfo {author} {\bibfnamefont {T.}~\bibnamefont
  {{Damour}}}\ and\ \bibinfo {author} {\bibfnamefont {G.}~\bibnamefont
  {{Esposito-Far{\`e}se}}},\ }\bibfield  {title} {\bibinfo {title}
  {{{Tensor-scalar gravity and binary-pulsar experiments}}},\ }\href
  {https://doi.org/10.1103/PhysRevD.54.1474} {\bibfield  {journal} {\bibinfo
  {journal} {Phys. Rev. D}\ }\textbf {\bibinfo {volume} {54}},\ \bibinfo
  {pages} {1474} (\bibinfo {year} {1996})}\BibitemShut {NoStop}%
\bibitem [{\citenamefont {{Turyshev}}\ \emph
  {et~al.}(2007{\natexlab{a}})\citenamefont {{Turyshev}}, \citenamefont
  {{Israelsson}}, \citenamefont {{Shao}}, \citenamefont {{Yu}}, \citenamefont
  {{Kusenko}}, \citenamefont {{Wright}}, \citenamefont {{Everitt}},
  \citenamefont {{Kasevich}}, \citenamefont {{Lipa}}, \citenamefont {{Mester}},
  \citenamefont {{Reasenberg}}, \citenamefont {{Walsworth}}, \citenamefont
  {{Ashby}}, \citenamefont {{Gould}},\ and\ \citenamefont
  {{Paik}}}]{Turyshev:2007}%
  \BibitemOpen
  \bibfield  {author} {\bibinfo {author} {\bibfnamefont {S.~G.}\ \bibnamefont
  {{Turyshev}}}, \bibinfo {author} {\bibfnamefont {U.~E.}\ \bibnamefont
  {{Israelsson}}}, \bibinfo {author} {\bibfnamefont {M.}~\bibnamefont
  {{Shao}}}, \bibinfo {author} {\bibfnamefont {N.}~\bibnamefont {{Yu}}},
  \bibinfo {author} {\bibfnamefont {A.}~\bibnamefont {{Kusenko}}}, \bibinfo
  {author} {\bibfnamefont {E.~L.}\ \bibnamefont {{Wright}}}, \bibinfo {author}
  {\bibfnamefont {C.~W.~F.}\ \bibnamefont {{Everitt}}}, \bibinfo {author}
  {\bibfnamefont {M.}~\bibnamefont {{Kasevich}}}, \bibinfo {author}
  {\bibfnamefont {J.~A.}\ \bibnamefont {{Lipa}}}, \bibinfo {author}
  {\bibfnamefont {J.~C.}\ \bibnamefont {{Mester}}}, \bibinfo {author}
  {\bibfnamefont {R.~D.}\ \bibnamefont {{Reasenberg}}}, \bibinfo {author}
  {\bibfnamefont {R.~L.}\ \bibnamefont {{Walsworth}}}, \bibinfo {author}
  {\bibfnamefont {N.}~\bibnamefont {{Ashby}}}, \bibinfo {author} {\bibfnamefont
  {H.}~\bibnamefont {{Gould}}},\ and\ \bibinfo {author} {\bibfnamefont {H.~J.}\
  \bibnamefont {{Paik}}},\ }\bibfield  {title} {\bibinfo {title} {{Space-Based
  Research in Fundamental Physics and Quantum Technologies}},\ }\href
  {https://doi.org/10.1142/S0218271807011760} {\bibfield  {journal} {\bibinfo
  {journal} {Int. J. Mod. Phys. D}\ }\textbf {\bibinfo {volume} {16}},\
  \bibinfo {pages} {1879} (\bibinfo {year} {2007}{\natexlab{a}})},\ \Eprint
  {https://arxiv.org/abs/arXiv:0711.0150 [gr-qc]} {arXiv:0711.0150 [gr-qc]}
  \BibitemShut {NoStop}%
\bibitem [{\citenamefont {{Turyshev}}(2009)}]{Turyshev:2009}%
  \BibitemOpen
  \bibfield  {author} {\bibinfo {author} {\bibfnamefont {S.~G.}\ \bibnamefont
  {{Turyshev}}},\ }\bibfield  {title} {\bibinfo {title} {{Experimental tests of
  general relativity: recent progress and future directions}},\ }\href
  {https://doi.org/10.3367/UFNe.0179.200901a.0003} {\bibfield  {journal}
  {\bibinfo  {journal} {Physics Uspekhi}\ }\textbf {\bibinfo {volume} {52}},\
  \bibinfo {pages} {1} (\bibinfo {year} {2009})},\ \Eprint
  {https://arxiv.org/abs/arXiv:0809.3730 [gr-qc]} {arXiv:0809.3730 [gr-qc]}
  \BibitemShut {NoStop}%
\bibitem [{\citenamefont {Battelier}\ \emph {et~al.}(2021)\citenamefont
  {Battelier} \emph {et~al.}}]{Battelier2021ExpAstron}%
  \BibitemOpen
  \bibfield  {author} {\bibinfo {author} {\bibfnamefont {B.}~\bibnamefont
  {Battelier}} \emph {et~al.},\ }\bibfield  {title} {\bibinfo {title}
  {{Exploring the Foundations of the Universe with Space-Based Atom
  Interferometry}},\ }\href {https://doi.org/10.1007/s10686-021-09723-1}
  {\bibfield  {journal} {\bibinfo  {journal} {Experimental Astronomy}\ }\textbf
  {\bibinfo {volume} {51}},\ \bibinfo {pages} {1695} (\bibinfo {year}
  {2021})}\BibitemShut {NoStop}%
\bibitem [{\citenamefont {{Gaaloul}}\ \emph {et~al.}(2022)\citenamefont
  {{Gaaloul}}, \citenamefont {{Ahlers}}, \citenamefont {{Badurina}},
  \citenamefont {{Bassi}}, \citenamefont {{Battelier}}, \citenamefont
  {{Beaufils}},\ and\ \citenamefont {et~al.}}]{Ahlers2022STEQUEST}%
  \BibitemOpen
  \bibfield  {author} {\bibinfo {author} {\bibfnamefont {N.}~\bibnamefont
  {{Gaaloul}}}, \bibinfo {author} {\bibfnamefont {H.}~\bibnamefont {{Ahlers}}},
  \bibinfo {author} {\bibfnamefont {L.}~\bibnamefont {{Badurina}}}, \bibinfo
  {author} {\bibfnamefont {A.}~\bibnamefont {{Bassi}}}, \bibinfo {author}
  {\bibfnamefont {B.}~\bibnamefont {{Battelier}}}, \bibinfo {author}
  {\bibfnamefont {Q.}~\bibnamefont {{Beaufils}}},\ and\ \bibinfo {author}
  {\bibnamefont {et~al.}},\ }\href {https://doi.org/10.48550/arXiv.2211.15412}
  {\bibinfo {title} {{STE-QUEST -- Space Time Explorer and QUantum Equivalence
  principle Space Test: The 2022 medium-class mission concept}}} (\bibinfo
  {year} {2022}),\ \Eprint {https://arxiv.org/abs/2211.15412} {arXiv:2211.15412
  [physics.space-ph]} \BibitemShut {NoStop}%
\bibitem [{\citenamefont {Turyshev}\ \emph {et~al.}(2024)\citenamefont
  {Turyshev}, \citenamefont {Chiow},\ and\ \citenamefont {Yu}}]{TetPRD2024}%
  \BibitemOpen
  \bibfield  {author} {\bibinfo {author} {\bibfnamefont {S.~G.}\ \bibnamefont
  {Turyshev}}, \bibinfo {author} {\bibfnamefont {S.-w.}\ \bibnamefont
  {Chiow}},\ and\ \bibinfo {author} {\bibfnamefont {N.}~\bibnamefont {Yu}},\
  }\bibfield  {title} {\bibinfo {title} {{Searching for New Physics in the
  Solar System with Tetrahedral Spacecraft Formations}},\ }\href
  {https://doi.org/10.1103/PhysRevD.109.084059} {\bibfield  {journal} {\bibinfo
   {journal} {Phys. Rev. D}\ }\textbf {\bibinfo {volume} {109}},\ \bibinfo
  {pages} {084059} (\bibinfo {year} {2024})},\ \Eprint
  {https://arxiv.org/abs/arXiv:2404.02096 [gr-qc]} {arXiv:2404.02096 [gr-qc]}
  \BibitemShut {NoStop}%
\bibitem [{\citenamefont {{Williams}}\ \emph {et~al.}(2016)\citenamefont
  {{Williams}}, \citenamefont {{Chiow}}, \citenamefont {{Yu}},\ and\
  \citenamefont {{M{\"u}ller}}}]{QTEST2015}%
  \BibitemOpen
  \bibfield  {author} {\bibinfo {author} {\bibfnamefont {J.}~\bibnamefont
  {{Williams}}}, \bibinfo {author} {\bibfnamefont {S.-w.}\ \bibnamefont
  {{Chiow}}}, \bibinfo {author} {\bibfnamefont {N.}~\bibnamefont {{Yu}}},\ and\
  \bibinfo {author} {\bibfnamefont {H.}~\bibnamefont {{M{\"u}ller}}},\
  }\bibfield  {title} {\bibinfo {title} {{Quantum test of the equivalence
  principle and space-time aboard the International Space Station}},\ }\href
  {https://doi.org/10.1088/1367-2630/18/2/025018} {\bibfield  {journal}
  {\bibinfo  {journal} {New J. Phys.}\ }\textbf {\bibinfo {volume} {18}},\
  \bibinfo {eid} {025018} (\bibinfo {year} {2016})}\BibitemShut {NoStop}%
\bibitem [{\citenamefont {Badurina}\ and\ \citenamefont
  {et~al.}(2021)}]{Badurina2021Prospects}%
  \BibitemOpen
  \bibfield  {author} {\bibinfo {author} {\bibfnamefont {L.}~\bibnamefont
  {Badurina}}\ and\ \bibinfo {author} {\bibnamefont {et~al.}},\ }\bibfield
  {title} {\bibinfo {title} {{Prospective Sensitivities of Atom Interferometers
  to Gravitational Waves and Ultralight Dark Matter}},\ }\href
  {https://doi.org/10.1007/JHEP05(2021)011} {\bibfield  {journal} {\bibinfo
  {journal} {JHEP}\ }\bibinfo  {number} { (05)},\ \bibinfo {pages}
  {011}}\BibitemShut {NoStop}%
\bibitem [{\citenamefont {Abe}\ and\ \citenamefont {et~al.}(2021)}]{MAGIS100}%
  \BibitemOpen
\bibfield  {number} {  }\bibfield  {author} {\bibinfo {author} {\bibfnamefont
  {M.}~\bibnamefont {Abe}}\ and\ \bibinfo {author} {\bibnamefont {et~al.}},\
  }\href@noop {} {\bibinfo {title} {{Matter-wave Atomic Gradiometer
  Interferometric Sensor (MAGIS-100)}}} (\bibinfo {year} {2021}),\ \bibinfo
  {note} {concept and science case for 100\,m AI; pathfinder for long-baseline
  sensors},\ \Eprint {https://arxiv.org/abs/2104.02835} {arXiv:2104.02835}
  \BibitemShut {NoStop}%
\bibitem [{\citenamefont {Tino}\ and\ \citenamefont {et~al.}(2019)}]{SAGE2019}%
  \BibitemOpen
  \bibfield  {author} {\bibinfo {author} {\bibfnamefont {G.~M.}\ \bibnamefont
  {Tino}}\ and\ \bibinfo {author} {\bibnamefont {et~al.}},\ }\bibfield  {title}
  {\bibinfo {title} {{SAGE: A Proposal for a Space Atomic Gravity Explorer}},\
  }\href {https://doi.org/10.1140/epjd/e2019-100061-5} {\bibfield  {journal}
  {\bibinfo  {journal} {Eur. Phys. J. D}\ }\textbf {\bibinfo {volume} {73}},\
  \bibinfo {pages} {228} (\bibinfo {year} {2019})}\BibitemShut {NoStop}%
\bibitem [{\citenamefont {Iess}\ and\ \citenamefont {et~al.}(2021)}]{Iess2021}%
  \BibitemOpen
  \bibfield  {author} {\bibinfo {author} {\bibfnamefont {L.}~\bibnamefont
  {Iess}}\ and\ \bibinfo {author} {\bibnamefont {et~al.}},\ }\bibfield  {title}
  {\bibinfo {title} {{Gravity, Geodesy and Fundamental Physics with
  BepiColombo's MORE Investigation}},\ }\bibfield  {journal} {\bibinfo
  {journal} {Space Sci. Rev.}\ }\textbf {\bibinfo {volume} {217}},\ \href
  {https://doi.org/10.1007/s11214-021-00800-3} {10.1007/s11214-021-00800-3}
  (\bibinfo {year} {2021})\BibitemShut {NoStop}%
\bibitem [{\citenamefont {di~Stefano}\ \emph
  {et~al.}(2021{\natexlab{a}})\citenamefont {di~Stefano}, \citenamefont
  {Cappuccio},\ and\ \citenamefont {Iess}}]{diStefano2021}%
  \BibitemOpen
  \bibfield  {author} {\bibinfo {author} {\bibfnamefont {I.}~\bibnamefont
  {di~Stefano}}, \bibinfo {author} {\bibfnamefont {P.}~\bibnamefont
  {Cappuccio}},\ and\ \bibinfo {author} {\bibfnamefont {L.}~\bibnamefont
  {Iess}},\ }\bibfield  {title} {\bibinfo {title} {{The BepiColombo solar
  conjunction experiments revisited}},\ }\href
  {https://doi.org/10.1088/1361-6382/abd301} {\bibfield  {journal} {\bibinfo
  {journal} {Classical and Quantum Gravity}\ }\textbf {\bibinfo {volume}
  {38}},\ \bibinfo {pages} {055002} (\bibinfo {year}
  {2021}{\natexlab{a}})}\BibitemShut {NoStop}%
\bibitem [{\citenamefont {{NASA Jet Propulsion
  Laboratory}}(2023)}]{JPL_DSOC_FirstData_2023}%
  \BibitemOpen
  \bibfield  {author} {\bibinfo {author} {\bibnamefont {{NASA Jet Propulsion
  Laboratory}}},\ }\href@noop {} {\bibinfo {title} {{NASA's Deep Space Optical
  Comm Demo Sends, Receives First Data}}},\ \bibinfo {howpublished}
  {\url{https://www.jpl.nasa.gov/news/nasas-deep-space-optical-comm-demo-sends-receives-first-data/}}
  (\bibinfo {year} {2023}),\ \bibinfo {note} {news release}\BibitemShut
  {NoStop}%
\bibitem [{\citenamefont {{NASA Jet Propulsion
  Laboratory}}(2024)}]{JPL_DSOC_Record_2024}%
  \BibitemOpen
  \bibfield  {author} {\bibinfo {author} {\bibnamefont {{NASA Jet Propulsion
  Laboratory}}},\ }\href@noop {} {\bibinfo {title} {{NASA's Laser Comms Demo
  Makes Deep Space Record, Completes First Phase}}},\ \bibinfo {howpublished}
  {\url{https://www.jpl.nasa.gov/news/nasas-laser-comms-demo-makes-deep-space-record-completes-first-phase/}}
  (\bibinfo {year} {2024}),\ \bibinfo {note} {news release}\BibitemShut
  {NoStop}%
\bibitem [{\citenamefont {{NASA}}(2025)}]{NASA_DSOC_Page}%
  \BibitemOpen
  \bibfield  {author} {\bibinfo {author} {\bibnamefont {{NASA}}},\ }\href@noop
  {} {\bibinfo {title} {{Deep Space Optical Communications (DSOC)}}},\ \bibinfo
  {howpublished}
  {\url{https://www.nasa.gov/mission/deep-space-optical-communications-dsoc/}}
  (\bibinfo {year} {2025}),\ \bibinfo {note} {mission page; page last updated
  2025-02-28}\BibitemShut {NoStop}%
\bibitem [{\citenamefont {Iess}\ \emph {et~al.}(2021)\citenamefont {Iess},
  \citenamefont {Asmar}, \citenamefont {Cappuccio}, \citenamefont {Cascioli},
  \citenamefont {De~Marchi}, \citenamefont {di~Stefano} \emph
  {et~al.}}]{Iess2021SSR}%
  \BibitemOpen
  \bibfield  {author} {\bibinfo {author} {\bibfnamefont {L.}~\bibnamefont
  {Iess}}, \bibinfo {author} {\bibfnamefont {S.~W.}\ \bibnamefont {Asmar}},
  \bibinfo {author} {\bibfnamefont {P.}~\bibnamefont {Cappuccio}}, \bibinfo
  {author} {\bibfnamefont {G.}~\bibnamefont {Cascioli}}, \bibinfo {author}
  {\bibfnamefont {F.}~\bibnamefont {De~Marchi}}, \bibinfo {author}
  {\bibfnamefont {I.}~\bibnamefont {di~Stefano}}, \emph {et~al.},\ }\bibfield
  {title} {\bibinfo {title} {Gravity, geodesy and fundamental physics with
  bepicolombo's more investigation},\ }\bibfield  {journal} {\bibinfo
  {journal} {Space Science Reviews}\ }\textbf {\bibinfo {volume} {217}},\ \href
  {https://doi.org/10.1007/s11214-021-00800-3} {10.1007/s11214-021-00800-3}
  (\bibinfo {year} {2021})\BibitemShut {NoStop}%
\bibitem [{\citenamefont {di~Stefano}\ \emph
  {et~al.}(2021{\natexlab{b}})\citenamefont {di~Stefano}, \citenamefont
  {Cappuccio},\ and\ \citenamefont {Iess}}]{diStefano2021Cqg}%
  \BibitemOpen
  \bibfield  {author} {\bibinfo {author} {\bibfnamefont {I.}~\bibnamefont
  {di~Stefano}}, \bibinfo {author} {\bibfnamefont {P.}~\bibnamefont
  {Cappuccio}},\ and\ \bibinfo {author} {\bibfnamefont {L.}~\bibnamefont
  {Iess}},\ }\bibfield  {title} {\bibinfo {title} {The bepicolombo solar
  conjunction experiments revisited},\ }\href
  {https://doi.org/10.1088/1361-6382/abd301} {\bibfield  {journal} {\bibinfo
  {journal} {Classical and Quantum Gravity}\ }\textbf {\bibinfo {volume}
  {38}},\ \bibinfo {pages} {055002} (\bibinfo {year} {2021}{\natexlab{b}})},\
  \Eprint {https://arxiv.org/abs/2201.05107} {arXiv:2201.05107 [gr-qc]}
  \BibitemShut {NoStop}%
\bibitem [{\citenamefont {Biskupek}\ \emph {et~al.}(2021)\citenamefont
  {Biskupek}, \citenamefont {M{\"u}ller},\ and\ \citenamefont
  {Torre}}]{Biskupek2021}%
  \BibitemOpen
  \bibfield  {author} {\bibinfo {author} {\bibfnamefont {L.}~\bibnamefont
  {Biskupek}}, \bibinfo {author} {\bibfnamefont {J.}~\bibnamefont
  {M{\"u}ller}},\ and\ \bibinfo {author} {\bibfnamefont {J.-M.}\ \bibnamefont
  {Torre}},\ }\bibfield  {title} {\bibinfo {title} {{Benefit of New
  High-Precision LLR Data for the Determination of Relativistic Parameters}},\
  }\href {https://doi.org/10.3390/universe7020034} {\bibfield  {journal}
  {\bibinfo  {journal} {Universe}\ }\textbf {\bibinfo {volume} {7}},\ \bibinfo
  {pages} {34} (\bibinfo {year} {2021})}\BibitemShut {NoStop}%
\bibitem [{\citenamefont {Fischer}\ \emph {et~al.}(2024)\citenamefont
  {Fischer}, \citenamefont {Kaeding},\ and\ \citenamefont
  {Pitschmann}}]{FischerKaedingPitschmann2024}%
  \BibitemOpen
  \bibfield  {author} {\bibinfo {author} {\bibfnamefont {H.}~\bibnamefont
  {Fischer}}, \bibinfo {author} {\bibfnamefont {C.}~\bibnamefont {Kaeding}},\
  and\ \bibinfo {author} {\bibfnamefont {M.}~\bibnamefont {Pitschmann}},\
  }\href@noop {} {\bibinfo {title} {{Screened Scalar Fields in the Laboratory
  and the Solar System}}} (\bibinfo {year} {2024}),\ \Eprint
  {https://arxiv.org/abs/2405.14638} {arXiv:2405.14638 [gr-qc]} \BibitemShut
  {NoStop}%
\bibitem [{\citenamefont {Benisty}(2022)}]{Benisty:2022Yukawa}%
  \BibitemOpen
  \bibfield  {author} {\bibinfo {author} {\bibfnamefont {D.}~\bibnamefont
  {Benisty}},\ }\bibfield  {title} {\bibinfo {title} {Testing modified gravity
  via yukawa potential in two body problem: Analytical solution and
  observational constraints},\ }\href
  {https://doi.org/10.1103/PhysRevD.106.043001} {\bibfield  {journal} {\bibinfo
   {journal} {Phys. Rev. D}\ }\textbf {\bibinfo {volume} {106}},\ \bibinfo
  {pages} {043001} (\bibinfo {year} {2022})},\ \Eprint
  {https://arxiv.org/abs/2207.08235} {arXiv:2207.08235 [gr-qc]} \BibitemShut
  {NoStop}%
\bibitem [{\citenamefont {Badurina}\ \emph {et~al.}(2020)\citenamefont
  {Badurina} \emph {et~al.}}]{Badurina2020AION}%
  \BibitemOpen
  \bibfield  {author} {\bibinfo {author} {\bibfnamefont {L.}~\bibnamefont
  {Badurina}} \emph {et~al.},\ }\bibfield  {title} {\bibinfo {title} {{AION: An
  atom interferometer observatory and network}},\ }\href
  {https://doi.org/10.1088/1475-7516/2020/05/011} {\bibfield  {journal}
  {\bibinfo  {journal} {JCAP}\ }\bibfield  {number} {\bibinfo  {number} {
  (5)},\ \bibinfo {pages} {011}},\ }\Eprint {https://arxiv.org/abs/1911.11755}
  {1911.11755} \BibitemShut {NoStop}%
\bibitem [{\citenamefont {Abou El-Neaj}\ \emph {et~al.}(2020)\citenamefont
  {Abou El-Neaj} \emph {et~al.}}]{AEDGE2020}%
  \BibitemOpen
  \bibfield  {author} {\bibinfo {author} {\bibfnamefont {Y.}~\bibnamefont {Abou
  El-Neaj}} \emph {et~al.},\ }\bibfield  {title} {\bibinfo {title} {{AEDGE:
  Atomic Experiment for Dark Matter and Gravity Exploration in Space}},\ }\href
  {https://doi.org/10.1140/epjqt/s40507-020-0080-0} {\bibfield  {journal}
  {\bibinfo  {journal} {EPJ Quantum Technology}\ }\textbf {\bibinfo {volume}
  {7}},\ \bibinfo {pages} {6} (\bibinfo {year} {2020})}\BibitemShut {NoStop}%
\bibitem [{\citenamefont {Van~Tilburg}\ \emph {et~al.}(2015)\citenamefont
  {Van~Tilburg}, \citenamefont {Leefer}, \citenamefont {Bougas},\ and\
  \citenamefont {Budker}}]{VanTilburg2015PRL}%
  \BibitemOpen
  \bibfield  {author} {\bibinfo {author} {\bibfnamefont {K.}~\bibnamefont
  {Van~Tilburg}}, \bibinfo {author} {\bibfnamefont {N.}~\bibnamefont {Leefer}},
  \bibinfo {author} {\bibfnamefont {L.}~\bibnamefont {Bougas}},\ and\ \bibinfo
  {author} {\bibfnamefont {D.}~\bibnamefont {Budker}},\ }\bibfield  {title}
  {\bibinfo {title} {{Search for Ultralight Scalar Dark Matter with Atomic
  Spectroscopy}},\ }\href {https://doi.org/10.1103/PhysRevLett.115.011802}
  {\bibfield  {journal} {\bibinfo  {journal} {Phys. Rev. Lett.}\ }\textbf
  {\bibinfo {volume} {115}},\ \bibinfo {pages} {011802} (\bibinfo {year}
  {2015})}\BibitemShut {NoStop}%
\bibitem [{\citenamefont {Stadnik}\ and\ \citenamefont
  {Flambaum}(2015{\natexlab{a}})}]{StadnikFlambaum2015PRL}%
  \BibitemOpen
  \bibfield  {author} {\bibinfo {author} {\bibfnamefont {Y.~V.}\ \bibnamefont
  {Stadnik}}\ and\ \bibinfo {author} {\bibfnamefont {V.~V.}\ \bibnamefont
  {Flambaum}},\ }\bibfield  {title} {\bibinfo {title} {{Searching for Dark
  Matter and Variation of Fundamental Constants with Laser and Maser
  Interferometry}},\ }\href {https://doi.org/10.1103/PhysRevLett.114.161301}
  {\bibfield  {journal} {\bibinfo  {journal} {Phys. Rev. Lett.}\ }\textbf
  {\bibinfo {volume} {114}},\ \bibinfo {pages} {161301} (\bibinfo {year}
  {2015}{\natexlab{a}})}\BibitemShut {NoStop}%
\bibitem [{\citenamefont {Arvanitaki}\ \emph {et~al.}(2015)\citenamefont
  {Arvanitaki}, \citenamefont {Huang},\ and\ \citenamefont
  {Van~Tilburg}}]{Arvanitaki2015PRD}%
  \BibitemOpen
  \bibfield  {author} {\bibinfo {author} {\bibfnamefont {A.}~\bibnamefont
  {Arvanitaki}}, \bibinfo {author} {\bibfnamefont {J.}~\bibnamefont {Huang}},\
  and\ \bibinfo {author} {\bibfnamefont {K.}~\bibnamefont {Van~Tilburg}},\
  }\bibfield  {title} {\bibinfo {title} {{Searching for Dilaton Dark Matter
  with Atomic Clocks}},\ }\href {https://doi.org/10.1103/PhysRevD.91.015015}
  {\bibfield  {journal} {\bibinfo  {journal} {Phys. Rev. D}\ }\textbf {\bibinfo
  {volume} {91}},\ \bibinfo {pages} {015015} (\bibinfo {year}
  {2015})}\BibitemShut {NoStop}%
\bibitem [{\citenamefont {{Turyshev}}(2025)}]{Turyshev2025-CCR}%
  \BibitemOpen
  \bibfield  {author} {\bibinfo {author} {\bibfnamefont {S.~G.}\ \bibnamefont
  {{Turyshev}}},\ }\href {https://doi.org/10.48550/arXiv.2504.06409} {\bibinfo
  {title} {{{High-Precision Lunar Corner-Cube Retroreflectors: A Wave-Optics
  Perspective}}}} (\bibinfo {year} {2025}),\ \Eprint
  {https://arxiv.org/abs/arXiv:2504.06409 [physics.optics]} {arXiv:2504.06409
  [physics.optics]} \BibitemShut {NoStop}%
\bibitem [{\citenamefont {Smith}\ \emph {et~al.}(2006)\citenamefont {Smith},
  \citenamefont {Zuber}, \citenamefont {Sun}, \citenamefont {Neumann},
  \citenamefont {Cavanaugh}, \citenamefont {McGarry},\ and\ \citenamefont
  {Zagwodzki}}]{Smith2006_Science}%
  \BibitemOpen
  \bibfield  {author} {\bibinfo {author} {\bibfnamefont {D.~E.}\ \bibnamefont
  {Smith}}, \bibinfo {author} {\bibfnamefont {M.~T.}\ \bibnamefont {Zuber}},
  \bibinfo {author} {\bibfnamefont {X.}~\bibnamefont {Sun}}, \bibinfo {author}
  {\bibfnamefont {G.~A.}\ \bibnamefont {Neumann}}, \bibinfo {author}
  {\bibfnamefont {J.~F.}\ \bibnamefont {Cavanaugh}}, \bibinfo {author}
  {\bibfnamefont {J.~F.}\ \bibnamefont {McGarry}},\ and\ \bibinfo {author}
  {\bibfnamefont {T.~W.}\ \bibnamefont {Zagwodzki}},\ }\bibfield  {title}
  {\bibinfo {title} {{Two-Way Laser Link over Interplanetary Distance}},\
  }\href {https://doi.org/10.1126/science.1120091} {\bibfield  {journal}
  {\bibinfo  {journal} {Science}\ }\textbf {\bibinfo {volume} {311}},\ \bibinfo
  {pages} {53} (\bibinfo {year} {2006})}\BibitemShut {NoStop}%
\bibitem [{\citenamefont {Murphy~Jr.}(2012)}]{Murphy2012_CQG}%
  \BibitemOpen
  \bibfield  {author} {\bibinfo {author} {\bibfnamefont {T.~W.}\ \bibnamefont
  {Murphy~Jr.}},\ }\bibfield  {title} {\bibinfo {title} {{Lunar laser ranging:
  the millimeter challenge}},\ }\href
  {https://doi.org/10.1088/0264-9381/29/18/184005} {\bibfield  {journal}
  {\bibinfo  {journal} {CQG}\ }\textbf {\bibinfo {volume} {29}},\ \bibinfo
  {pages} {184005} (\bibinfo {year} {2012})}\BibitemShut {NoStop}%
\bibitem [{\citenamefont {Zhang}\ \emph {et~al.}(2024)\citenamefont {Zhang},
  \citenamefont {M{\"u}ller},\ and\ \citenamefont {Biskupek}}]{Zhang2024_AA}%
  \BibitemOpen
  \bibfield  {author} {\bibinfo {author} {\bibfnamefont {M.}~\bibnamefont
  {Zhang}}, \bibinfo {author} {\bibfnamefont {J.}~\bibnamefont {M{\"u}ller}},\
  and\ \bibinfo {author} {\bibfnamefont {L.}~\bibnamefont {Biskupek}},\
  }\bibfield  {title} {\bibinfo {title} {{Advantages of combining Lunar Laser
  Ranging and Differential Lunar Laser Ranging}},\ }\href
  {https://doi.org/10.1051/0004-6361/202347643} {\bibfield  {journal} {\bibinfo
   {journal} {Astronomy \& Astrophysics}\ }\textbf {\bibinfo {volume} {681}},\
  \bibinfo {pages} {A5} (\bibinfo {year} {2024})}\BibitemShut {NoStop}%
\bibitem [{\citenamefont {Turyshev}\ \emph {et~al.}(2004)\citenamefont
  {Turyshev}, \citenamefont {Shao},\ and\ \citenamefont
  {Nordtvedt}}]{Turyshev2004_CQG}%
  \BibitemOpen
  \bibfield  {author} {\bibinfo {author} {\bibfnamefont {S.~G.}\ \bibnamefont
  {Turyshev}}, \bibinfo {author} {\bibfnamefont {M.}~\bibnamefont {Shao}},\
  and\ \bibinfo {author} {\bibfnamefont {K.}~\bibnamefont {Nordtvedt}},\
  }\bibfield  {title} {\bibinfo {title} {{The Laser Astrometric Test of
  Relativity (LATOR) Mission}},\ }\href
  {https://doi.org/10.1088/0264-9381/21/12/004} {\bibfield  {journal} {\bibinfo
   {journal} {CQG}\ }\textbf {\bibinfo {volume} {21}},\ \bibinfo {pages} {2773}
  (\bibinfo {year} {2004})},\ \Eprint
  {https://arxiv.org/abs/arXiv:gr-qc/0311020} {arXiv:gr-qc/0311020}
  \BibitemShut {NoStop}%
\bibitem [{\citenamefont {{Turyshev}}\ \emph
  {et~al.}(2007{\natexlab{b}})\citenamefont {{Turyshev}}, \citenamefont
  {{Shao}},\ and\ \citenamefont {{Nordtvedt}}}]{Turyshev2007_LATOR}%
  \BibitemOpen
  \bibfield  {author} {\bibinfo {author} {\bibfnamefont {S.~G.}\ \bibnamefont
  {{Turyshev}}}, \bibinfo {author} {\bibfnamefont {M.}~\bibnamefont {{Shao}}},\
  and\ \bibinfo {author} {\bibfnamefont {K.}~\bibnamefont {{Nordtvedt}}},\
  }\bibfield  {title} {\bibinfo {title} {{{Mission design for the laser
  astrometric test of relativity}}},\ }\href
  {https://doi.org/10.1016/j.asr.2005.07.003} {\bibfield  {journal} {\bibinfo
  {journal} {Adv. Space Res.}\ }\textbf {\bibinfo {volume} {39}},\ \bibinfo
  {pages} {297} (\bibinfo {year} {2007}{\natexlab{b}})},\ \Eprint
  {https://arxiv.org/abs/gr-qc/0409111} {gr-qc/0409111} \BibitemShut {NoStop}%
\bibitem [{\citenamefont {{Turyshev}}\ \emph {et~al.}(2009)\citenamefont
  {{Turyshev}}, \citenamefont {{Shao}}, \citenamefont {{Girerd}},\ and\
  \citenamefont {{Lane}}}]{Turyshev2009-BEACON}%
  \BibitemOpen
  \bibfield  {author} {\bibinfo {author} {\bibfnamefont {S.~G.}\ \bibnamefont
  {{Turyshev}}}, \bibinfo {author} {\bibfnamefont {M.}~\bibnamefont {{Shao}}},
  \bibinfo {author} {\bibfnamefont {A.}~\bibnamefont {{Girerd}}},\ and\
  \bibinfo {author} {\bibfnamefont {B.}~\bibnamefont {{Lane}}},\ }\bibfield
  {title} {\bibinfo {title} {{{Search for New Physics with the Beacon
  Mission}}},\ }\href {https://doi.org/10.1142/S0218271809014893} {\bibfield
  {journal} {\bibinfo  {journal} {IJMPD}\ }\textbf {\bibinfo {volume} {18}},\
  \bibinfo {pages} {1025} (\bibinfo {year} {2009})},\ \Eprint
  {https://arxiv.org/abs/arXiv:0805.4033 [gr-qc]} {arXiv:0805.4033 [gr-qc]}
  \BibitemShut {NoStop}%
\bibitem [{\citenamefont {Bekenstein}(1993)}]{Bekenstein1993PRD}%
  \BibitemOpen
  \bibfield  {author} {\bibinfo {author} {\bibfnamefont {J.~D.}\ \bibnamefont
  {Bekenstein}},\ }\bibfield  {title} {\bibinfo {title} {{Relation Between
  Physical and Gravitational Geometry}},\ }\href
  {https://doi.org/10.1103/PhysRevD.48.3641} {\bibfield  {journal} {\bibinfo
  {journal} {Phys. Rev. D}\ }\textbf {\bibinfo {volume} {48}},\ \bibinfo
  {pages} {3641} (\bibinfo {year} {1993})}\BibitemShut {NoStop}%
\bibitem [{\citenamefont {Koivisto}\ \emph {et~al.}(2012)\citenamefont
  {Koivisto}, \citenamefont {Mota},\ and\ \citenamefont
  {Zumalac{\'a}rregui}}]{KoivistoMotaZumalacarregui2012PRL}%
  \BibitemOpen
  \bibfield  {author} {\bibinfo {author} {\bibfnamefont {T.~S.}\ \bibnamefont
  {Koivisto}}, \bibinfo {author} {\bibfnamefont {D.~F.}\ \bibnamefont {Mota}},\
  and\ \bibinfo {author} {\bibfnamefont {M.}~\bibnamefont
  {Zumalac{\'a}rregui}},\ }\bibfield  {title} {\bibinfo {title} {{Screening
  Modifications of Gravity Through Disformally Coupled Fields}},\ }\href
  {https://doi.org/10.1103/PhysRevLett.109.241102} {\bibfield  {journal}
  {\bibinfo  {journal} {Phys. Rev. Lett.}\ }\textbf {\bibinfo {volume} {109}},\
  \bibinfo {pages} {241102} (\bibinfo {year} {2012})}\BibitemShut {NoStop}%
\bibitem [{\citenamefont {Brax}\ \emph {et~al.}(2013)\citenamefont {Brax},
  \citenamefont {Burrage}, \citenamefont {Davis},\ and\ \citenamefont
  {Gubitosi}}]{BraxBurrageDavisGubitosi2013JCAP}%
  \BibitemOpen
  \bibfield  {author} {\bibinfo {author} {\bibfnamefont {P.}~\bibnamefont
  {Brax}}, \bibinfo {author} {\bibfnamefont {C.}~\bibnamefont {Burrage}},
  \bibinfo {author} {\bibfnamefont {A.-C.}\ \bibnamefont {Davis}},\ and\
  \bibinfo {author} {\bibfnamefont {G.}~\bibnamefont {Gubitosi}},\ }\bibfield
  {title} {\bibinfo {title} {{Cosmological Tests of the Disformal Coupling to
  Radiation}},\ }\href {https://doi.org/10.1088/1475-7516/2013/11/001}
  {\bibfield  {journal} {\bibinfo  {journal} {JCAP}\ }\textbf {\bibinfo
  {volume} {2013}}\bibinfo  {number} { (11)},\ \bibinfo {pages}
  {001}}\BibitemShut {NoStop}%
\bibitem [{\citenamefont {Sakstein}(2014)}]{Sakstein2014JCAP}%
  \BibitemOpen
\bibfield  {number} {  }\bibfield  {author} {\bibinfo {author} {\bibfnamefont
  {J.}~\bibnamefont {Sakstein}},\ }\bibfield  {title} {\bibinfo {title}
  {{Disformal Theories of Gravity: From the Solar System to Cosmology}},\
  }\href {https://doi.org/10.1088/1475-7516/2014/12/012} {\bibfield  {journal}
  {\bibinfo  {journal} {JCAP}\ }\textbf {\bibinfo {volume} {2014}}\bibinfo
  {number} { (12)},\ \bibinfo {pages} {012}}\BibitemShut {NoStop}%
\bibitem [{\citenamefont {Brax}\ \emph {et~al.}(2018)\citenamefont {Brax},
  \citenamefont {Burrage},\ and\ \citenamefont
  {Davis}}]{BraxBurrageDavis2018PRD}%
  \BibitemOpen
\bibfield  {number} {  }\bibfield  {author} {\bibinfo {author} {\bibfnamefont
  {P.}~\bibnamefont {Brax}}, \bibinfo {author} {\bibfnamefont {C.}~\bibnamefont
  {Burrage}},\ and\ \bibinfo {author} {\bibfnamefont {A.-C.}\ \bibnamefont
  {Davis}},\ }\bibfield  {title} {\bibinfo {title} {{Gravitational Effects of
  Disformal Couplings}},\ }\href {https://doi.org/10.1103/PhysRevD.98.063531}
  {\bibfield  {journal} {\bibinfo  {journal} {Phys. Rev. D}\ }\textbf {\bibinfo
  {volume} {98}},\ \bibinfo {pages} {063531} (\bibinfo {year}
  {2018})}\BibitemShut {NoStop}%
\bibitem [{\citenamefont {Benisty}\ and\ \citenamefont
  {Davis}(2022)}]{BenistyDavis:2022GC}%
  \BibitemOpen
  \bibfield  {author} {\bibinfo {author} {\bibfnamefont {D.}~\bibnamefont
  {Benisty}}\ and\ \bibinfo {author} {\bibfnamefont {A.-C.}\ \bibnamefont
  {Davis}},\ }\bibfield  {title} {\bibinfo {title} {Dark energy interactions
  near the galactic centre},\ }\href
  {https://doi.org/10.1103/PhysRevD.105.024052} {\bibfield  {journal} {\bibinfo
   {journal} {Phys. Rev. D}\ }\textbf {\bibinfo {volume} {105}},\ \bibinfo
  {pages} {024052} (\bibinfo {year} {2022})},\ \Eprint
  {https://arxiv.org/abs/2108.06286} {arXiv:2108.06286 [astro-ph.CO]}
  \BibitemShut {NoStop}%
\bibitem [{\citenamefont {{Chakraborty}}\ \emph {et~al.}(2025)\citenamefont
  {{Chakraborty}}, \citenamefont {{Chanda}}, \citenamefont {{Das}},\ and\
  \citenamefont {{Dutta}}}]{Chakraborty:2025}%
  \BibitemOpen
  \bibfield  {author} {\bibinfo {author} {\bibfnamefont {A.}~\bibnamefont
  {{Chakraborty}}}, \bibinfo {author} {\bibfnamefont {P.~K.}\ \bibnamefont
  {{Chanda}}}, \bibinfo {author} {\bibfnamefont {S.}~\bibnamefont {{Das}}},\
  and\ \bibinfo {author} {\bibfnamefont {K.}~\bibnamefont {{Dutta}}},\ }\href
  {https://doi.org/10.48550/arXiv.2503.10806} {\bibinfo {title} {{{DESI
  results: Hint towards coupled dark matter and dark energy}}}} (\bibinfo
  {year} {2025}),\ \Eprint {https://arxiv.org/abs/arXiv:2503.10806
  [astro-ph.CO]} {arXiv:2503.10806 [astro-ph.CO]} \BibitemShut {NoStop}%
\bibitem [{\citenamefont {Graham}\ and\ \citenamefont
  {Rajendran}(2013)}]{GrahamRajendran2013}%
  \BibitemOpen
  \bibfield  {author} {\bibinfo {author} {\bibfnamefont {P.~W.}\ \bibnamefont
  {Graham}}\ and\ \bibinfo {author} {\bibfnamefont {S.}~\bibnamefont
  {Rajendran}},\ }\bibfield  {title} {\bibinfo {title} {{New Observables for
  Light Dark Matter}},\ }\href@noop {} {\bibfield  {journal} {\bibinfo
  {journal} {Phys. Rev. D}\ }\textbf {\bibinfo {volume} {88}},\ \bibinfo
  {pages} {035023} (\bibinfo {year} {2013})},\ \Eprint
  {https://arxiv.org/abs/1306.6088} {arXiv:1306.6088} \BibitemShut {NoStop}%
\bibitem [{\citenamefont {Budker}\ \emph {et~al.}(2014)\citenamefont {Budker},
  \citenamefont {Graham}, \citenamefont {Ledbetter}, \citenamefont
  {Rajendran},\ and\ \citenamefont {Sushkov}}]{Budker2014}%
  \BibitemOpen
  \bibfield  {author} {\bibinfo {author} {\bibfnamefont {D.}~\bibnamefont
  {Budker}}, \bibinfo {author} {\bibfnamefont {P.~W.}\ \bibnamefont {Graham}},
  \bibinfo {author} {\bibfnamefont {M.}~\bibnamefont {Ledbetter}}, \bibinfo
  {author} {\bibfnamefont {S.}~\bibnamefont {Rajendran}},\ and\ \bibinfo
  {author} {\bibfnamefont {A.~O.}\ \bibnamefont {Sushkov}},\ }\bibfield
  {title} {\bibinfo {title} {{Proposal to Search for Axionlike Dark Matter
  Using Nuclear Magnetic Resonance}},\ }\href@noop {} {\bibfield  {journal}
  {\bibinfo  {journal} {Phys. Rev. X}\ }\textbf {\bibinfo {volume} {4}},\
  \bibinfo {pages} {021030} (\bibinfo {year} {2014})},\ \Eprint
  {https://arxiv.org/abs/1306.6089} {arXiv:1306.6089} \BibitemShut {NoStop}%
\bibitem [{\citenamefont {Holdom}(1986)}]{Holdom1986}%
  \BibitemOpen
  \bibfield  {author} {\bibinfo {author} {\bibfnamefont {B.}~\bibnamefont
  {Holdom}},\ }\bibfield  {title} {\bibinfo {title} {{Two U(1)'s and Epsilon
  Charge Shifts}},\ }\href@noop {} {\bibfield  {journal} {\bibinfo  {journal}
  {Phys. Lett. B}\ }\textbf {\bibinfo {volume} {166}},\ \bibinfo {pages} {196}
  (\bibinfo {year} {1986})}\BibitemShut {NoStop}%
\bibitem [{\citenamefont {Arias}\ \emph {et~al.}(2012)\citenamefont {Arias},
  \citenamefont {Cadamuro}, \citenamefont {Goodsell}, \citenamefont {Jaeckel},
  \citenamefont {Redondo},\ and\ \citenamefont {Ringwald}}]{Arias2012}%
  \BibitemOpen
  \bibfield  {author} {\bibinfo {author} {\bibfnamefont {P.}~\bibnamefont
  {Arias}}, \bibinfo {author} {\bibfnamefont {D.}~\bibnamefont {Cadamuro}},
  \bibinfo {author} {\bibfnamefont {M.}~\bibnamefont {Goodsell}}, \bibinfo
  {author} {\bibfnamefont {J.}~\bibnamefont {Jaeckel}}, \bibinfo {author}
  {\bibfnamefont {J.}~\bibnamefont {Redondo}},\ and\ \bibinfo {author}
  {\bibfnamefont {A.}~\bibnamefont {Ringwald}},\ }\bibfield  {title} {\bibinfo
  {title} {{WISPy Cold Dark Matter}},\ }\href
  {https://doi.org/10.1088/1475-7516/2012/06/013} {\bibfield  {journal}
  {\bibinfo  {journal} {JCAP}\ }\textbf {\bibinfo {volume} {2012}}\bibfield
  {number} {\bibinfo  {number} { (06)},\ \bibinfo {pages} {013}},\ }\Eprint
  {https://arxiv.org/abs/1201.5902} {arXiv:1201.5902} \BibitemShut {NoStop}%
\bibitem [{\citenamefont {Derevianko}\ and\ \citenamefont
  {Pospelov}(2014)}]{DereviankoPospelov2014}%
  \BibitemOpen
  \bibfield  {author} {\bibinfo {author} {\bibfnamefont {A.}~\bibnamefont
  {Derevianko}}\ and\ \bibinfo {author} {\bibfnamefont {M.}~\bibnamefont
  {Pospelov}},\ }\bibfield  {title} {\bibinfo {title} {{Hunting for Topological
  Dark Matter with Atomic Clocks}},\ }\href@noop {} {\bibfield  {journal}
  {\bibinfo  {journal} {Nat. Phys.}\ }\textbf {\bibinfo {volume} {10}},\
  \bibinfo {pages} {933} (\bibinfo {year} {2014})},\ \Eprint
  {https://arxiv.org/abs/1311.1244} {arXiv:1311.1244} \BibitemShut {NoStop}%
\bibitem [{\citenamefont {Khmelnitsky}\ and\ \citenamefont
  {Rubakov}(2014)}]{KhmelnitskyRubakov2014}%
  \BibitemOpen
  \bibfield  {author} {\bibinfo {author} {\bibfnamefont {A.}~\bibnamefont
  {Khmelnitsky}}\ and\ \bibinfo {author} {\bibfnamefont {V.}~\bibnamefont
  {Rubakov}},\ }\bibfield  {title} {\bibinfo {title} {{Pulsar timing signal
  from ultralight scalar dark matter}},\ }\href
  {https://doi.org/10.1088/1475-7516/2014/02/019} {\bibfield  {journal}
  {\bibinfo  {journal} {JCAP}\ }\textbf {\bibinfo {volume} {2014}}\bibfield
  {number} {\bibinfo  {number} { (02)},\ \bibinfo {pages} {019}},\ }\Eprint
  {https://arxiv.org/abs/1309.5888} {arXiv:1309.5888} \BibitemShut {NoStop}%
\bibitem [{\citenamefont {Stadnik}\ and\ \citenamefont
  {Flambaum}(2015{\natexlab{b}})}]{StadnikFlambaum2015HiggsPortal}%
  \BibitemOpen
  \bibfield  {author} {\bibinfo {author} {\bibfnamefont {Y.~V.}\ \bibnamefont
  {Stadnik}}\ and\ \bibinfo {author} {\bibfnamefont {V.~V.}\ \bibnamefont
  {Flambaum}},\ }\bibfield  {title} {\bibinfo {title} {{Can Dark Matter Induce
  Cosmological Evolution of the Fundamental Constants of Nature?}},\
  }\href@noop {} {\bibfield  {journal} {\bibinfo  {journal} {Phys. Rev. Lett.}\
  }\textbf {\bibinfo {volume} {115}},\ \bibinfo {pages} {201301} (\bibinfo
  {year} {2015}{\natexlab{b}})},\ \Eprint {https://arxiv.org/abs/1503.08540}
  {arXiv:1503.08540} \BibitemShut {NoStop}%
\bibitem [{\citenamefont {Damour}\ and\ \citenamefont
  {Donoghue}(2010)}]{DamourDonoghue2010a}%
  \BibitemOpen
  \bibfield  {author} {\bibinfo {author} {\bibfnamefont {T.}~\bibnamefont
  {Damour}}\ and\ \bibinfo {author} {\bibfnamefont {J.~F.}\ \bibnamefont
  {Donoghue}},\ }\bibfield  {title} {\bibinfo {title} {{Equivalence Principle
  Violations and Couplings of a Light Dilaton}},\ }\href@noop {} {\bibfield
  {journal} {\bibinfo  {journal} {Phys. Rev. D}\ }\textbf {\bibinfo {volume}
  {82}},\ \bibinfo {pages} {084033} (\bibinfo {year} {2010})},\ \Eprint
  {https://arxiv.org/abs/1007.2792} {arXiv:1007.2792} \BibitemShut {NoStop}%
\bibitem [{\citenamefont {Centers}\ and\ \citenamefont {{et
  al.}}(2021)}]{Centers2021ClockDMNetwork}%
  \BibitemOpen
  \bibfield  {author} {\bibinfo {author} {\bibfnamefont {G.~P.}\ \bibnamefont
  {Centers}}\ and\ \bibinfo {author} {\bibnamefont {{et al.}}},\ }\bibfield
  {title} {\bibinfo {title} {{Stochastic variations of fundamental constants
  using clock networks}},\ }\href@noop {} {\bibfield  {journal} {\bibinfo
  {journal} {Nature Communications}\ }\textbf {\bibinfo {volume} {12}},\
  \bibinfo {pages} {7321} (\bibinfo {year} {2021})},\ \Eprint
  {https://arxiv.org/abs/1905.13650} {arXiv:1905.13650} \BibitemShut {NoStop}%
\end{thebibliography}

%

\appendix

\section{Beyond universal conformal couplings: disformal terms}
\label{app:disformal}

A general matter metric may include a disformal term,
\begin{equation}
\tilde g_{\mu\nu} = C(\phi)\, g_{\mu\nu} + D(\phi)\,\partial_\mu\phi\,\partial_\nu\phi\,,
\end{equation}
with $C(\phi)>0$ and $D(\phi)$ analytic (see
\cite{Bekenstein1993PRD} for the original construction and \cite{KoivistoMotaZumalacarregui2012PRL} for ``disformal screening.'')
In the non-relativistic limit around \emph{static} sources, the disformal term does not generate a fifth force at leading order, so PPN bounds primarily constrain the conformal slope $C'(\phi_\star)/C(\phi_\star)$.
However, for time-dependent backgrounds ($\dot\phi\neq 0$), disformal effects can enter light propagation and cosmology, modifying distance duality and CMB spectral distortions, and inducing Solar System signatures suppressed by $\dot\phi$ (see \cite{BraxBurrageDavisGubitosi2013JCAP,Sakstein2014JCAP,BraxBurrageDavis2018PRD}.)
Our guardrails extend verbatim: (i) impose EEP/PPN nulls; (ii) map DESI/Euclid posteriors on $\mu(z,k),\Sigma(z,k)$ to $\{C(\phi),D(\phi)\}$ consistent with $c_T\!\simeq\!c$; (iii) pursue dedicated Solar System tests only when a specified $\{C,D\}$ predicts at least one local residual above credible thresholds. For observational constraints on conformal and disformal couplings from S2 orbits near Sgr~A*, see~\cite{BenistyDavis:2022GC}.

\section{A toy joint-likelihood across regimes}
\label{app:joint}

Let $\theta$ denote cosmology-level MG parameters (e.g., $\theta=\{\mu_0,\Sigma_0\}$) with posterior
\begin{equation}
p(\theta|C)\propto \exp\!\left[-\tfrac12(\theta-\bar\theta)^{\!\top}\,\mathbf{C}_C^{-1}\,(\theta-\bar\theta)\right].
\end{equation}
Let $r(\theta)$ be a local residual (e.g., $r=\gamma-1$) predicted via the screening map of Sec.~II (thin shell or Vainshtein).
Given a Solar System measurement $d_S$ with covariance $\mathbf{C}_S$, define
\begin{equation}
p(d_S|\theta)\propto \exp\!\left[-\tfrac12\bigl(d_S-r(\theta)\bigr)^{\!\top}\mathbf{C}_S^{-1}\bigl(d_S-r(\theta)\bigr)\right].
\end{equation}
The joint posterior is
\begin{equation}
p(\theta|C,S)\propto p(\theta|C)\,p(d_S|\theta)\,.
\end{equation}
For small excursions one may linearize $r(\theta)\simeq r(\bar\theta)+\mathbf{J}(\theta-\bar\theta)$ with Jacobian $\mathbf{J}$, which yields a closed-form Gaussian update of $\bar\theta$ and covariance. A null Solar System result tightens $\theta$ along the rows of $\mathbf{J}$; a detection triggers a non-linear refit with the same microphysical parameterization.

\section{Reproducibility recipe (cosmology $\to$ Solar System)}
\label{sec:recipe}

Here we present a recipe on how to relate cosmological conditions to those of the Solar System: 
{}
\begin{enumerate}
\item Pick a linear-response parameter $\mulin \equiv \mu(z{=}0,k{\sim}0.1\,h\,\mathrm{Mpc}^{-1})-1$ and map to the local coupling via estublished expression 
$\chi \simeq \sqrt{\mulin/2}$ (unscreened limit).
\item For chameleon-like models with $V(\phi)=\Lambda^{4+n}\phi^{-n}$ and $A(\phi)=e^{\chi\phi/M_{\rm Pl}}$, compute the density minimum using
\[
\phi_\star(\rho)=\Big(\frac{n\,\Lambda^{4+n} M_{\rm Pl}}{\chi\,\rho}\Big)^{\!1/(n+1)} \quad \text{[Eq.~(\ref{eq:phi_min_powerlaw})].}
\]
\item In the Sun-screened regime ($\rho_c\!\gg\!\rho_\infty$), use
\[
\frac{\Delta R}{R} \simeq \frac{\phi_\star(\rho_\infty)}{6\,\chi\,M_{\rm Pl}\,\Phi_{N\odot}} \quad \text{[Eqs.~(\ref{eq:thin-shell}),(\ref{eq:PhiN-numbers})].}
\]
\item Translate a null Shapiro test at sensitivity $|\gamma-1|_{\max}$ into the thin-shell requirement
\[
\frac{\Delta R}{R} \lesssim \frac{1}{3\,\chi}\sqrt{\frac{|\gamma-1|_{\max}}{2}}\!,
\]
valid when the solar source is screened.
\end{enumerate}

\begin{table}[t]
\centering
\caption{Quick-look thin-shell bounds at the Sun for representative
$\mulin$ and $|\,\gamma-1\,|_{\max}$, using
$(\Delta R/R)_{\max} = (1/3\chi)\sqrt{|\,\gamma-1\,|_{\max}/2}$ with $\chi=\sqrt{\mulin/2}$.}
\begin{tabular}{cccc}
\hline
$\mulin$ & $\chi$ & $|\,\gamma-1\,|_{\max}$ & $(\Delta R/R)_{\max}$ \\
\hline\hline
0.05 & 0.158 & $1\times 10^{-6}$ & $1.49\times 10^{-3}$\\
0.05 & 0.158 & $5\times 10^{-6}$ & $3.33\times 10^{-3}$\\
0.10 & 0.224 & $1\times 10^{-6}$ & $1.05\times 10^{-3}$\\
0.10 & 0.224 & $5\times 10^{-6}$ & $2.36\times 10^{-3}$\\
0.15 & 0.274 & $1\times 10^{-6}$ & $8.61\times 10^{-4}$\\
0.15 & 0.274 & $5\times 10^{-6}$ & $1.92\times 10^{-3}$\\
\hline
\end{tabular}
\end{table}

\begin{table}[t]
\centering
\caption{{Quick-look Earth EEP guardrail.}
A null bound $\eta<\eta_{\max}$ on the E\"otv\"os parameter in (\ref{eq:eta-AIS-E}) limits
the product $\chi_\oplus \min\{1,3\Delta R_\oplus/R_\oplus\}$ to
$\eta_{\max}/(2|\Delta K_{\rm eff}|)$ for the chosen species/equipment combination
summarized by $\Delta K_{\rm eff}$. Two regimes follow directly:
(i) if $3\Delta R_\oplus/R_\oplus<1$ (screened Earth), the thin-shell relations, \eqref{eq:thin-shell}--\eqref{eq:thin_shell_powerlaw},
imply the \emph{ambient field excursion} bound
$\phi_\star(\rho_\infty)\le (\eta_{\max}/|\Delta K_{\rm eff}|)\,M_{\rm Pl}\,\Phi_{N,\oplus}$
(which inherits the same $\rho_\infty^{-1/(n+1)}$ scaling as Table~\ref{tab:ambient-density});
(ii) if $3\Delta R_\oplus/R_\oplus\ge 1$ (unscreened Earth), one has
$|\chi_\oplus|\le \eta_{\max}/(2|\Delta K_{\rm eff}|)$.
Numerical columns assume $\Phi_{N,\oplus}=6.96\times10^{-10}$ (Table~\ref{tab:potentials}) and
$\chi=\sqrt{\mu_{\rm lin,0}/2}$ (\ref{eq:eta-AIS-clock}).}
\label{tab:quick-lookup-eep}
\small
\begin{tabular}{cccccc}
\hline
$\mu_{\rm lin,0}$ & $\chi$ & $\eta_{\max}$ & $|\Delta K_{\rm eff}|$ &
$(\Delta R_\oplus/R_\oplus)_{\max}$\;\;({screened case}) &
$\big[\phi_\star/M_{\rm Pl}\big]_{\max}$ \\
\hline\hline
0.05 & 0.158 & $10^{-16}$ & 1.0 & $1.06\times10^{-16}$ & $6.96\times10^{-26}$ \\
0.10 & 0.224 & $10^{-16}$ & 1.0 & $7.46\times10^{-17}$ & $6.96\times10^{-26}$ \\
0.15 & 0.274 & $10^{-16}$ & 1.0 & $6.09\times10^{-17}$ & $6.96\times10^{-26}$ \\
\hline
0.10 & 0.224 & $10^{-16}$ & 0.3 & $2.49\times10^{-16}$ & $2.32\times10^{-25}$ \\
0.10 & 0.224 & $10^{-17}$ & 1.0 & $7.46\times10^{-18}$ & $6.96\times10^{-27}$ \\
\hline
\end{tabular}
\end{table}

\section{Sun thin shell and PPN \texorpdfstring{$\gamma$}{gamma} for \texorpdfstring{$(n,\chi)=(0.16,0.28)$}{(n,chi)=(0.16,0.28)}}
\label{sec:crosscheck}

This section considers the Sun thin-shell fraction $\Delta R/R$ and the associated PPN deviation $\gamma-1$ for the \emph{illustrative} choice $(n,\chi)=(0.16,0.28)$ using the framework defined in Secs.~\ref{sec:theory}--\ref{sec:cosmo}, considering the low $n$ scenarios  \cite{Chakraborty:2025}. We refer to the thin-shell relation and density minimum in Eqs.~\eqref{eq:thin-shell} and \eqref{eq:phi_min_powerlaw}, the Sun’s surface potential in Eq.~\eqref{eq:PhiN-numbers}/Table~\ref{tab:potentials}, the $\gamma$--shell map in Eq.~\eqref{eq:gamma_from_shell}, and the null-test guardrail in Eq.~\eqref{eq:thin-shell-bound}. The ambient-density prior along a solar-conjunction ray is summarized in Table~\ref{tab:ambient-density}. Starting from \eqref{eq:thin-shell} and \eqref{eq:phi_min_powerlaw},
\begin{equation}
\frac{\Delta R}{R}\simeq \frac{\phi_\infty-\phi_c}{6\,\chi\,\Mpl\,\Phi_N},
\qquad
\phi_\star(\rho)=\Big(\frac{n\,\Lambda^{4+n} \Mpl}{\chi\,\rho}\Big)^{\!\frac{1}{n+1}},
\end{equation}
we fix the cosmological normalization by requiring $V(\phi_\star(\rho_{\rm cos}))=\rho_{\rm DE}$ for the power-law potential $V(\phi)=\Lambda^{4+n}\phi^{-n}$. Using $V'(\phi_\star)+\rho_{\rm cos}A'(\phi_\star)=0$ one obtains
\begin{equation}
\Lambda^{4+n}=\rho_{\rm DE}^{\,n+1}\,n^{\,n}\,M_{\rm Pl}^{n}\,(\chi\,\rho_{\rm cos})^{-n}.
\label{eq:cross-Lambda}
\end{equation}

In the screened-Sun limit ($\phi_c\ll\phi_\infty$), $\Delta R/R\simeq \phi_\star(\rho_\infty)/(6\chi \Mpl \Phi_N)$. Substituting \eqref{eq:cross-Lambda} into $\phi_\star(\rho_\infty)$ yields the closed form
\begin{equation}
\frac{\Delta R}{R}=\frac{n}{6\,\chi^{2}\,\Phi_N}\;
\frac{\rho_{\rm DE}}{\rho_{\rm cos}^{\,n/(n+1)}\,\rho_\infty^{\,1/(n+1)}}.
\label{eq:cross-DeltaR-master}
\end{equation}
For $\rho_{\rm cos}=\rho_{\rm DE}$ this reduces to
\begin{equation}
\frac{\Delta R}{R}=\frac{n}{6\,\chi^{2}\,\Phi_N}\left(\frac{\rho_{\rm DE}}{\rho_\infty}\right)^{\!1/(n+1)},
\label{eq:cross-DeltaR-simple}
\end{equation}
which makes the $\rho_\infty^{-1/(n+1)}$ scaling explicit, consistent with Table~\ref{tab:ambient-density}. 

The Sun-sourced PPN mapping used in the main text is \eqref{eq:gamma_from_shell}, repeated here for convenience:
\begin{equation}
|\gamma-1|\;\simeq\;18\,\chi^2\!\left(\frac{\Delta R}{R}\right)^{\!2}.
\label{eq:D2}
\end{equation}
In the small-coupling limit one has $\gamma-1\simeq -2\alpha_\odot^2$; we thus compare to data using $|\gamma-1|=2\alpha_\odot^2$ throughout.

Numerical surface potentials needed below are given in Table~\ref{tab:potentials}. We adopt $\Phi_{N\odot}=2.12\times10^{-6}$ as per Eq.~\eqref{eq:PhiN-numbers} and Table~\ref{tab:potentials}, $(n,\chi)=(0.16,0.28)$, $\rho_{\rm DE}=10^{-29}\,{\rm g\,cm^{-3}}$, and the near-conjunction ambient-density prior given as $\rho_\infty\in[10^{-22},10^{-19}]\,{\rm g\,cm^{-3}}$ (Table~\ref{tab:ambient-density}). With $\rho_{\rm cos}=\rho_{\rm DE}$, Eq.~\eqref{eq:cross-DeltaR-simple} gives, for the central choice $\rho_\infty=10^{-20}\,{\rm g\,cm^{-3}}$,
\begin{align}
\frac{\Delta R}{R}
&=\underbrace{\frac{0.16}{6(0.28)^2(2.12\times10^{-6})}}_{1.604\times10^{5}}
\Big(10^{-9}\Big)^{1/(1+0.16)}
=2.80\times10^{-3},\\
\gamma-1&=18(0.28)^2\,(2.80\times10^{-3})^2
=1.10\times10^{-5}.
\end{align}
Across the prior range of \(\rho_\infty\) used in the paper (Table~\ref{tab:ambient-density}; ``a few \(R_\odot\)'' along the conjunction ray),
\[
\begin{array}{lcl}
\rho_\infty=10^{-22}\;{\rm g\,cm^{-3}}:\! & \Delta R/R=1.482\times10^{-1}, & \gamma-1=3.10\times10^{-2},\\[2pt]
\rho_\infty=10^{-19}\;{\rm g\,cm^{-3}}:\! & \Delta R/R=3.842\times10^{-4}, & \gamma-1=2.08\times10^{-7}.
\end{array}
\]
The thin-shell approximation requires $\Delta R/R\ll1$; when this fails, the screened-source premise is not satisfied and such points are excluded by Solar System bounds.

Using Eq.~\eqref{eq:gamma_from_shell}, the null-test guardrail Eq.~\eqref{eq:thin-shell-bound} is
\begin{equation}
\Big(\frac{\Delta R}{R}\Big)_{\!\max}
=\frac{1}{3\chi}\sqrt{\frac{|\gamma-1|_{\max}}{2}}.
\end{equation}
For \(\chi=0.28\),
\[
\begin{array}{lll}
|\gamma-1|_{\max}=2.3\times10^{-5}\ \text{(Cassini, Eq.~(\ref{eq:cassini-gamma}))}: & (\Delta R/R)_{\max}=4.04\times10^{-3},\\[2pt]
|\gamma-1|_{\max}=5\times10^{-6}: & (\Delta R/R)_{\max}=1.88\times10^{-3},\\[2pt]
|\gamma-1|_{\max}=1\times10^{-6}: & (\Delta R/R)_{\max}=8.42\times10^{-4}.
\end{array}
\]
Our central result \(\Delta R/R=2.80\times10^{-3}\) (at \(\rho_\infty=10^{-20}\,{\rm g\,cm^{-3}}\)) is \emph{below} the Cassini bound and consistent with the mapping shown in  Fig.~\ref{fig:mapping}.

\section{Broader ULDM model space and Solar System observables.}
\label{sec:ULDM-broader}

Eqs.~\eqref{eq:clock-dm}--\eqref{eq:tc-correct} summarize the narrowband signal in clock/interferometer channels from a coherently oscillating \emph{scalar} with linear couplings. Here we enlarge the model set to include: (i) scalars with \emph{quadratic} couplings; (ii) \emph{pseudoscalars} (axion-like) with derivative and gauge couplings; (iii) \emph{vectors} (dark photon or gauged $B\!-\!L$); (iv) \emph{transient} ULDM (topological defects); and, in addition, four classes often discussed in the ULDM literature and relevant to Solar System tests:
(v) \emph{gravity-only (“metric”) ULDM}, (vi) \emph{Higgs-portal scalars} (predictive coupling pattern), (vii) \emph{spin‑2 ULDM} (coherent tidal fields), and (viii) \emph{substructure transits} (soliton cores/miniclusters). We keep the notation of Sec.~\ref{sec:DM-clocks}, normalize to \eqref{eq:clock-dm}--\eqref{eq:tc-correct}, and use Fig.~\ref{fig:solar-system-tests}(d) for the mass--frequency band picture. Compact derivations appear in Appendix~\ref{app:ULDM}.

\subsection{Scalars with linear \& quadratic couplings.}
Allow both $d_i$ and $d_i^{(2)}$ in \eqref{eq:clock-dm}:
\[
\frac{\delta\nu}{\nu}(t)=\sum_i K_i\!\left(d_i\,\phi(t)+\tfrac{1}{2}d_i^{(2)}\,\phi^2(t)\right),\qquad \phi(t)=\phi_0\cos(m_\phi t),
\]
so, beyond the $m_\phi$ carrier, there is a DC offset $\propto\langle\phi^2\rangle$ and a line at $2m_\phi$ with amplitude 
$A_{2m_\phi}\!\simeq\!\tfrac{1}{4}\big|\sum_i K_i d_i^{(2)}\big|\phi_0^2=\tfrac{1}{2}\big|\sum_i K_i d_i^{(2)}\big|\rho_{\rm DM}/m_\phi^2$, and the same ${\sim}\sqrt{T/t_c}$ stacking law from \eqref{eq:tc-correct}.

\subsection{Pseudoscalars (axion-like).}
With $aF\tilde F$, $aG\tilde G$, and $\partial_\mu a\,\bar e\gamma^\mu\gamma^5 e$, the axial-electron term induces
\[
\delta\omega_{\rm spin}(t)\simeq \frac{C_e}{f_a}\,m_a a_0\,\hat{\mathbf s}\!\cdot\!\hat{\mathbf n}\,\cos(m_a t)
= \frac{C_e}{f_a}\sqrt{2\rho_{\rm DM}}\,\hat{\mathbf s}\!\cdot\!\hat{\mathbf n}\,\cos(m_a t),
\]
with the coherence bandwidth set by \eqref{eq:tc-correct}; Zeeman/hyperfine-sensitive transitions inherit a directional template (daily/annual sidebands). See Appendix~\ref{app:ULDM}.

\subsection{Vectors (dark photon or $B\!-\!L$).}

A massive $A'_\mu$ with kinetic mixing $\varepsilon$ and/or $g_{B-L}$ yields a coherent background with $|\mathbf{E}'_0|=\sqrt{2\rho_{\rm DM}}$, independent of $m_{A'}$. Two channels are clean: (a) \textit{oscillatory EEP (AIS)}, giving $\eta^{\rm osc}$ at $m_{A'}$ and complementing the static guardrail \eqref{eq:eta-AIS}; and (b) \textit{Zeeman/magnetometer anisotropy} via the lab‑frame EM response induced by kinetic mixing (Appendix~\ref{app:ULDM}).

\subsection{Topological defects.}
\label{sec:top-def}

A wall/string of thickness $\ell\!\sim\!(m_\phi v)^{-1}$ produces
\[
\left.\frac{\delta\nu}{\nu}\right|_{\rm defect}\!\simeq\! \sum_i K_i\!\left(d_i\,\Delta\phi+\tfrac{1}{2}d_i^{(2)}\,\Delta\phi^2\right),
\]
with crossing time $\tau_{\rm cross}\!\sim\!\ell/v$ and a correlated time‑of‑arrival across a spatially separated network.

\subsection{Gravity-only (“metric”) ULDM.}

Even if the field has \emph{no direct} couplings to SM operators ($d_i\!=\!d_i^{(2)}\!=\!0$), its oscillating stress--energy sources a coherent, universal gravitational potential modulation at $2m_\phi$, $\delta\Phi(t)$, which imprints on separated clocks as a redshift,
\[
\left.\frac{\delta\nu}{\nu}\right|_{\rm grav}(t)\simeq \frac{\delta\Phi(t)}{c^2},\qquad \delta\Phi(t)\sim \mathcal{O} \Big(\frac{4\pi G\,\rho_{\rm DM}}{m_\phi^2}\Big)\cos(2m_\phi t),
\]
favoring low $m_\phi$ through the $m_\phi^{-2}$ scaling. The detection template is the same narrowband carrier with coherence from \eqref{eq:tc-correct}; the observable is a \emph{differential} redshift between separated stations with known tidal response (Appendix~\ref{app:ULDM}). This channel is EEP\,-preserving and complementary to \eqref{eq:clock-dm}. 

\subsection{Higgs-portal scalars (predictive coupling pattern).}
If the light scalar mixes with the Higgs, the low‑energy couplings to fermion masses are \emph{aligned} and proportional to mass, inducing a fixed pattern among $\{d_{m_e},d_{m_q},d_g,d_\alpha\}$ rather than independent coefficients. In clock networks this produces \emph{correlated} responses across species and transitions, enabling over‑constrained fits that break degeneracies in $\Delta K$ combinations (Appendix~\ref{app:ULDM}). Practically, with the same link stability and $t_c$ from \eqref{eq:tc-correct}, the SNR scaling is that of \eqref{eq:clock-dm} with $d_{\rm eff}$ replaced by the one‑parameter Higgs‑pattern.

\subsection{Spin‑2 ULDM (coherent tidal fields).}
A light spin‑2 field behaves as a coherent, quadrupolar tidal background oscillating at $m_2$. For two stations separated by baseline $\mathbf{L}$ with line‑of‑sight unit vector $\hat{\mathbf n}$, the leading clock redshift/tidal phase carries a quadrupolar angular pattern $\propto \hat{n}_i \hat{n}_j T_{ij}(t)$ with $T_{ij}$ the (traceless) tidal tensor set by the local energy density. The resulting narrowband signal is again coherence‑limited by \eqref{eq:tc-correct} and is best extracted by \emph{networks} (Appendix~\ref{app:ULDM}).

\subsection{Substructure (soliton/minicluster) transits.}
ULDM substructure enhances $\rho_{\rm DM}$ over a crossing time $\tau_{\rm cross}\!\sim\!R_{\rm sub}/v$ (days--months, depending on model), producing \emph{longer} transients than thin defects. Network geometry yields correlated, staggered arrivals; matched banks generalize the defect templates in Appendix~\ref{app:ULDM}. 

\subsection{Near-term impact.}

Tables~\ref{tab:benchmarks} and \ref{tab:systematics-quant} summarize the near-term impacts of the models above.
Using the same stability/link targets and the coherence law \eqref{eq:tc-correct}, the coupling reaches scale as
\[
|d_{\rm eff}|\ \lesssim\ \frac{\sigma_y}{|\Delta K|\,\phi_0}\sqrt{\frac{t_c}{T}},\qquad 
|d^{(2)}_{\rm eff}|\ \lesssim\ \frac{2\,\sigma_y}{|\Delta K|}\,\frac{m_\phi^2}{\rho_{\rm DM}}\sqrt{\frac{t_c}{T}},
\]
for scalars; for $B\!-\!L$ vectors in AIS,
\[
g_{B-L}\ \lesssim\ \frac{g\,S_a^{1/2}}{\Delta(Q_{B-L}/M)\,\sqrt{2\rho_{\rm DM}}}\sqrt{\frac{t_c}{T}}.
\]
For the gravity‑only channel, the redshift amplitude entering the same coherence/stacking logic is $A_{\rm grav}\!\equiv\!|\delta\Phi|/c^2 \sim (4\pi G\rho_{\rm DM}/m_\phi^2)/c^2$, emphasizing improvements at low $m_\phi$. For spin‑2 ULDM and substructure transits, the SNR scalings in Table~\ref{tab:ULDM-landscape} apply. Overall, the programmatic improvements collected in Tables~\ref{tab:benchmarks} and \ref{tab:systematics-quant} propagate to $(3$--$10)\times$ gains across $m\!\sim\!10^{-24}$--$10^{-15}$~eV in clock/AIS channels, while AU‑scale reprocessing tied to \eqref{eq:yukawa} and Table~\ref{tab:mass-range} tightens long‑range tails in the $10^9$--$10^{13}$\,m window.

\begin{table*}[t]
\setlength{\tabcolsep}{4pt}
\renewcommand{\arraystretch}{1.00}
\caption{{ULDM models, couplings, and Solar System observables.}
Notation follows Sec.~\ref{sec:DM-clocks}. The ULDM coherence time $t_c$ and narrowband build‑up follow \eqref{eq:tc-correct};
for scalar clocks and AIS the baseline response is \eqref{eq:clock-dm}; AIS composition dependence maps through \eqref{eq:eta-AIS}.
Carrier mass--frequency conversion is summarized in Fig.~\ref{fig:solar-system-tests}(d). SNR expressions assume coherent
matching within a coherence bin ($\tau\!\lesssim\!t_c$) and incoherent stacking over $T/t_c$. Here $\phi_0=\sqrt{2\rho_{\rm DM}}/m_\phi$,
$\sigma_y$ is the Allan deviation for the averaging time used per bin, $S_\omega^{1/2}$ the spin‑frequency ASD, $S_a^{1/2}$
the differential‑acceleration ASD, and $\Delta K$ the relevant clock/AI sensitivity combination.}
\label{tab:ULDM-landscape}
\centering
\begin{tabular}{p{3.0cm}p{3.0cm}p{5.3cm}p{5.3cm}}
\hline\hline
{Model \& coupling} & {Primary channel(s)} & {Signal template (carrier/sidebands)} & {SNR scaling (schematic)} \\
\hline
Scalar, linear $d_i$ \cite{Arvanitaki2015PRD,GrahamRajendran2013} &
Clocks (\ref{eq:clock-dm}); AI &
$\delta\nu/\nu=\sum_i K_i d_i \phi_0\cos(m_\phi t)$; line at $m_\phi$, coherence $Q\!\sim\!1/v^2$;
daily/annual sidebands on moving baselines &
$\displaystyle {\rm SNR}\!\sim\!\frac{|\Delta K\,d_{\rm eff}|\,\phi_0}{\sigma_y}\sqrt{\frac{T}{t_c}}$ \\[2pt]

Scalar, quadratic $d^{(2)}_i$ \cite{Arvanitaki2015PRD} &
Clocks; AI &
DC offset $+\,$line at $2m_\phi$ with amplitude $\simeq \frac{1}{4}|\sum_i K_i d^{(2)}_i|\,\phi_0^2$ &
$\displaystyle {\rm SNR}_{2f}\!\sim\!\frac{|\Delta K\,d^{(2)}_{\rm eff}|\,\rho_{\rm DM}/m_\phi^2}{\sigma_y}\sqrt{\frac{T}{t_c}}$ \\[2pt]

Pseudoscalar ($a$), axial $C_e/f_a$ \cite{Budker2014} &
Co‑magnetometers; Zeeman‑sensitive clocks &
Spin‑precession modulation at $m_a$: $\delta\omega_{\rm spin}\!\propto\!(C_e/f_a)\sqrt{2\rho_{\rm DM}}\cos(m_a t)$; directional daily/annual sidebands &
$\displaystyle {\rm SNR}\!\sim\!\frac{|C_e|}{f_a}\frac{\sqrt{2\rho_{\rm DM}}}{S_\omega^{1/2}}\sqrt{\frac{T}{t_c}}$ \\[2pt]

Vector, kinetic mixing $\varepsilon$ \cite{Holdom1986,Arias2012} &
Zeeman clocks; magnetometers &
EM‑like field at $m_{A'}$ with $|\mathbf{E}'_0|\!=\!\sqrt{2\rho_{\rm DM}}$; daily/annual sidebands from lab motion &
$\displaystyle {\rm SNR}\!\sim\!\varepsilon\,\frac{\sqrt{2\rho_{\rm DM}}}{S_\omega^{1/2}}\sqrt{\frac{T}{t_c}}$ \\[2pt]

Vector, gauged $B\!-\!L$ ($g_{B-L}$) \cite{Arias2012} &
AIS (oscillatory EEP via \eqref{eq:eta-AIS}) &
$\eta^{\rm osc}(t) \simeq \frac{g_{B-L}}{g}\Delta(Q_{B-L}/M)\,|\mathbf{E}'_0|\cos(m_{A'} t)$ &
$\displaystyle {\rm SNR}\!\sim\!\frac{g_{B-L}}{g}\,\Delta\Big(\frac{Q_{B-L}}{M}\Big)\frac{\sqrt{2\rho_{\rm DM}}}{S_a^{1/2}}\sqrt{\frac{T}{t_c}}$ \\[2pt]

Topological defects (walls/strings) \cite{DereviankoPospelov2014} &
Clock network; AIS array &
Transient step/pulse; crossing time $\tau_{\rm cross}\!\sim\!(m_\phi v)^{-1}\!/v$; correlated TOA across stations &
$\displaystyle {\rm SNR}\!\sim\!\frac{(\Delta\nu/\nu)_{\rm step}}{\sigma_y}\sqrt{N_{\rm stat}}$ (matched transient bank) \\[2pt]
\hline
{Gravity‑only (“metric”) ULDM} \cite{KhmelnitskyRubakov2014} &
Separated clocks (redshift) &
Universal potential modulation at $2m_\phi$: $\delta\nu/\nu\simeq \delta\Phi/c^2$, with $|\delta\Phi|\!\sim\!4\pi G\rho_{\rm DM}/m_\phi^2$ &
$\displaystyle {\rm SNR}\!\sim\!\frac{A_{\rm grav}}{\sigma_y}\sqrt{\frac{T}{t_c}},\ \
A_{\rm grav}\!\equiv\!\frac{|\delta\Phi|}{c^2}$ \\[2pt]

{Higgs‑portal scalar (aligned $d_i$)} \cite{StadnikFlambaum2015HiggsPortal,Arvanitaki2015PRD,DamourDonoghue2010a} &
Multi‑species clocks &
One‑parameter, correlated pattern across $d_{m_e},d_{m_q},d_g,d_\alpha$ at $m_\phi$; internal consistency across species breaks degeneracies &
$\displaystyle {\rm SNR}\!\sim\!\frac{|\Delta K\,d_H|\,\phi_0}{\sigma_y}\sqrt{\frac{T}{t_c}}$ (global 1‑parameter fit) \\[2pt]

{Spin‑2 ULDM (tidal)} &
Clock network; gradiometers &
Quadrupolar tidal tensor $T_{ij}(t)$ at $m_2$; redshift $\propto \hat n_i\hat n_j T_{ij}$; network angular weighting &
$\displaystyle {\rm SNR}\!\sim\!\frac{A_2}{\sigma_y}\sqrt{\frac{T}{t_c}}$ (network mode separation) \\[2pt]

{Substructure transits (soliton/minicluster)} \cite{Centers2021ClockDMNetwork} &
Clocks/AIS network &
Longer transient with density enhancement $\rho_{\rm sub}\!\gg\!\rho_{\rm DM}$ over $\tau_{\rm cross}\!\sim\!R_{\rm sub}/v$; correlated/staggered arrivals &
$\displaystyle {\rm SNR}\!\sim\!\frac{(\Delta\nu/\nu)_{\rm sub}}{\sigma_y}\sqrt{N_{\rm stat}}$ (matched long‑transient bank) \\
\hline
\end{tabular}
\end{table*}

\section{ULDM model space and detailed mapping to observables}
\label{app:ULDM}

We collect formulas supporting Sec.~\ref{sec:DM-clocks} and Table~\ref{tab:ULDM-landscape}, using the clock/AI normalization \eqref{eq:clock-dm}--\eqref{eq:tc-correct}. We adopt natural units $c=\hbar=1$, and use $\rho_{\rm DM}\simeq 0.3~{\rm GeV\,cm^{-3}}$ for estimates, and the coherence law \eqref{eq:tc-correct}.
 When comparing with Secs.~\ref{sec:observables}--\ref{sec:program}, restore factors of $c$ and $\hbar$ using standard dimensional analysis.

\subsection{
Scalars with linear and quadratic couplings}
With $\phi(t)=\phi_0\cos(m_\phi t)$ and $\phi_0=\sqrt{2\rho_{\rm DM}}/m_\phi$, the fractional shift generalizing \eqref{eq:clock-dm} is
\begin{align}
\frac{\delta\nu}{\nu}(t) &= \sum_i K_i\!\left(d_i\,\phi(t) + \tfrac{1}{2} d^{(2)}_i\,\phi^2(t)\right), \label{eq:E-scalar-start}\\
&= \underbrace{\tfrac{1}{4}\Big(\sum_i K_i d^{(2)}_i\Big)\phi_0^2}_{\text{DC}}
+ \underbrace{\Big(\sum_i K_i d_i\Big)\phi_0\cos m_\phi t}_{m_\phi}
+ \underbrace{\tfrac{1}{4}\Big(\sum_i K_i d^{(2)}_i\Big)\phi_0^2\cos 2m_\phi t}_{2m_\phi}. \label{eq:E-scalar-lines}
\end{align}
For a matched-filter search, per-bin SNR and incoherent stacking across $T/t_c$ give
\begin{equation}
{\rm SNR}_{f}\simeq \frac{A_f}{\sigma_y}\sqrt{\frac{T}{t_c}},
\quad 
A_{m_\phi}=\Big|\sum_i K_i d_i\Big|\phi_0,\quad 
A_{2m_\phi}=\tfrac14\Big|\sum_i K_i d^{(2)}_i\Big|\phi_0^2. 
\label{eq:E-snr}
\end{equation}

\subsection{
Pseudoscalars (axion-like)}
Low-energy couplings of $a$,
\begin{equation}
\mathcal{L}\supset \frac{C_\gamma}{4f_a}aF\tilde F + \frac{C_g}{f_a}aG\tilde G + \frac{C_e}{2f_a}\partial_\mu a\,\bar e\gamma^\mu\gamma^5 e,
\label{eq:E-axion-L}
\end{equation}
induce a spin-precession modulation
\begin{equation}
\delta\omega_{\rm spin}(t) \simeq \frac{C_e}{f_a}\,m_a a_0\,\hat{\mathbf s}\!\cdot\!\hat{\mathbf n}\,\cos(m_a t)
= \frac{C_e}{f_a}\sqrt{2\rho_{\rm DM}}\,\hat{\mathbf s}\!\cdot\!\hat{\mathbf n}\,\cos(m_a t),
\label{eq:E-axion-spin}
\end{equation}
with relative bandwidth set by \eqref{eq:tc-correct}. Zeeman clock templates follow by inserting \eqref{eq:E-axion-spin} in the hyperfine/Zeeman response.

\subsection{
Vectors (dark photon, $B\!-\!L$)}
For a massive vector $A'_\mu$ with kinetic mixing $\varepsilon$ and/or $B\!-\!L$ coupling $g_{B-L}$,
\begin{equation}
\mathcal{L}\supset -\tfrac{1}{4}F'_{\mu\nu}F'^{\mu\nu}+\tfrac{1}{2}m_{A'}^2A'^2 + \tfrac{\varepsilon}{2}F'_{\mu\nu}F^{\mu\nu} + g_{B-L}A'_\mu J_{B-L}^\mu.
\label{eq:E-vector-L}
\end{equation}
In a virialized halo, $A'_i(t)=A'_{0,i}\cos(m_{A'}t)$ so the effective dark electric field is
\begin{equation}
\mathbf{E}'(t)=-\dot{\mathbf{A}}' = m_{A'}\mathbf{A}'_0\sin(m_{A'}t),\qquad |\mathbf{E}'_0|=\sqrt{2\rho_{\rm DM}}.
\label{eq:E-Edp}
\end{equation}

\paragraph*{Oscillatory EEP (AIS):}
A body with $B\!-\!L$ charge $Q_{B-L}$ and mass $M$ experiences the following acceleration $a_{B-L}(t)= (g_{B-L}Q_{B-L}/M)\,|\mathbf{E}'_0|\cos(m_{A'} t)$; the Eötvös parameter for $A,B$ then reads
\begin{equation}
\eta_{A,B}^{\rm osc}(t)
\simeq \frac{g_{B-L}}{g}\Big(\frac{Q_{B-L}}{M}\Big|_A - \frac{Q_{B-L}}{M}\Big|_B\Big) |\mathbf{E}'_0|\cos(m_{A'} t),
\label{eq:E-eta-osc}
\end{equation}
to be compared with the static AIS mapping \eqref{eq:eta-AIS}. The same $\sqrt{T/t_c}$ stacking applies with $t_c$ from \eqref{eq:tc-correct}.

\paragraph*{Clock/magnetometer channel:}
Kinetic mixing induces lab-frame EM fields at $m_{A'}$ that modulate Zeeman transitions with daily/annual sidebands; the narrowband carrier follows \eqref{eq:E-Edp}.

\subsection{
Transient (topological-defect) ULDM}

For a domain wall or string of thickness $\ell\sim (m_\phi v)^{-1}$ and field excursion $\Delta\phi$, a station sees
\begin{equation}
\left.\frac{\delta\nu}{\nu}\right|_{\rm defect}
\simeq \sum_i K_i\!\left(d_i\,\Delta\phi + \tfrac{1}{2}d^{(2)}_i\,\Delta\phi^2\right),\qquad \tau_{\rm cross}\sim \frac{\ell}{v},
\label{eq:E-defect}
\end{equation}
and a spatially correlated arrival pattern across a network; matched-transient banks yield $\mathrm{SNR}\sim (\Delta\nu/\nu)\,(\sqrt{N_{\rm stat}}/\sigma_y)$.

\subsection{
Scaling of near-term bounds}

Using \eqref{eq:tc-correct} and the system targets in Table~\ref{tab:systematics-quant} one obtains
\begin{align}
|d_{\rm eff}| &\lesssim 
\frac{\sigma_y}{|\Delta K|\,\phi_0}\sqrt{\frac{t_c}{T}}
= \frac{\sigma_y}{|\Delta K|}\sqrt{\frac{m_\phi}{2\rho_{\rm DM}}}\sqrt{\frac{t_c}{T}}, \label{eq:E-d-scaling}\\
|d^{(2)}_{\rm eff}| &\lesssim \frac{2\,\sigma_y}{|\Delta K|}\,\frac{m_\phi^2}{\rho_{\rm DM}}\sqrt{\frac{t_c}{T}}, \label{eq:E-d2-scaling}\\
g_{B-L} &\lesssim
\frac{g\,S_a^{1/2}}{\Delta(Q_{B-L}/M)\,\sqrt{2\rho_{\rm DM}}}\,\sqrt{\frac{t_c}{T}}, \label{eq:E-gBL-scaling}
\end{align}
which are the relations quoted in Sec.~\ref{sec:DM-clocks}.

\subsection{
Gravity-only (“metric”) ULDM}

Even with $d_i\!=\!d_i^{(2)}\!=\!0$ in \eqref{eq:clock-dm}, a coherent bosonic field sources an oscillating stress--energy that drives a universal gravitational potential modulation at $2m_\phi$,
\begin{equation}
\left.\frac{\delta\nu}{\nu}\right|_{\rm grav}(t)\simeq \frac{\delta\Phi(t)}{c^2},\qquad 
\delta\Phi(t)\sim \frac{4\pi G\,\rho_{\rm DM}}{m_\phi^2}\cos(2m_\phi t),
\label{eq:E-metric}
\end{equation}
which is read by \emph{separated} clocks as a differential redshift. The carrier is narrowband with coherence set by \eqref{eq:tc-correct}; the per‑bin SNR follows ${\rm SNR}\!\sim\!(A_{\rm grav}/\sigma_y)\sqrt{T/t_c}$ with $A_{\rm grav}\!=\!|\delta\Phi|/c^2$ and network baselines entering the geometrical response.

\subsection{
Higgs-portal scalars}

For a light scalar mixing with the Higgs, low‑energy couplings align with fermion masses (and induce correlated gluon/EM coefficients via loops), producing a constrained pattern among $\{d_{m_e},d_{m_q},d_g,d_\alpha\}$. In the language of \eqref{eq:clock-dm},
\begin{equation}
\frac{\delta\nu}{\nu}(t) \;=\; \Big[\Delta K_{m}\,d_{H}^{(m)} + \Delta K_{g}\,d_{H}^{(g)} + \Delta K_{\alpha}\,d_{H}^{(\alpha)}\Big]\phi(t),
\label{eq:E-Hportal}
\end{equation}
with the three $d_{H}^{(X)}$ tied to a \emph{single} portal parameter (mixing angle or effective scale). Hence multi‑species comparisons overconstrain \eqref{eq:E-Hportal}, enabling powerful internal consistency tests. The SNR scaling follows the linear‑scalar case with $d_{\rm eff}\!\to\!d_{H}$.

\subsection{
Spin-2 ULDM}

A coherent spin‑2 background yields an oscillating, traceless tidal tensor $T_{ij}(t)$; a two‑clock link along $\hat{\mathbf n}$ accumulates a differential redshift $\propto \hat{n}_i\hat{n}_j T_{ij}(t)$,
\begin{equation}
\frac{\delta\nu}{\nu}\bigg|_{2}(t)\;=\; A_{2}\, \hat{n}_i\hat{n}_j\,\mathcal{Q}_{ij}\cos(m_2 t) \qquad\text{with}\qquad 
A_{2}\propto \frac{\rho_{\rm DM}}{M_*^2\,m_2^2},
\label{eq:E-spin2}
\end{equation}
where $\mathcal{Q}_{ij}$ encodes the polarization content and $M_*$ the (model‑dependent) coupling scale. The response carries a characteristic quadrupolar angular pattern; network geometry is advantageous for mode separation.

\subsection{
Substructure (soliton/minicluster) transits}

For a transient overdensity $\rho(t)$ of size $R_{\rm sub}$ and relative speed $v$, the enhancement over a duration $\tau_{\rm cross}\!\sim\!R_{\rm sub}/v$ boosts the amplitudes in \eqref{eq:clock-dm} and \eqref{eq:E-metric} by $\rho\!\to\!\rho_{\rm sub}$. The network observable is a long‑duration, band‑limited transient with correlated, staggered arrivals; matched filters generalize the defect templates (Sec.~\ref{sec:top-def}), retaining the $\sqrt{N_{\rm stat}}$ gain for $N_{\rm stat}$ stations. Composition dependence (if present) can be separated from universal (metric) components via species/baseline diversity.

\end{document}